\documentclass[11pt,a4paper]{article}
\usepackage{hyperref}
\usepackage[numbers, square, comma, sort&compress]{natbib}
\usepackage{amssymb}
\usepackage{amsthm}
\usepackage{amstext}
\usepackage{amsmath}
\usepackage{amsfonts}
\usepackage{bbold}
\usepackage{empheq}
\usepackage{color}
\usepackage{xcolor}

\usepackage{verbatim}
\usepackage{geometry}
\usepackage{latexsym}
\usepackage{t1enc}
\usepackage{graphicx}
\usepackage[all]{xy}
\usepackage{makeidx}
\usepackage{slashed}
\usepackage{multicol}
\usepackage{alltt}
\usepackage{multirow}

\allowdisplaybreaks

\newcommand{\be}{\begin{equation}}
\newcommand{\ee}{\end{equation}}
\newcommand{\bea}{\begin{eqnarray}}
\newcommand{\eea}{\end{eqnarray}}
\newcommand{\eps}{\epsilon}

\newcommand{\e}{\epsilon}

\newcommand{\cFi}[3]{{\cal F}_{(#1);#2}^{#3}}

\def\beq{\begin{equation}}
\def\eeq{\end{equation}}
\def\beqa{\begin{eqnarray}}
\def\eeqa{\end{eqnarray}}

\def\eqn#1{eq.~(\ref{#1})}

\def\spa#1.#2{\left\langle#1\,#2\right\rangle}
\def\spb#1.#2{\left[#1\,#2\right]}
\def\spash#1.#2{\spa{\smash{#1}}.{\smash{#2}}}
\def\spbsh#1.#2{\spb{\smash{#1}}.{\smash{#2}}}
\def\sand#1.#2.#3{%
  \left\langle\smash{#1^{-}}{\vphantom1}\right|{#2}%
  \left|\smash{#3^{-}}{\vphantom1}\right\rangle}
\def\sandp#1.#2.#3{%
  \left\langle\smash{#1^{-}}{\vphantom1}\right|{#2}%
  \left|\smash{#3^{+}}{\vphantom1}\right\rangle}
\def\sandpp#1.#2.#3{%
  \left\langle\smash{#1^{+}}{\vphantom1}\right|{#2}%
  \left|\smash{#3^{+}}{\vphantom1}\right\rangle}
\def\sandpm#1.#2.#3{%
  \left\langle\smash{#1^{+}}{\vphantom1}\right|{#2}%
  \left|\smash{#3^{-}}{\vphantom1}\right\rangle}
\def\sandmp#1.#2.#3{%
  \left\langle\smash{#1^{-}}{\vphantom1}\right|{#2}%
  \left|\smash{#3^{+}}{\vphantom1}\right\rangle}

\def\ssand#1.#2.#3{%
  \left\langle\smash{#1}{\vphantom1}\right|{#2}%
  \left|\smash{#3}{\vphantom1}\right]}
\def\ssandp#1.#2.#3{%
  \left\langle\smash{#1}{\vphantom1}\right|{#2}%
  \left|\smash{#3}{\vphantom1}\right\rangle}
\def\ssandpp#1.#2.#3{%
  \left\langle\smash{#1}{\vphantom1}\right|{#2}%
  \left|\smash{#3}{\vphantom1}\right\rangle}

\def\proj{\flat}
\def\projdot#1.#2{k_{#1}^\proj\cdot k_{#2}^\proj}
\def\sandff#1.#2.#3{%
  \left\langle\smash{#1^{\proj,-}}{\vphantom1}\right|{#2}%
  \left|\smash{#3^{\proj,-}}{\vphantom1}\right\rangle}
\def\sandnf#1.#2.#3{%
  \left\langle\smash{#1^{-}}{\vphantom1}\right|{#2}%
  \left|\smash{#3^{\proj,-}}{\vphantom1}\right\rangle}
\def\sandfn#1.#2.#3{%
  \left\langle\smash{#1^{\proj,-}}{\vphantom1}\right|{#2}%
  \left|\smash{#3^{-}}{\vphantom1}\right\rangle}

\def\spa#1.#2{\left\langle#1\,#2\right\rangle}
\def\spb#1.#2{\left[#1\,#2\right]}

\numberwithin{equation}{section}

\begin{document}

\begin{titlepage}

\begin{center}
\Large{\sc{Boomerang webs up to three-loop order}}
\end{center}

\vskip 8mm

\begin{center}

Einan~Gardi$^a$\footnote{einan.gardi@ed.ac.uk},
Mark Harley,
Rebecca Lodin$^b$\footnote{rebecca.lodin@physics.uu.se},
Martina Palusa$^a$\footnote{martina.palusa@ed.ac.uk},
Jennifer M.~Smillie$^a$\footnote{j.m.smillie@ed.ac.uk},
Chris~D.~White$^c$\footnote{christopher.white@qmul.ac.uk}
and Stephanie Yeomans  \\ [6mm]

\vspace{6mm}

\textit{$^a$ Higgs Centre for Theoretical Physics, School of Physics and Astronomy, \\
The University of Edinburgh, Edinburgh EH9 3FD, Scotland, UK}\\
\vspace{1mm}

\textit{$^b$ Department of Physics and Astronomy, Uppsala University,\\ Box 516, SE-75120 Uppsala, Sweden}\\
\vspace{1mm}

\textit{$^c$Centre for Theoretical Physics, School of Physical and Chemical Sciences, \\
Queen Mary University of London, 327 Mile End Road, London E1 4NS, UK} \\
\vspace{1mm}

\end{center}

\vspace{5mm}

\begin{abstract}
\noindent Webs are sets of Feynman diagrams which manifest soft gluon exponentiation in gauge theory scattering amplitudes: individual webs contribute to the logarithm of the amplitude and their ultraviolet renormalization encodes its infrared structure. In this paper, we consider the particular class of {\it boomerang webs}, consisting
of multiple gluon exchanges, but where at least one gluon has both
of its endpoints on the same Wilson line.
First, we use the replica trick to prove that diagrams involving self-energy insertions along the Wilson line do not contribute to the web, i.e. their exponentiated colour factor vanishes. Consequently boomerang webs effectively involve only integrals where boomerang gluons straddle one or more gluons that connect to other Wilson lines.
Next we classify and calculate
all boomerang webs involving semi-infinite non-lightlike Wilson lines up to three-loop order, including a detailed discussion of how to regulate and renormalize them. Furthermore, we show that they can be written using a basis of specific harmonic polylogarithms, that has been
conjectured to be sufficient for expressing all multiple gluon exchange webs. However, boomerang webs  differ from other gluon-exchange webs by featuring a lower and non-uniform transcendental weight.
We cross-check our results by showing
how certain boomerang webs can be determined by the so-called {\it
  collinear reduction} of previously calculated webs.
Our results are a necessary ingredient of the soft anomalous dimension for non-lightlike Wilson lines at three loops.
\end{abstract}

\end{titlepage}

\tableofcontents

\section{Introduction}
\label{sec:introduction}

\noindent
%\Einan{Bare coupling vs renormalized coupling}\\
%\noindent
%\Einan{Feynman gauge and gauge dependence}

The structure of perturbative scattering amplitudes in non-Abelian
gauge theories continues to be an important research area due to a wide range of phenomenological and formal applications. Of particular
interest are those universal quantities in field theory that govern
the all-order behaviour of amplitudes. One such quantity is the {\it
  soft anomalous dimension}, which controls the long-distance singularities of on-shell form factors and amplitudes. These singularities give rise to logarithms of kinematic invariants in perturbative cross-sections, which reflect incomplete cancellation between real and virtual correction, and dominate the perturbative expansion in many instances.

The soft anomalous dimension can also be determined from 
ultraviolet renormalization properties of correlators of Wilson-line
operators~\cite{Polyakov:1980ca,Arefeva:1980zd,Dotsenko:1979wb,Brandt:1981kf,Korchemsky:1985xj,Korchemsky:1985xu,Korchemsky:1987wg}.
In calculating it one must make a distinction between the colour singlet case, relevant for example for an on-shell form factor, where the singularity structure is known in full to three loops (in particular the angle-dependent cusp anomalous dimension was computed to three loops in QCD in~\cite{Grozin:2015kna,Grozin:2014hna} and to four loops in QED in \cite{Bruser:2020bsh}), and the more complicated case of multi-leg scattering amplitudes, which is of interest here, where the soft anomalous dimension is matrix-valued in the space of possible colour flows
in the underlying hard process.
One must make a further distinction between lightlike Wilson lines, corresponding to the scattering of massless particles, as discussed for example in~\cite{Korchemsky:1993hr,Korchemskaya:1996je,Korchemskaya:1994qp,Catani:1996vz,Catani:1998bh,Sterman:2002qn,Dixon:2008gr,Kidonakis:1998nf,Bonciani:2003nt,Dokshitzer:2005ig,Aybat:2006mz,Gardi:2009qi,Becher:2009cu,Becher:2009qa,Gardi:2009zv,
  Dixon:2009gx,Dixon:2009ur,DelDuca:2011xm,DelDuca:2011ae,Caron-Huot:2013fea,Ahrens:2012qz,Naculich:2013xa,Erdogan:2014gha,Gehrmann:2010ue,Agarwal:2021zft} and non-lightlike Wilson lines, corresponding to the scattering of heavy (coloured) particles, such as top quarks, see e.g. refs.~\cite{Kidonakis:2009ev,Mitov:2009sv,Becher:2009kw,Beneke:2009rj,Czakon:2009zw,Ferroglia:2009ii,Chiu:2009mg,Mitov:2010xw,Gardi:2013saa,Falcioni:2014pka,Henn:2013fah}.
In massless scattering, the soft anomalous dimension is highly constrained~\cite{Gardi:2009qi,Becher:2009cu,Becher:2009qa,Gardi:2009zv} and it was computed in full at three-loop
order~\cite{Almelid:2015jia,Gardi:2016ttq}. Furthermore, it was shown~\cite{Almelid:2017qju} that its precise form can be deduced from general considerations and special kinematic limits. These considerations do not apply
directly to the massive case, and so the state-of-the-art knowledge of this quantity remains two loops~\cite{Becher:2009kw}. 
While specific three-loop contributions have been directly computed in refs.~\cite{Gardi:2013saa,Falcioni:2014pka}, a complete calculation is beyond the reach of present methods. In this paper, we continue the calculation of the three-loop massive soft
anomalous dimension, by focusing on a particular class of
contributions that have not been previously obtained.

A particularly convenient language for organising calculations
involving multiple Wilson lines is that of {\it webs}, first developed
in the classic work of
refs.~\cite{Gatheral:1983cz,Frenkel:1984pz,Sterman:1981jc} for the
two-line case. The starting point for this formalism is the fact that
vacuum expectation values of Wilson lines are known to
exponentiate. Crucially, the logarithm of the Wilson-line correlator
can be given a Feynman diagram interpretation by itself, where the
term ``webs'' refers to the relevant diagrams. In the two-line case in
QCD, webs can be conveniently characterised by the fact that they are
two-particle irreducible. Furthermore, their colour factors are modified in the
logarithm of the amplitude, such that all colour factors have the
property of being maximally non-Abelian, i.e. akin to the colour
factors of fully connected gluon graphs. Perhaps unsurprisingly,
things are more complicated in the multiparton case, and a number of
formalisms have been
developed~\cite{Mitov:2010rp,Laenen:2008gt,Gardi:2010rn,Gardi:2011wa,Gardi:2011yz,Gardi:2013ita,Falcioni:2014pka,Vladimirov:2014wga,Vladimirov:2015fea,Vladimirov:2016qkd,Vladimirov:2016dll,Vladimirov:2017ksc}. Here
we will adopt the approach originated in ref.~\cite{Gardi:2010rn} (see
also~\cite{White:2015wha} for a review and references~\cite{Agarwal:2020nyc,Agarwal:2021him} for recent progress beyond three loops), in which webs are closed sets
of diagrams related by permutations of gluon attachments on the Wilson
lines. Each such web is associated with a {\it web mixing matrix}
describing how the colour and kinematic degrees of freedom are
entangled in the logarithm of the amplitude. These matrices have a
combinatorial definition that has been studied from a purely
mathematical point of
view~\cite{Dukes:2013wa,Dukes:2013gea,Dukes:2016ger}, but in this
paper simply provide a convenient way to organise the combination of
different Feynman diagrams. The renormalization of multiparton webs
has been spelled out in ref.~\cite{Gardi:2011yz}, and involves
combining diagrams at a given perturbative order with an intricate set
of lower-order information. Furthermore, it is known that only certain
combinations of diagrams survive in the logarithm of the amplitude,
where each is accompanied by a fully connected colour
factor~\cite{Gardi:2013ita}, in direct analogy with the two-parton
case.

Previously calculated three-loop webs involving massive lines include
the broad class of {\it multiple gluon exchange webs (MGEWs)}, defined
such that the Wilson lines are connected by multiple gluon
emissions, with no three- or four-gluon vertices located off the
Wilson lines. Such diagrams involving four lines (the maximal number
that can be connected at this order) were calculated in
ref.~\cite{Gardi:2013saa}. Those involving three lines were calculated
in refs.~\cite{Falcioni:2014pka}, where an interesting relationship
with previous results was developed. Namely, it is possible to
generate parts of webs connecting $n-1$ Wilson lines from those
connecting $n$ lines, by taking two lines in a given $n$-line web to
be collinear. The procedure can then be iterated to generate parts of
webs with even fewer lines, and was dubbed {\it collinear
  reduction} in ref.~\cite{Falcioni:2014pka}. It provides a highly
nontrivial and useful consistency check of higher-loop computations,
and we will encounter this idea in what follows.

References~\cite{Gardi:2013saa,Falcioni:2014pka} initiated an ongoing
programme of work, to calculate all relevant diagrams for the massive
three-loop soft anomalous dimension. Our aim in this study is to
consider the next natural class of diagrams, namely MGEWs
in which at least one gluon has both its endpoints on the {\it same} Wilson
line. We shall refer to such gluons as {\it boomerang gluons}, and to
the corresponding sets of diagrams containing them as {\it boomerang
  webs}. These were not considered in the above three-loop references,
as they present additional complications related to the presence of
ultraviolet divergences when the ends of a gluon meet at the same
spacetime point (possibly with another gluon in between). Such
complications were already present at lower orders (see
e.g.~\cite{Korchemsky:1987wg} for a non-trivial two-loop example), but
must be reconsidered here. Firstly,
references~\cite{Gardi:2013saa,Falcioni:2014pka} have developed a
regulator that is well-suited to isolating ultraviolet divergences
in the web approach, and we will need to see how to generalise this
regulator to boomerang webs. Secondly, we must account for these
additional ultraviolet divergences within the general scheme developed in
ref.~\cite{Gardi:2011yz} for renormalizing multiparton webs. We will
deal with these issues in the following, and in turn present explicit
results for all boomerang webs up to three-loop order.

Our final expressions form an important contribution to the three-loop
soft anomalous dimension. In addition, we will also see a number of
interesting results along the way. In particular, a large
class of individual diagrams entering boomerang webs -- namely those
containing self-energy insertions alongside gluons which straddle multiple Wilson lines -- can be proven not to appear at all, at any order in the logarithm of the Wilson-line correlator.
Consequently boomerang webs spanning two or more Wilson lines effectively involve only integrals where boomerang gluons straddle one or more gluons that connect to other Wilson lines.
Of course, this greatly reduces the number of integrals that need to be evaluated and simplifies the work required to assemble all contributions. Another
important feature is that our final results can be written in terms
of a special class of basis functions that have appeared already for MGEWs connecting four lines or
fewer~\cite{Gardi:2013saa,Falcioni:2014pka}, and that have been conjectured to hold for MGEWs more generally.
Nevertheless, it is not a priori obvious that this class of functions would be sufficient to express boomerang webs. Indeed, while
for non-boomerang MGEWs all ultraviolet divergences are associated with the renormalization of the multi-Wilson-line vertex, boomerang webs feature other divergences as well. We will see that while the former have a uniform, maximal transcendental weight of $(2n-1)$ at $n$ loops, the latter feature a lower and non-uniform weight. Despite this, we will find that the above-mentioned function basis suffices to express all boomerang webs to three loops, bolstering the expectation that it applies to this class of webs to all orders.

The structure of the paper is as follows. In section~\ref{sec:review},
we review necessary properties regarding the soft anomalous dimension and webs and their renormalization. In
section~\ref{sec:mushroom}, we consider boomerang webs at one- and two-loop order and discuss their regularisation and
renormalization, preparing the grounds for the rest of the paper. In section~\ref{sec:selfenergy} we prove the decoupling of  self-energy contributions from boomerang webs to all orders in perturbation theory. In section~\ref{sec:calculate}, we calculate complete
expressions for all three-loop boomerang webs. In
section~\ref{sec:collinear}, we describe how collinear reduction can
be used to check the consistency of parts of the results of
section~\ref{sec:calculate}. Finally, we discuss our results and
conclude in section~\ref{sec:discuss}. Technical details are contained in six appendices.

\section{The soft anomalous dimension from webs}
\label{sec:review}

In this section, we review salient details regarding the web formalism
that we need for the rest of the paper. We will be brief, referring
the reader to
refs.~\cite{Gardi:2010rn,Gardi:2011yz,Gardi:2013ita,Gardi:2013saa,Falcioni:2014pka}
for more details.

\subsection{Wilson lines and the soft anomalous dimension}
\label{sec:wilson}
Let us first consider a Wilson-line operator associated with a
semi-infinite straight-line contour:
\begin{equation}
\Phi_{\beta_i}\equiv{\cal P}\exp\left\{i g_s\int_0^\infty d\lambda
\beta_i^\mu\,A_\mu(\lambda\,\beta_i)\right\},
\label{Wilson1}
\end{equation}
where $A_\mu$ is the gauge field, ${\cal P}$ denotes
path ordering of colour generators along the Wilson-line contour,
$\lambda$ is a distance parameter, and $\beta_i$ the 4-velocity
tangent to the curve (n.b. throughout, we will be concerned with
non-null Wilson lines). Our aim is to study the vacuum expectation
value of a product of Wilson-line operators, and to examine its
renormalization properties, for which we will use dimensional
regularisation in $d=4-2\epsilon$ dimensions. However, as is
well-known, Feynman diagrams involving Wilson lines vanish in
dimensional regularisation, as scaleless integrals. This can be understood as an exact cancellation between ultraviolet
divergences associated with the vertex (at the origin) at which the Wilson lines meet
and infrared (long-distance) divergences associated with gluons emitted and
absorbed at infinity. To remove the latter, we follow
refs.~\cite{Gardi:2013saa,Falcioni:2014pka} in modifying each Wilson
line, defining instead
\begin{equation}
\Phi_{\beta_i}^{(m)}\equiv{\cal P}\exp\left\{i g_s\int_0^\infty d\lambda
\beta_i^\mu\,A_\mu(\lambda\,\beta_i)
e^{-im\lambda\sqrt{\beta_i^2-i\varepsilon}}
\right\},
\label{Wilson2}
\end{equation}
where $\varepsilon$ is the infinitesimal quantity appearing in the
Feynman $i\varepsilon$ prescription. Here $m$ is an additional
regulator that has the effect of dampening emissions with increasing
distance along the Wilson line, thus smoothly removing long-distance
behaviour. As has been found for previous MGEWs, and as we will see in
what follows, this regulator is well-suited to the
practical calculation of higher-loop webs. Armed with this regulator,
we define the {\it soft function} of $L$ Wilson lines with velocities $\{\beta_k\}$ as
\begin{equation}
{\cal S}\left(\gamma_{ij},\alpha_s(\mu^2),\epsilon,\frac{m}{\mu}\right)
\equiv \left\langle 0 \left| \Phi_{\beta_1}^{(m)}\otimes
\Phi_{\beta_2}^{(m)}\otimes\ldots\otimes \Phi_{\beta_L}^{(m)}\right|
0\right\rangle,
\label{Sdef}
\end{equation}
which depends on the $d$-dimensional coupling $\alpha_s(\mu^2)$ satisfying
\begin{equation}
\frac{d\alpha_s}{d\ln\mu^2}=-\alpha_s\left[\epsilon+b_0\alpha_s
+b_1\alpha_s^2+\ldots\right]\,,
\label{b0def}
\end{equation}
where $\mu$ is the dimensional regularisation scale, and on the {\it cusp
  angles}\footnote{More precisely, $\gamma_{ij}/2=\cosh(\phi_{ij})$ where 
$\phi_{ij}$ is the Minkowski-space angle between lines $i$ and $j$. In a timelike process, when $\phi_{ij}$ is real, $\gamma_{ij}>2$ (or $-1<\alpha_{ij}<0$) while in a spacelike one $\gamma_{ij}<-2$ (or $0<\alpha_{ij}<1$).}
\begin{equation}
\gamma_{ij}=\frac{2\beta_i\cdot\beta_j}
{\sqrt{\beta_i^2-i\varepsilon}\sqrt{\beta_j^2-i\varepsilon}}
=-\left(\alpha_{ij}+\frac{1}{\alpha_{ij}}\right),
\label{gammaij}
\end{equation}
where we have defined the parameter $\alpha_{ij}$ associated with each
pair of lines $i$ and $j$ for later use. We shall always pick $|\alpha_{ij}|\leq 1$. 
Due to the additional
regulator, all singularities as $\epsilon\rightarrow0$ are ultraviolet in
origin, and the fact that multiple Wilson-line operators are
multiplicatively renormalizable~\cite{Dotsenko:1979wb} means that we
can then define the renormalized soft function
\begin{equation}
{\cal S}_{\rm ren.}\left(\gamma_{ij},\alpha_s(\mu^2),\epsilon,
\frac{m}{\mu}\right)={\cal S}\left(\gamma_{ij},\alpha_s(\mu^2),\epsilon,
\frac{m}{\mu}\right)Z(\gamma_{ij},\alpha_s(\mu^2),\epsilon),
\label{Srendef}
\end{equation}
where the factor $Z$ collects all singularities associated with the renormalization of the vertex at which the Wilson lines meet.
This leads to the
renormalization group equation
\beq
  \mu \frac{d }{d \mu} Z \left( \gamma_{ij}, \alpha_s(\mu^2), \e \right) \, = \, - \,
  Z \left( \gamma_{ij}, \alpha_s(\mu^2), \e \right)
  \Gamma \left(\gamma_{ij}, \alpha_s(\mu^2) \right) \, ,
\label{Zeq}
\eeq
where $\Gamma$ is the soft anomalous dimension referred to above: it
is a finite quantity that encapsulates the ultraviolet singularities of $Z$ and
${\cal S}$. 
Each of the Wilson lines $\Phi_{\beta_k}^{(m)}$ in eq.~(\ref{Sdef}) carries independent colour indices in a tensor product, and thus all quantities appearing in
eqs.~(\ref{Srendef}, \ref{Zeq}) must be interpreted
as matrix-valued in the space of possible colour
flows between the Wilson lines. 
As such, the order in which quantities
appear on the right-hand side is important. Defining the perturbative
expansion\footnote{Throughout, we will define the perturbative
  expansion of other quantities similarly to eq.~(\ref{Gammaexp})
  unless otherwise stated.}
\begin{equation}
\Gamma\left(\gamma_{ij}, \alpha_s(\mu^2) \right)=\sum_{n=1}^\infty \left(\alpha_s(\mu^2)\right)^n\,\Gamma^{(n)}\left(\gamma_{ij}\right),
\label{Gammaexp}
\end{equation}
we may write the solution of eq.~(\ref{Zeq}) (suppressing the
dependence on the cusp angles and the scale) as
\beqa Z (\gamma_{ij},\alpha_s(\mu^2),\epsilon)&
= & \exp \Bigg\{ \alpha_s \, \frac{1}{2 \e} \, \Gamma^{(1)} + \,
\alpha_s^2 \left( \frac{1}{4 \e} \Gamma^{(2)} - \frac{b_0}{4 \e^2}
\Gamma^{(1)} \right) \\ & & + \, \alpha_s^3 \left( \frac{1}{6 \e}
\Gamma^{(3)} + \frac{1}{48 \e^2} \left[ \Gamma^{(1)}, \Gamma^{(2)}
  \right] - \frac{1}{6 \e^2} \left(b_0 \Gamma^{(2)} + b_1 \Gamma^{(1)}
\right) + \frac{b_0^2}{6 \e^3} \Gamma^{(1)} \right) \nonumber \\ & & +
\, \alpha_s^4 \left( \frac{1}{8 \e} \Gamma^{(4)} + \frac{1}{48 \e^2}
\left[ \Gamma^{(1)}, \Gamma^{(3)} \right] - \frac{b_0}{8 \e^2}
\Gamma^{(3)} + \frac{1}{8 \e^2} \left( \frac{b_0^2}{\e} - b_1 \right)
\Gamma^{(2)} \right. \nonumber \\ & & \hspace{5mm} \left. - \frac{1}{8
  \e^2} \left( \frac{b_0^3}{\e^2} - \frac{2 b_0 b_1}{\e} + b_2 \right)
\Gamma^{(1)} - \frac{b_0}{48 \e^3} \left[ \Gamma^{(1)}, \Gamma^{(2)}
  \right] \right) + {\cal O} \left( \alpha_s^5 \right) \Bigg\} \, ,
\nonumber
\label{Zexp}
\eeqa
where the $\beta$-function coefficients of the $d$-dimensional  coupling are defined in eq.~(\ref{b0def}). The unrenormalized soft function
also has an exponential form, which
for now we may write as
\beq
  {\cal S} \left(\alpha_s, \e \right) \, = \, \exp \Big[ w \left( \alpha_s, \e \right) \Big]
  \, = \, \exp \left[ \, \sum_{n = 1}^\infty \, \sum_{k = - n}^\infty  \left(\alpha_s(\mu^2) \right)^n
 \, \e^k \,
  w^{(n, k)} \right] \, ,
\label{Sunren}
\eeq
%\Einan{A separate issue to consider is the following: if we want the relation between the initial scale we introduce and the final scale to coincide with the standard and very familiar relation between the MS and ${\overline{\text{MS}}}$ scales in momentum space, then there is a way to do that: write the logs as $\log\left(\frac{\mu^2}{4m^2}\right)$. I do not mind either way.}
%
i.e. $w^{(n,k)}$ collects all contributions to the logarithm of the
soft function at a given order in the coupling $\alpha_s$, and
dimensional regularisation parameter $\epsilon$. Equations~(\ref{Zeq})
and~(\ref{Sunren}), together with the requirement that $\Gamma^{(n)}$
be finite as $\epsilon\rightarrow 0$ imply~\cite{Gardi:2011yz}
\beqa
\label{Gamres}
  \Gamma^{(1)} & = & - 2 w^{(1,-1)} \, ,\nonumber \\
  \Gamma^{(2)} & = & - 4 w^{(2,-1)} - 2 \left[ w^{(1,-1)}, w^{(1,0)} \right] \, ,\nonumber \\
  \Gamma^{(3)} & = & - 6 w^{(3,-1)} + \frac{3}{2} b_0 \left[ w^{(1,-1)}, w^{(1,1)} \right]
  + 3 \left[ w^{(1,0)}, w^{(2,-1)} \right] + 3 \left[ w^{(2,0)}, w^{(1,-1)} \right] \nonumber \\
  & & \hspace{-1cm}
  + \left[ w^{(1,0)}, \left[w^{(1,-1)}, w^{(1,0)} \right] \right] - \left[ w^{(1,-1)},
  \left[w^{(1,-1)}, w^{(1,1)} \right] \right] \, .
\eeqa
That is, the coefficients of the soft anomalous dimension are fixed
from the simple pole in $\epsilon$ of the logarithm of the soft
function at a given order in $\alpha_s=g_s^2/(4\pi)$, together with
commutators (in colour space) of various coefficients at lower
order.  Note that the anomalous dimension coefficients must be strictly independent of the infrared cutoff scale $m$.
%Note that the calculation of the anomalous dimension is based on taking a derivative of the $Z$ factor with respect to the ultraviolet renormalization scale $\mu$ (see eq. (2.7) in ref.~\cite{Gardi:2011yz}) so at this point any dependence on the infrared cutoff scale $m$, which appears in the soft function in eq.~(\ref{Srendef}), is removed.
%For the higher-order poles in $\epsilon$, one also obtains the
%consistency constraints
%\begin{align}
%w^{(2,-2)}&=-\frac12 b_0 w^{(1,-1)};\notag\\
%w^{(3,-3)}&=\frac13 b_0^2 w^{(1,-1)};\notag\\
%w^{(3,-2)}&=\left(-\frac23 w^{(2,-1)}-\frac{1}{12}\left[w^{(1,-1)},
%w^{(1,0)}\right]\right)b_0-\frac13 b_1 w^{(1,-1)}+\frac16\left[
%w^{(2,-1)},w^{(1,-1)}\right].
%\label{wconstraints}
%\end{align}
The problem of calculating the soft anomalous dimension has now been
reduced to finding the coefficients $w^{(n,k)}$ appearing in
eq.~(\ref{Sunren}). This is the subject of the following section.

\subsection{Webs and their kinematic and colour factors}
\label{sec:webs}

Equation~(\ref{Gamres}) relates the perturbative coefficients of the
soft anomalous dimension to the coefficients appearing in the
logarithm of the soft function, eq.~(\ref{Sunren}). As explained in
ref.~\cite{Gardi:2010rn}, we may write the total contribution $w^{(n)}$ at each
loop order $n$ as a sum of {\it webs}:
\begin{equation}
\label{web_decomposition}
w^{(n)} =\sum_{(n_1,n_2,\ldots n_L)} w_{(n_1,n_2,\ldots n_L)}^{(n)}\,,
%\quad\qquad w^{(n,k)} =\sum_{(n_1,n_2,\ldots n_L)} w_{(n_1,n_2,\ldots n_L)}^{(n,k)}\,,
\end{equation}
where each web 
\begin{equation}
W_{(n_1,n_2,\ldots n_L)}\,=\, \alpha_s^n \, w_{(n_1,n_2,\ldots n_L)}^{(n)}=\alpha_s^n\sum_{k=-n}^\infty \epsilon^k w_{(n_1,n_2,\ldots n_L)}^{(n,k)}\,,
\label{Wdef1}
\end{equation}
consists of a closed set of diagrams connecting $L$
Wilson lines, with a fixed number of gluon attachments
$(n_1,n_2,\ldots, n_L)$ on each line, where $n_i\geq 0$. The
diagrams in a single web are interrelated by all possible permutations
of the gluon attachments along each Wilson line.\footnote{Note however that the set of numbers $(n_1,n_2,\ldots, n_L)$ does \emph{not} uniquely identify a given web, even at a given order in perturbation theory. For example $W_{(1,2,3)}$ webs can be formed at three loops by multiple-gluon exchanges, with or without a boomerang gluon. Of course, three and four gluon vertices off the Wilson lines also distinguish between webs. 
We refer the reader to ref.~\cite{Gardi:2013ita} for a full classification of all webs at three loops, and to refs.~\cite{Agarwal:2020nyc,Agarwal:2021him} for a classification at four loops using correlator webs. } 
Each diagram $D\in W_{(n_1,\ldots,n_L)}$
has a colour factor $C(D)$ and kinematic part ${\cal F}(D)$, such that
the contribution of the web to $w^{(n)}$ may be written\footnote{From now on, we
  will suppress the attachment indices on a given web
  $W_{(n_1,\ldots,n_L)}$ where this is unimportant.} in the form
\begin{equation}
W=\sum_{D,D'\in W}{\cal F}(D)R_{D D'}{\cal C}(D').
\label{Wnidef}
\end{equation}
The quantity $R_{D D'}$ (a matrix in the space of diagrams) is
called a {\it web mixing matrix}, and has a purely combinatorial
definition. An algorithm to calculate the mixing matrix for a given
web was given in ref.~\cite{Gardi:2010rn}, further combinatorial
aspects have been explored in
refs.~\cite{Dukes:2013wa,Dukes:2013gea,Dukes:2016ger}, and recent
progress beyond three loops was reported in refs.~\cite{Agarwal:2020nyc,Agarwal:2021him}.  
Physically, the web mixing matrix describes how colour and kinematic factors are
entangled in the logarithm of the soft function.

Although a full understanding of web mixing matrices remains elusive,
some general properties have been well-established. Chief among these
is the fact that web mixing matrices are idempotent, and thus act as
projection operators, with eigenvalues $\lambda_i\in\{0,1\}$. The rank
$r$ of a $p$-dimensional web mixing matrix is the number of unit
eigenvalues. Let $Y$ be the matrix that diagonalises the web mixing
matrix:
\begin{equation}
YRY^{-1}={\rm diag}(\lambda_1,\lambda_2,\ldots \lambda_p),\qquad
\lambda_i=\begin{cases} 1,\quad i\leq r; \\ 0,\quad r+1\leq i\leq p.
\end{cases}
\label{Ydef}
\end{equation}
Then we can write the contribution of a single web as
\begin{equation}
W=\sum_{j=1}^r\left(\sum_D {\cal F}(D) Y^{-1}_{D,j}\right)
\left(\sum_{D'} Y_{j,D'} C(D')\right)\equiv\sum_{j=1}^r
{\cal F}_{W;j}^{(n)} c_j^{[n,L]}\,.
\label{WY}
\end{equation}
As expressed by the second equality, this has the form of a sum over
combinations of kinematic factors ${\cal F}_{W;\,j}^{(n)}$ (one for each unit
eigenvalue), each accompanied by a corresponding colour factor~$c_j^{[n,L]}$, where $n$ indicates the loop order  and~$L$ denotes the number of Wilson lines.
It has now been proven~\cite{Gardi:2013ita} that each
such colour factor is equivalent to the colour factor of a fully
connected soft gluon graph. As mentioned above, this is the
appropriate generalisation of the maximally non-Abelian property of
two-line webs~\cite{Gatheral:1983cz,Frenkel:1984pz,Sterman:1981jc} to
the multiparton case. We will briefly discuss our basis of these connected colour factors in section~\ref{sec:bases} below.

Having introduced the colour decomposition of each web in eq.~(\ref{WY}), we may write the corresponding Laurent expansion in $\epsilon$, 
eq.~(\ref{Wdef1}), more explicitly as
\begin{equation}
W_{(n_1,n_2,\ldots n_L)}=\alpha_s^n w_{(n_1,n_2,\ldots n_L)}^{(n)}=\alpha_s^n\sum_{k=-n}^\infty \epsilon^k w_{(n_1,n_2,\ldots n_L)}^{(n,k)}
=
\sum_{j=1}^r
{\cal F}^{(n)}_{(n_1,n_2,\ldots n_L);j} c_j^{[n,L]}
\,.
\label{Wdef2}
\end{equation}
To obtain the contributions of a given $n$-loop web to $w^{(n,k)}$ of eq.~(\ref{web_decomposition}) we must therefore expand its kinematic function ${\cal F}^{(n)}_{(n_1,n_2,\ldots n_L);j}$ in $\epsilon$:
\begin{equation}
{\cal F}^{(n)}_{(n_1,n_2,\ldots n_L);j}=\sum_{k=-n}^{\infty}\epsilon^k
{\cal F}^{(n,k)}_{(n_1,n_2,\ldots n_L);j} \,,
\label{calF_eps_expansion}
\end{equation}
and then recast the result as
\begin{equation}
w_{(n_1,n_2,\ldots n_L)}^{(n,k)}= \sum_{j=1}^r
{\cal F}^{(n,k)}_{(n_1,n_2,\ldots n_L);j} c_j^{[n,L]}\,.
\label{w_calF}
\end{equation}
In order to express the anomalous dimension in eq.~(\ref{Gamres}) at order $n$ in the loop expansion we need, specifically, the single pole terms ($k=-1$) of each web. It is convenient to write
\begin{equation}
\Gamma^{(n)}= -2 n\, \overline{w}^{(n,-1)},\qquad\quad \overline{w}^{(n,-1)} =\!\!\sum_{(n_1,n_2,\ldots n_L)}\!\! \overline{w}_{(n_1,n_2,\ldots n_L)}^{(n,-1)}
\label{Gamma_wbar}
\end{equation}
where we followed refs.~\cite{Gardi:2013saa,Falcioni:2014pka} in defining \emph{subtracted webs} $\overline{w}$ which include, \emph{for each web}, the commutators of the relevant web-subdiagrams taken at ${\cal O}(\epsilon^{-1})$, according  to eq.~(\ref{Gamres}).
For example, for three-line webs at two loops, according to the second relation in eq.~(\ref{Gamres}) the subtracted $(1,1,2)$ web is defined as
\begin{equation}
\overline{w}_{(1,1,2)}^{(2,-1)}=w_{(1,1,2)}^{(2,-1)} +\frac12 \left[w_{(1,0,1)}^{(1,-1)},\, w_{(0,1,1)}^{(1,0)}\right]+\frac12 \left[w_{(1,1,0)}^{(1,-1)},\, w_{(1,0,1)}^{(1,0)}\right]\,.
\end{equation}
The colour decomposition of each subtracted web in eq.~(\ref{Gamma_wbar}) readily follows from eq.~(\ref{w_calF}):
\begin{equation}
\overline{w}_{(n_1,n_2,\ldots n_L)}^{(n,-1)}\,=\,
\sum_{j=1}^r \overline{w}_{(n_1,n_2,\ldots n_L);j}^{(n,-1)}
\,=\,
\left(\frac{1}{4\pi}\right)^n\,
\sum_{j=1}^r
 F^{(n)}_{(n_1,n_2,\ldots n_L);j} c_j^{[n,L]}\,,
\label{wbar_calF_relation}
\end{equation}
where the $\{F^{(n)}_{(n_1,n_2,\ldots n_L);j}\}$ carry the kinematic dependence on the Wilson-line velocities associated with the colour structure $j$. These kinematic functions are independent of both the infrared cutoff scale $m$ and the dimensional regulator and they directly contribute to the anomalous dimension, eq.~(\ref{Gamma_wbar}). Their calculation -- for the case of boomerang webs -- will be a central goal of the present paper.

\subsection{Web colour bases}
\label{sec:bases}

Given that any superposition of degenerate eigenvectors of the web
mixing matrix is also an eigenvector, the matrix $Y$ in eq.~(\ref{WY})
is not unique. Put another way, the basis of colour factors
$c_j^{[n,L]}$ is also not unique, and one must choose a suitable basis
before calculating all webs at a given order. One such basis was
presented in ref.~\cite{Gardi:2013ita}, which developed an alternative
language for the logarithm of the soft function. That is, one may
think of the latter as consisting of diagrams composed of effective
vertices $\{V_K^{(l)}\}$, describing the emission of $K$ gluons from
the specific Wilson line~$l$. In general there can be several such vertices on a given line, but such that their respective position along the line is fully symmetrised.  The colour factor associated with each
such vertex is that of a fully connected gluon configuration. For
example, the case of two gluons has only the single possibility
\begin{equation}
C_{2,1}^{ab}=\left[T^a,T^b\right]=if^{abc}T^c,
\label{V2colour}
\end{equation}
which is the same as the colour factor associated with a gluon emitted
from the Wilson line, that then splits into two via a three-gluon
vertex. For three gluons, there are two independent connected colour
factors, namely
\begin{align}
C_{3,1}^{ab,c}=\left[\left[T^a,T^b\right],T^c\right]=f^{abd}f^{ecd} T^e;\notag\\
C_{3,2}^{ac,b}=\left[\left[T^a,T^c\right],T^b\right]=f^{acd}f^{ebd} T^e.
\label{V3colour}
\end{align}
Ref.~\cite{Gardi:2013ita} showed that any \emph{connected} diagram -- i.e. one that remains connected when the Wilson lines themselves are removed --  composed of such vertices on the Wilson lines, and ordinary QCD vertices off the Wilson lines, has a connected (``maximally non-Abelian'') colour factor. In this way the effective-vertex formalism was used in establishing the non-Abelian exponentiation theorem for multiple Wilson lines.
Furthermore, this formalism provides a neat
way to fix a suitable colour basis for webs. For a given web
$W_{(n_1,n_2,\ldots,n_L)}$, the possible connected colour factors are
generated by the possible assignments of effective vertices on each
Wilson line, commensurate with the gluon attachment numbers
$\{n_i\}$. As explained in ref.~\cite{Gardi:2013ita}, if more than one
effective vertex is present on a given line, one determines the
contribution of this line to the overall colour factor by fully
symmetrising over the individual vertex colour factors $\{C_i\}$:
\begin{equation}
\left\{C_1\,C_2\ldots C_n\right\}_+\equiv \frac{1}{n!}
\sum_{\pi\in S_n}C_{\pi_1}\,C_{\pi_2}\ldots C_{\pi_n}.
\label{Csym}
\end{equation}
We can use this to formulate a basis for the overall connected colour
factors of webs connecting~$L$ lines as follows. Firstly, let us
denote by $\{C_{K,j}(l)\}$ the set of $(K-1)!$ independent colour
factors associated with a given effective vertex $V_K^{(l)}$ on Wilson line $l$ (examples
are given in eqs.~(\ref{V2colour}, \ref{V3colour})). Then a fully
general web colour basis is provided by the colour factors
\begin{equation}
c_j^{[n,L]}=\prod_{l=1}^L\left\{ C_{K_1,j_1}{(l)}\,C_{K_2,j_2}{(l)}\ldots
C_{K_{n_l},j_{n_l}}{(l)}\right\}_+,
\label{colbasis}
\end{equation}
consisting of the different choices of effective vertex factors
multiplied together on each line, and symmetrised according to
eq.~(\ref{Csym}).
The effective vertex colour matrix $C_{K,j}{(l)}$ carries $K$ adjoint indices, which may be contracted in (\ref{colbasis}) with those of other colour matrices on any of the Wilson lines. In particular, we will be interested in this paper in boomerang webs where there are contractions between the adjoint indices of pairs of effective vertex colour matrices on the same line. As noted already in ref.~\cite{Gardi:2013ita}, in this case the basis defined by eq.~(\ref{colbasis}) is expected to be over-complete: there may be linear relations between $c_j^{[n,L]}$ consisting of different sets of vertices $C_{K,j}{(l)}$, all having the same total number of gluons emitted from line $l$ (out of which some pairs are contracted to form boomerang gluons).
This will become important in section~\ref{coll_reduction_boomerang} (see eqs. (\ref{two_ways_of_writing_c3}) and (\ref{two_ways_of_writing_c2}) there) where we will study a related, highly non-trivial relation between webs spanning a different number of Wilson lines upon taking collinear limits.

\subsection{Kinematic factors of MGEWs}
\label{sec:MGEWkin}

Having addressed the colour structure of webs in the previous
sections, we must also describe how to calculate the kinematic part
${\cal F}(D)$ of a web diagram
$D$. References~\cite{Gardi:2013saa,Falcioni:2014pka} developed a
systematic procedure for calculating the kinematic parts of multiple
gluon exchange webs, that will provide a highly useful starting point
for what follows. First, we will use the Feynman gauge gluon
propagator in configuration space, which in $d=4-2\epsilon$ dimensions
is
\begin{equation}
{\cal D}_{\mu\nu}(x)=-{\cal N}\eta_{\mu\nu}
(-x^2+i\varepsilon)^{\epsilon-1},
\label{propdef}
\end{equation}
where
\begin{equation}
{\cal N}=\frac{\Gamma(1-\epsilon)}{4\pi^{2-\epsilon}}.
\label{Ndef}
\end{equation}
Furthermore, eq.~(\ref{Wilson2}) implies that the Feynman rule for
emission of a gluon from a Wilson line is
\begin{equation}
ig_s\bar{\mu}^\epsilon
\, \int_0^\infty d\lambda \beta_i^\mu
e^{-im\lambda\sqrt{\beta_i^2-i\varepsilon}}.
\label{FRemit}
\end{equation}
These results are sufficient to calculate any MGEW, given (by
definition) the absence of three- or four-gluon vertices located off the
Wilson lines. Let us now consider such a web, consisting of~$n$
individual gluon exchanges, where the~$k^{\rm th}$ such gluon
straddles the Wilson lines $i(k)$ and $j(k)\neq i(k)$ (i.e. we do not
yet allow for the possibility of boomerang gluons). Letting $s_k$
and~$t_k$ denote the distance parameters of the gluon along these two
Wilson lines, the expression for a given web diagram $D$ is given by
\begin{align}
{\cal F}^{(n)}(D)&=(g_s^2\,\bar{\mu}^{2\epsilon}\,{\cal N})^n
\prod_{k=1}^n\left(\beta_{i(k)}\cdot \beta_{j(k)}
\int_0^\infty ds_k dt_k\right)
\,\prod_{k=1}^n\left[-(\beta_{i(k)}s_k-\beta_{j(k)}t_k)^2+i\varepsilon\right]^{-1+\epsilon}
\notag\\
&\quad\times\,\Theta_D[\{s_k,t_k\}]\,
 \exp\left[-im\sum_{k=1}^n\left(s_k\sqrt{\beta_{i(k)}^2-i\varepsilon}
+t_k\sqrt{\beta_{j(k)}^2-i\varepsilon}\right)\right].
\label{FDform}
\end{align}
Here $\Theta_D[\{s_k,t_k\}]$ consists of a product of Heaviside
functions involving the distance parameters, that implements the
ordering of the gluons on each Wilson line. To carry out the integrals
in eq.~(\ref{FDform}), one may first rescale to
\begin{equation}
\sigma_k= s_k\sqrt{\beta_{i(k)}^2},\quad
\tau_k= t_k\sqrt{\beta_{j(k)}^2},
\label{strescale}
\end{equation}
before changing variables according to
\begin{equation}
\sigma_k=x_k\lambda_k,\quad \tau_k=(1-x_k)\lambda_k;\qquad
0\leq \lambda_k\leq \infty, \quad 0\leq x_k\leq 1,
\label{sigmatau}
\end{equation}
where $\lambda_k$ measures how far a given gluon is from the origin (the hard interaction vertex, where the Wilson lines meet), and $x_k$ is an ``angular'' variable, which tends to 0 or 1 in the limits where the gluon is collinear with line $i(k)$ or $j(k)$ respectively. Equation~(\ref{FDform}) then
becomes
\beqa
\label{gendiag2}
  {\cal F}^{(n)} \left( D \right) & = &
 \,  \left(\frac{1}{2} \, g_s^2 \bar{\mu}^{2\e} \,
  \frac{\Gamma(1 - \e)}{4 \pi^{2 - \e}}\right)^n\,
  \prod_{k = 1}^n \Bigg[ \int_0^\infty d \lambda_k \, \lambda_k^{- 1 + 2 \e} \,
  {\rm e}^{ - i (m -i \varepsilon) \lambda_k }
  \\ & &  \hspace{-5mm} \nonumber \times \, \,
  \int_0^1  d x_k \, \gamma_k\,
  \Big[ - x_k^2 - (1 - x_k)^2 +  \gamma_k \, x_k (1 - x_k) +i\varepsilon\Big]^{- 1 + \e} \Bigg]
  \, \Theta_D \big[ \left\{ x_k, \lambda_k \right\} \big]  \, ,
\eeqa
where $\gamma_k\equiv \gamma_{i(k)j(k)}$ is the cusp angle between
lines $i(k)$ and $j(k)$, as defined in eq.~(\ref{gammaij}). To proceed, one may define
\beq
\lambda_k \, = \, \left( 1 - y_{k
  - 1} \right) \, \prod_{p = k}^n y_p \, , \quad k=1,\ldots,n,\quad
y_0=0,
\label{defy}
\eeq
so that the exponential-regulator factor simplified to ${\rm e}^{ - i (m -i \varepsilon) y_n}$, and after integrating over  $y_n$, eq.~(\ref{gendiag2}) becomes (see ref.~\cite{Falcioni:2014pka} for more details):
\beqa
  {\cal F}^{(n)} \left( D \right) & = &\kappa^n \,  \Gamma(2n\e)
\, \, \nonumber \prod_{k = 1}^n \Bigg[ \int_0^1 d x_k\,
  \gamma_k \, \Big[  x_k^2  +(1 - x_k)^2 -  \gamma_k \, x_k (1 - x_k) \Big]^{- 1 + \e} \Bigg]
  \\ & &  \hspace{+15mm} \times \, \, \nonumber \prod_{k' = 1}^{n - 1} \Bigg[\int_0^1 d y_{k'}
  \left( 1 - y_{k'} \right)^{- 1 + 2 \e} \, y_{k'} ^{- 1+ 2 k' \e} \Bigg] \, \,\,
  \Theta_D \big[ \left\{ x_k, y_{k'} \right\}\big]
  \nonumber  \\  & = & \kappa^n \, \, \Gamma( 2 n \e )
  \, \prod_{k = 1}^n \bigg[ \int_0^1  d x_k  \,\gamma_k \, P_\epsilon
  \left( x_k, \gamma_k \right) \bigg]  \, \phi_D^{(n)} \left( x_i; \e \right)
   \, ,\label{gendiag3}
\eeqa
where we have defined the expansion parameter
\beq
 \kappa \, \equiv - \frac{1}{2} \, g_s^2 \left(\frac{\bar{\mu}^2}{m^2}\right)^{\e} \,
  \frac{\Gamma(1 - \e)}{4 \pi^{2 - \e}} \, ,
\label{prefac}
\eeq
which is convenient at intermediate stages of the calculation.
In the second line in eq.~(\ref{gendiag3}) we also defined the propagator-related function
\beq P_\epsilon \left( x, \gamma \right) \, \equiv \, \left[
  x^2 +(1 - x)^2 - x (1 - x) \gamma) \right]^{-1 + \e} \,
\label{propafu}
\eeq
and the {\it kernel of diagram $D$}
\beq
  \phi_D^{(n)}  \left( x_i; \e \right) \, = \, \prod_{k = 1}^{n - 1} \Bigg[ \int_0^1 d y_k
  \left( 1 - y_k \right)^{- 1 + 2 \e} \, y_k ^{- 1+ 2 k \e} \Bigg] \, \,
  \Theta_D \big[ \left\{ x_i, y_i \right\} \big] \, ,
\label{phi_D}
\eeq
consisting of integrals over Heaviside functions originating from the
ordering of gluon attachments. At this point it is natural to perform the
integrals defining the kernel  for each diagram, expanded as a Laurent series in $\epsilon$, obtaining $\phi_D^{(n)}$ in terms of logarithms and polylogarithms of the variables $\{x_i\}$. In eq.~(\ref{gendiag3}), the kernel will eventually be integrated over the variables $\{x_i\}$ after multiplying it with the functions $P_{\epsilon}(x_i,\gamma_i)$ related to the gluon propagators.
The overall divergence in the factor $\Gamma(2n\epsilon)$ in eq.~(\ref{gendiag3}) is associated with the ultraviolet divergence one obtains upon shrinking the \emph{entire} soft gluon diagram $D$ to the
origin~\cite{Gardi:2013saa,Falcioni:2014pka}.

All diagrams within a given web (i.e. with the same numbers of gluon
attachments at a given perturbative order) will have an integral
expression of the form of eq.~(\ref{gendiag3}). The only difference
between such diagrams will be the kernel of eq.~(\ref{phi_D}), which
is the only part sensitive to the ordering of gluons on the Wilson
lines. It then follows from section~\ref{sec:webs} that the
contribution of a web $W$ to the colour structure $c_j^{[n,L]}$ in our
chosen basis is given by
\beqa
  {\cal F}_{W; \, j}^{(n)} \left( \gamma_{i j}, \e \right) \, = \,  \kappa^n \, \Gamma( 2 n \e )
  \,  \prod_{k = 1}^n \bigg[ \int_0^1 d x_k  \,\gamma_k \, P_\epsilon \left( x_k, \gamma_k
  \right) \bigg]  \, \phi_{W, \, j}^{(n)} \left( x_i; \e \right)  \, ,
\label{genweb}
\eeqa
where, following eq.~(\ref{WY}), the {\it web kernel} is defined by
\beq
  \phi^{(n)}_{W, \, j} \left( x_i; \e \right) \, = \, \sum_{D \in W} Y_{D,j}^{-1} \, \phi^{(n)}_D
  \left( x_i; \e \right) \, .
\label{webkernel}
\eeq
As an example, we collect in appendix \ref{app:lowerorder} the final results for the kinematic factors $\{{\cal F}^{(n)}_W\}$ of one- and two-loop MGEWs, after integration over the $\{y_k\}$ variables of eq.~(\ref{phi_D}). Similar three-loop results can be found in refs.~\cite{Gardi:2013saa,Falcioni:2014pka}.
The integrals over the variables $\{x_k\}$ in (\ref{webkernel}) could in principle also be
carried out at this stage. However, in forming the soft anomalous
dimension, one must combine the result for each web with commutators
of its web-subdiagrams, as prescribed by eq.~(\ref{Gamres}),
leading to the definition of {\it subtracted webs} in eq.~(\ref{Gamma_wbar}).
It turns out that performing the integrals over the~$\{x_k\}$ variables at the level of the subtracted webs
is also much easier to carry out than for the web itself. This was explained in refs.~\cite{Gardi:2013saa,Falcioni:2014pka}, showing
that for subtracted webs this integration yields a highly restricted class of functions, which we briefly recall below.

Following refs.~\cite{Gardi:2013saa,Falcioni:2014pka}, we write each~${\cal
  O}(\alpha_s^n)$ subtracted web as in eq.~(\ref{wbar_calF_relation}), namely
\beq
  \overline{w}^{(n, -1)} \left( \alpha_k \right) \, = \,
  \left(\frac{1}{4\pi}\right)^n \, \sum_{j=1}^{r}  \, c_j^{[n,L]} \, F^{(n)}_{W; \, j}
  \big( \alpha_k \big) \, ,
\label{subtracted_web_form}
\eeq
where from eqs.~(\ref{gammaij}) we define
$\gamma_k=-\alpha_k-1/\alpha_k$. %At this point we express the prefactor $\Gamma(2n\e) \kappa^n$ in (\ref{genweb}) in terms of the $n$-th power of the strong coupling $\alpha_s(\mu^2_{\rm \overline{\scriptscriptstyle{MS}}})$  using the standard replacement $\mu^2=\mu^2_{\rm \overline{\scriptscriptstyle{MS}}}\frac{e^{\gamma_E}}{4\pi}$.
Essential to deriving the subtracted web of eq.~(\ref{subtracted_web_form}) is the fact that the commutators in eq.~(\ref{Gamres}) build up the same fully connected colour factors as in the chosen basis of
section~\ref{sec:bases}. 
The kinematic function multiplying each colour
structure, $F^{(n)}_{W;\,j}$, contains integrals
over the variables~$\{x_k\}$, as well as the propagator functions of
eq.~(\ref{propafu}), rewritten in terms of~$\alpha$:
\begin{align}
\label{propafu2}
\begin{split}
 & p_{\epsilon} \left(x, \alpha \right) \, \equiv \gamma\, P_\epsilon \left( x, \gamma \right)= \, - \, \left( \alpha + \frac{1}{\alpha} \right) \,
  \Big[ q ( x, \alpha) \Big]^{- 1 + \e} \, \\&
  q \left(x, \alpha \right) \, \equiv \, x^2 + (1 - x)^2 +\left(\alpha+\frac{1}{\alpha}\right) \,
  x (1 - x) \, = -\frac{(1 - \alpha)^2}{\alpha} \left(x - \frac{1}{1 - \alpha}\right)\left(x + \frac{\alpha}{1 - \alpha}\right)\,.
\end{split}
\end{align}
The factorization property of $q \left(x, \alpha \right)$ clarifies the advantage of using the variable~$\alpha$ over using~$\gamma$ (see also ref.~\cite{Henn:2013wfa}). This ultimately amounts to rationalising the symbol alphabet.
The integrals may be carried out after expansion in the dimensional
regularisation parameter~$\epsilon$, for which eq.~(\ref{propafu2})
becomes
\beq
  p_\e \left( x, \alpha \right) \, = \, p_0 \left( x, \alpha \right) \,
  \sum_{n = 0}^\infty \frac{\e^n}{n!} \Big[ \log  q \left( x, \alpha \right) \Big]^n \, ,
\label{expP}
\eeq
where
\beq
  p_0 \left( x, \alpha \right) \, = \, - \, \left( \alpha + \frac{1}{\alpha} \right) \,
  \frac{1}{q(x,\alpha)} \, = \, r (\alpha) \left[ \frac{1}{x - \frac{1}{1 - \alpha}} -
  \frac{1}{x + \frac{\alpha}{1 - \alpha}} \right] \, ,
\label{p0_part_fracs}
\eeq
is the leading part of the propagator function as $\epsilon\rightarrow
0$, and we have defined the rational prefactor
\beq
  r (\alpha)\, = \, \frac{1 + \alpha^2}{1 - \alpha^2} \, .
\label{r_def}
\eeq
Finally, the subtracted web kinematic factor can be written as
\beqa
  F^{(n)}_{W; \, j} \big( \alpha_i \big) & = &
     \prod_{k = 1}^n \left[\int_0^1\, d x_k \, p_0 (x_k, \alpha_k) \right]  \,
  {\cal G}^{(n)}_{W; \, j} \Big(x_i, q(x_i, \alpha_i) \Big) \, \nonumber
  \\ & = &  \, \left(\prod_{k = 1}^{n} r (\alpha_k) \right)
    \prod_{k = 1}^n
    \left[ \int_0^1 d x_k \left( \frac{1}{x_k - \frac{1}{1 - \alpha_k}}
  - \frac{1}{x_k + \frac{\alpha_k}{1 - \alpha_k}} \right) \right] \,
  {\cal G}^{(n)}_{W; \, j} \Big(x_i, q(x_i, \alpha_i) \Big)\, \nonumber
  \\ & \equiv &  \, \left(\prod_{k = 1}^{n} r (\alpha_k) \right)
  G^{(n)}_{W; \, j} \big(\alpha_i \big) \, ,
\label{subtracted_web_mge_kin}
\eeqa
which defines the {\it subtracted web kernel} ${\cal G}_{W;j}^{(n)}$,
and its fully integrated counterpart $G_{W;j}^{(n)}$.
In all previously studied MGEWs~\cite{Gardi:2013saa,Falcioni:2014pka}, the
subtracted web kernel consists exclusively of powers of logarithms of
certain rational functions of $x_k$ and $\alpha_k$ (details will follow).
% $x_k$, $(1-x_k)$ or $q(x_k,\alpha_k)$, and on Heaviside functions containing ratios of the variables $\{x_k\}$.
The integrals in the middle line of
eq.~(\ref{subtracted_web_mge_kin}) are then in so-called $d\log$ form\footnote{Similar observations regarding the $d\log$ form have been made
  in the context of the calculation of the cusp anomalous dimension in
  ref.~\cite{Henn:2013wfa}.},
and can be carried out explicitly to give $G^{(n)}_{W; \, j} \big(\alpha_i \big) $ as a pure transcendental function of weight $(2n-1)$, consisting of a sum of products of harmonic polylogarithms, where a given polylogarithm depends on a single angle $\alpha_{ij}$. More than this, the functions appearing
in the final answer are of a special type, as we review in the
following section. As stated  above, we have considered here 
only webs that do not contain boomerang gluons i.e. all gluon
exchanges begin and end on {\it different} Wilson lines. We will need
to generalise the above results to cope with the case when boomerang
gluons are indeed present.

\subsection{A basis of functions for MGEWs}
\label{sec:basisfunctions}

Upon integrating the subtracted web kernel for a given MGEW, 
one obtains a pure transcendental function $G^{(n)}_{W; \, j}$
taking the form of a sum of products of harmonic polylogarithms of $\alpha_{ij}$, where each polylogarithm depends on a single $\alpha_{ij}$.  The analytic properties of such functions can
be efficiently encoded by means of the symbol
map~\cite{Goncharov.A.B.:2009tja,
  Goncharov:2010jf,Duhr:2011zq,Duhr:2012fh}. It was argued already in ref.~\cite{Gardi:2013saa} that the symbol of
(integrated) subtracted MGEWs has the highly restricted alphabet
\begin{equation}
\left\{\alpha_{ij},\quad \eta_{ij}\equiv\frac{\alpha_{ij}}{1-\alpha_{ij}^2}
\right\}\,.
\label{alphabet}
\end{equation}
This structure realises the two symmetries
\begin{equation}
  \label{alphasymms}
  \alpha \to -\alpha \qquad {\rm and} \qquad \alpha \to \frac{1}{\alpha}
\end{equation}
at symbol-level.  Reference~\cite{Falcioni:2014pka} then proposed a set of basis functions
consistent with this symbol alphabet, and in terms of which all currently
calculated MGEWs can be expressed. To quote the basis, we may use the functions
defined in the previous section, as well as the additional function
$\tilde{q}(x,\alpha)$ given by
\beq
  \ln \tilde{q} (x, \alpha) \, \equiv \, \frac{1}{r (\alpha)} \int_0^1 dy \, p_0 (y, \alpha)
  \theta(x > y) \, = \ln \left(\frac{1}{x} + \alpha - 1 \right) - \ln \left(\frac{1}{x} +
  \frac{1}{\alpha} - 1 \right) \, .
\label{lnqtil}
\eeq
The proposed basis for 
$G^{(n)}_{W; \, j}$ in eq.~(\ref{subtracted_web_mge_kin})
can then be written as
\beq
  M_{k,l,n} (\alpha) \, = \, \frac{1}{r (\alpha)} \int_0^1 dx \, p_0 (x, \alpha) \ln^k
  \bigg( \frac{q (x, \alpha)}{x^2} \bigg) \ln^{l} \bigg( \frac{x}{1 - x} \bigg) \ln^n
  \tilde{q}(x, \alpha) \, ,
\label{eq:Mbasis}
\eeq
where each function in the set has uniform weight ${\rm
  w}=k+l+n+1$. Defined in this manner, the basis is actually
overcomplete, as the functions satisfy the relations
\beq
  M_{k,l,n} (\alpha) \, = \, (-1)^{l + n} \sum_{r = 0}^k \sum_{s = 0}^n
  \left( \begin{array}{c} k \\ r \end{array} \right)
  \left( \begin{array}{c} n \\ s \end{array} \right)
  2^{s + r} (- 1)^s  \log^s (\alpha) \, M_{k - r,l + r,n - s} (\alpha) \, .
\label{con2}
\eeq
For completeness, we quote the symbols of basis functions which occur up
to three-loop order -- as well as explicit forms for the functions
themselves -- in appendix~\ref{app:functions}. There is currently much
evidence that this basis is sufficient for describing MGEWs
to all orders in perturbation theory. Up to three-loop order, it
covers all such webs that do not involve boomerang
gluons~\cite{Gardi:2013saa,Falcioni:2014pka}, including those two-line
webs that involve intricate patterns of crossed gluon
exchanges. Furthermore, a certain special diagram type, called the
{\it Escher staircase} in ref.~\cite{Falcioni:2014pka}, can be
calculated for arbitrary numbers of gluon exchanges, and is fully
expressible in terms of the basis of eq.~(\ref{eq:Mbasis}). It remains
to be seen whether or not the basis will cope if boomerang gluons are
indeed present, and it is one of the aims of the present paper to
explore this.

Note that one of the simplifying features of subtracted web kernels,
discussed in detail in refs.~\cite{Gardi:2013saa,Falcioni:2014pka}, is
that higher weight polylogarithm functions (such as dilogs) are absent,
whereas they are present in the web kernel itself. This made it
particularly straightforward to formulate the above basis of
functions. However, there is nothing to forbid the possibility that
such dilogs are indeed present in the subtracted web kernel for more
general webs. If so, they threaten to undermine our basis of functions
for integrated webs. Another possibility is that polylogarithmic functions are present, but that after integration one still requires only the
restricted set of basis functions defined above. We will return to
this point later in the paper.

Finally, we point out that neither the simple rational structure of eq.~(\ref{subtracted_web_mge_kin}), consisting exclusively of powers of $r(\alpha_{ij})$, nor the highly restricted transcendental function basis are expected to hold for non-MGEWs. In particular, a richer structure was found in the full angle-dependent cusp anomalous dimension in QCD at three loops in~ref.~\cite{Grozin:2015kna,Grozin:2014hna} and also in QED at four 
loops~\cite{Bruser:2020bsh}.

\section{Boomerang webs up to two-loop order}
\label{sec:mushroom}

Having reviewed the properties of MGEWs and their calculation, we now turn to the main subject of this paper, which is to calculate
boomerang webs, namely MGEWs containing at least one gluon whose two
endpoints are attached to the same Wilson line. These were not
considered in refs.~\cite{Gardi:2013saa,Falcioni:2014pka} due to the
fact that they present an additional complication, namely the presence
of ultraviolet singularities associated with shrinking a boomerang
gluon to a point on its Wilson line that is not at the origin. These
extra singularities must be regulated and removed, where necessary,
via renormalization of the coupling $g_s$. This possibly involves
modifying the regulator of eq.~(\ref{Wilson2}). As a
warm-up exercise, we may consider boomerang webs up to two-loop order,
even though these have been calculated before using different
regulators~\cite{Korchemsky:1987wg}. 
The lessons drawn may
tell us how to generalise the results of 
section~\ref{sec:MGEWkin} and then apply them at three loops. 
We begin with the simplest boomerang web.

\subsection{The self-energy graph}
\label{sec:selfenergygraph}

\begin{figure}[h]
\begin{center}
\scalebox{0.5}{\includegraphics{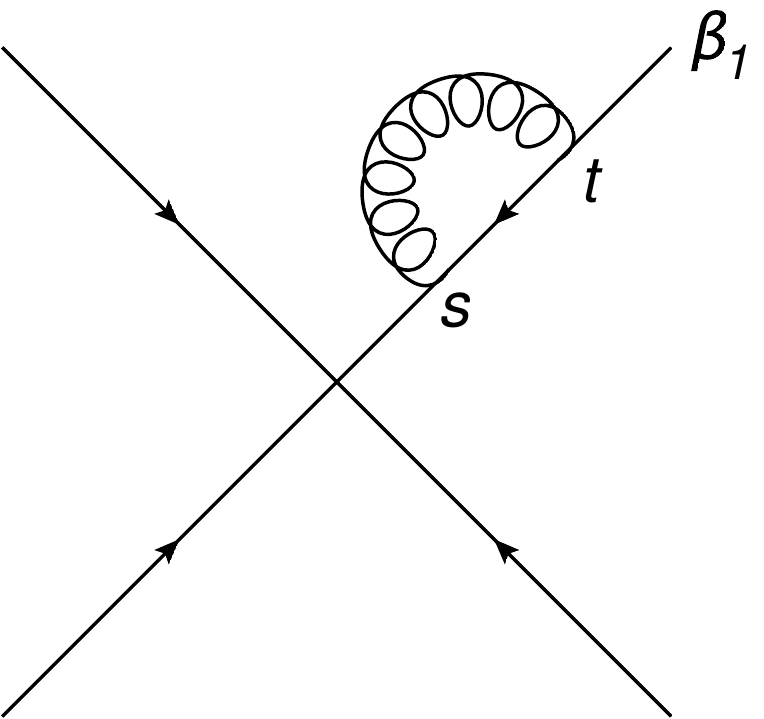}}
\caption{The self-energy web at one-loop order.}
\label{fig:selfenergy}
\end{center}
\end{figure}

The simplest boomerang web one can consider consists of the
self-energy graph of figure~\ref{fig:selfenergy}. This diagram forms a web by
itself, given that permutation of the two gluon attachments sends the
diagram to itself.
We take the 4-velocity of the Wilson line to which the gluons attach
to be $\beta_1$, and label the distance parameters of the gluon
emission vertices as shown in the figure. Note that the colour factor
of this graph is simply given by
\begin{equation}
{\cal C}_{\rm SE}=T_1^a\,T_1^a=C_{R_1},
\label{CSE}
\end{equation}
where the right-hand side is a quadratic Casimir in the appropriate
representation $R_1$ of the Wilson line. Thus, the colour factor of
this graph commutes with the colour factors of all other graphs or
webs, a fact that will be useful later on.

For the kinematic part of the self-energy diagram, we may apply the
results of eq.~(\ref{FDform}), together with the transformations of
eqs.~(\ref{strescale}, \ref{sigmatau}), to get
\begin{align}
{\cal F}_{\rm SE}^{(1)}&=g_s^2\bar{\mu}^{2\epsilon}{\cal N}\beta_1^2
\int_0^\infty ds \int_0^\infty
dt\left[-(t\beta_1-s\beta_1)^2+i\varepsilon
\right]^{\epsilon-1}\,e^{-im(t+s)\sqrt{\beta_1^2-i\varepsilon}}\,
\theta(t>s)\notag\\
&=\,-g_s^2\left(\frac{\bar{\mu}^2}{m^2}\right)^{\epsilon}{\cal N}
\int_0^1 dx\left[(2x-1)^2\right]^{\epsilon-1}
\,\theta\left(x>\frac12\right)
\int_0^\infty d\lambda \,\lambda^{2\epsilon-1}\, e^{-\lambda},
\label{FSE1}
\end{align}
where  the ``cusp angle'' in this case is
simply $\gamma_{11}=2$, according to the definition of
eq.~(\ref{gammaij}). The $\lambda$ integral is easily carried out to
give
\begin{equation}
{\cal F}_{\rm SE}^{(1)}=2\kappa \Gamma(2\epsilon)\int_{\frac12}^1
\frac{dx}{[(2x-1)^2]^{1-\epsilon}}\,,
\label{FSE2}
\end{equation}
where we expressed the prefactor in terms of $\kappa$ using eq.~(\ref{prefac}).
As discussed in section~\ref{sec:MGEWkin}, the pole in $\epsilon$ that
arises upon performing the $\lambda$ integration is an ultraviolet singularity
associated with shrinking the entire diagram to the origin. It is thus
associated with renormalization of the cusp vertex at which the Wilson
lines meet, and indeed appears in the soft anomalous dimension at
one-loop
order~\cite{Dotsenko:1979wb,Brandt:1981kf,Korchemsky:1985xj,Korchemskaya:1996je,Korchemskaya:1994qp}. We
are left with the integral over the $x$ variable, whose integration region from
$x=1/2$ to $x=1$ is dictated by the $\theta(t>s)$ in eq.~(\ref{FSE1}). There is of course a symmetry in the propagation of the gluon between the points
of emission and absorption, and swapping the two corresponds to transforming $x\to 1-x$.
The $x$ integral in eq.~(\ref{FSE2}) is divergent at the lower limit, for
$\epsilon\leq 1/2$. Physically, this corresponds to shrinking the
self-energy loop to a point away from the origin, and the fact that
the critical value of $\epsilon$ is $1/2$ rather than zero indicates a power-like, 
rather than logarithmic, singularity in four space-time dimensions.
We will follow the conventional procedure of focussing on logarithmic
divergences, and therefore only expand about $\epsilon=0$.
 Firstly, one carries out the integral to obtain
\begin{equation}
{\cal F}_{\rm SE}^{(1)}=2\kappa \Gamma(2\epsilon)\frac12\frac{1}
{2\epsilon-1},
\label{FSE3}
\end{equation}
assuming $\epsilon>1/2$. Next, one may analytically continue to near
$\epsilon=0$. In practice, this simply means expanding
eq.~(\ref{FSE3}) about $\epsilon=0$ to obtain
\begin{equation}
{\cal F}_{\rm SE}^{(1)}=-\kappa \Gamma(2\epsilon)\left[1+{\cal O}(\epsilon)
\right]
=  \frac{1}{4\epsilon} \, \frac{g_s^2}{4 \pi^2 } +{\cal O}(\epsilon^0)\,.
\label{FSE4}
\end{equation}
We see that there is in fact no additional $\epsilon\to 0$ divergence in this case
from shrinking the loop to a point. Nor indeed can there be: it is
known that the only ultraviolet singularities that affect Wilson lines are
associated with renormalization of the cusp at which the Wilson lines
meet, or with the coupling. There are no singularities associated with
field redefinitions of the Wilson lines themselves. Shrinking the self
energy to a point would indeed correspond to a renormalization of the
Wilson line itself, and is hence forbidden.

Here, we have seen that the regulator of eq.~(\ref{Wilson2}) is
sufficient to calculate the self-energy web at one-loop order. The situation will be different at two loops, as we describe in the following section.

\subsection{The mushroom (3,1) web}
\label{sec:13}

\begin{figure}[h]
\begin{center}
\scalebox{.9}{\includegraphics{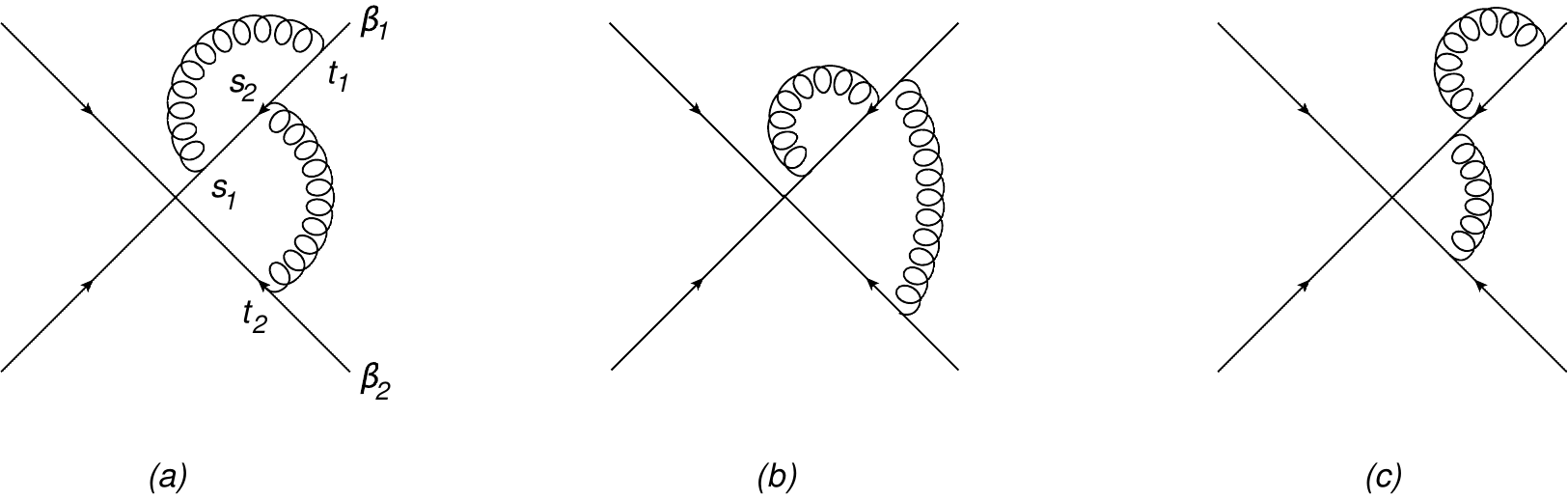}}
\caption{The (3,1) web.}
\label{fig:13}
\end{center}
\end{figure}

We now move to the calculation of the two-loop (3,1) web of
figure~\ref{fig:13}.
%Here we focus on the former.  
%  \Einan{Actually I do not think that it is absent -- the exponetiated colour factor of diagram 6$(a)$ is non-zero by itself. The proof of section 4 only implies that ECF of diagrams in which diagram 6$(a)$ is embeded as a subdiagram do not need to be considered.
%  Moreover, our prescription for the regulator does not extend in an obvious way for the (4) web and similar pure-self-energy boomerang webs: if we do not regularize any of the boomerang gluons, we have an infrared divergence and the result is scaleless. We could of course regularise one of the cluster of boomerangs and not the others, but we have not studies how this works. 
%  }
  Diagrams ($b$) and ($c$) in this web contain self-energy loops, and can be
calculated using the methods of the previous section. However, we will
see in due course that, although the kinematic factors of the
individual diagrams are non-zero, they do not in fact contribute to
the overall result after combination with the colour factors and web
mixing matrix,\footnote{A similar mechanism
  does not lead to the vanishing of the self-energy web at one-loop
  (figure~\ref{fig:selfenergy}), as there is nothing for this to
  cancel against.} as in eq.~(\ref{Wnidef}). We thus do not consider them further. More
interesting is diagram $(a)$, which has been previously called the {\it
  mushroom diagram} due to its resemblance to said fungus.
This diagram was of course computed, along with all other two-loop diagrams, in the original computation of the two-loop angle-dependent cusp anomalous dimension in ref.~\cite{Korchemsky:1987wg}. We repeat the calculation here, albeit using a different  regulator, preparing the ground for the evaluation of
higher-order diagrams.

Notably, diagram $(a)$ contains a boomerang gluon that straddles an extra emission. Here we
again expect an ultraviolet singularity as the boomerang gluon is shrunk to a
point. Furthermore, at least part of this singularity will not be
associated with renormalization of the cusp, as it will instead have
to do with the renormalization of the coupling of the gluon to the
Wilson line.

We can again apply the calculational methods of
section~\ref{sec:review} to obtain a result for the kinematic part of
diagram $(a)$. However, there is a subtlety in how to apply the
exponential regulator of eq.~(\ref{Wilson2}) for the specific case in
which a boomerang gluon straddles an extra emission. The exponential
regulator dampens the emission of a gluon that is emitted further from
the origin along the Wilson line. This in turn means that the
endpoints of the boomerang gluon on either side of the extra emission
are not treated equally. The latter is not a problem when shrinking
the entire diagram to the origin i.e. when obtaining those ultraviolet
singularities associated with renormalization of the cusp. However,
there is indeed a problem when trying to cleanly isolate the ultraviolet
singularity associated with shrinking the boomerang gluon to a point
around the extra gluon, and which contributes to the renormalization
of the coupling $g_s$. The safest and simplest way to proceed is to remove the
exponential regulator for the boomerang gluon, leaving it in place
only for the gluon exchange that links two different Wilson
lines. As we will see explicitly below, the regulation of the exchanged gluon will be sufficient to dampen the emission of the boomerang gluon at large distances. Given the rather subtle nature of the problem, we will present
here the calculation of the mushroom diagram in detail.

From figure~\ref{fig:13}$(a)$, the colour factor of the mushroom graph
is given by
\begin{equation}
  C(a) = T_1^b T_1^a T_1^b T_2^a = \left(C_{R_1}-\frac{1}{2}N_c\right)T_1\cdot T_2,
\label{Cmushroom}
\end{equation}
where $C_{R_1}$ denotes a quadratic Casimir in the
representation of line $1$, and the kinematic factor (excluding the exponential regulator for the boomerang gluon) is
\begin{equation}
\begin{split}
\mathcal{F}_a^{(2)} (\alpha_{12} , \epsilon)
= g_s^4 \bar{\mu}^{4\epsilon} \mathcal{N}^2(\beta_1) ^2(\beta_1 \cdot \beta_2) \int_{0}^{\infty} ds_1 ds_2 dt_1 dt_2 (-(t_1 \beta_1 -s_1 \beta_1)^2 + i\varepsilon)^{\epsilon -1} \\
\times(-(s_2 \beta_1 -t_2 \beta_2)^2 + i\varepsilon)^{\epsilon -1} e^{-ims_2\sqrt{\beta_1^2-i\varepsilon}}e^{-imt_2\sqrt{\beta_2^2-i\varepsilon}} \theta(t_1>s_2)\theta(s_2>s_1) .
\end{split}
\label{kinmushroom}
\end{equation}
Upon rescaling the parameters:
\begin{align*}
&s_1\sqrt{\beta_1^2-i\varepsilon} =\sigma_1
&s_2\sqrt{\beta_1^2-i\varepsilon} =\sigma_2 \\
&t_1\sqrt{\beta_1^2-i\varepsilon} =\tau_1
&t_2\sqrt{\beta_2^2-i\varepsilon} =\tau_2
\end{align*}
we get:
\begin{equation}
\label{FMcalc2}
\begin{split}
\mathcal{F}_a^{(2)} (\alpha_{12} , \epsilon)
= g_s^4 \bar{\mu}^{4\epsilon} \mathcal{N}^2\frac{\gamma_{12}}{2} \int_{0}^{\infty} d\sigma_1 d\sigma_2 d\tau_1 d\tau_2 (-\sigma_1^2 -\tau_1^2 +2\sigma_1\tau_1 + i\varepsilon)^{\epsilon -1} \\
\times
(-\sigma_2^2 -\tau_2^2 +\gamma_{12}\sigma_2\tau_2 + i\varepsilon)^{\epsilon -1}
e^{-i(m- i\varepsilon)(\sigma_2 + \tau_2)} \theta(\tau_1>\sigma_2)\theta(\sigma_2>\sigma_1) .
\end{split}
\end{equation}
We now perform another change of variables,
\begin{align*}
&\lambda_1 = \sigma_1 + \tau_1
&x = \frac{\tau_1}{\sigma_1 + \tau_1} \\
&\lambda_2 = \sigma_2 + \tau_2
&y = \frac{\sigma_2}{\sigma_2 + \tau_2}
\end{align*}
from which one finds
\begin{equation}
d\sigma_1 d\tau_1=\lambda_1 d\lambda_1 dx,\qquad
d\sigma_2 d\tau_2=\lambda_2 d\lambda_2 dy.
\label{Jacobian}
\end{equation}
% Extracting the infrared regulating parameter by $i(m- i\varepsilon) \lambda_k\rightarrow \lambda_k$ yields
% \begin{align}
% \mathcal{F}_a^{(2)} (\alpha_{12} , \mu^2/m^2 , \epsilon)
% &= g_s^4 \left(\frac{\mu^2}{m^2}\right)^{2\epsilon} \mathcal{N}^2\frac{\gamma_{12}}{2}  \int_{0}^{1} dx dy\, ((2x-1)^2)^{\epsilon -1} \,P_{\epsilon}(y, \gamma_{12}) \notag\\
% &\quad\times
% \int_{0}^{\infty} d\lambda_1 d\lambda_2
% \lambda_1^{2\epsilon-1}\lambda_2^{2\epsilon-1}e^{-\lambda_2}
% \theta \left(\lambda_1 >\frac{\lambda_2y}{x} \right)\theta \left(\lambda_1 <\frac{\lambda_2y}{1-x}\right) .
% \label{FMcalc2}
% \end{align}
At this point the integrals over $\lambda_k$ are straightforward:
the $\lambda_1$ integral is bounded from both ends by Heaviside functions, which imply
\begin{equation}
\label{theta31}
\frac{\lambda_2 y}{x}\leq \lambda_1 \leq \frac{\lambda_2 y}{1-x}\,,
\end{equation}
while the $\lambda_2$ integral is regulated by the exponential damping in the infrared, and by dimensional regularization in the ultraviolet. We thus obtain:
%\begin{align}
%\mathcal{F}_a^{(2)} (\alpha_{12} , \mu^2/m^2 , \epsilon)
%&= g_s^4 \mu^{4\epsilon} \mathcal{N}^2\frac{\gamma_{12}}{4\epsilon}
%\int_{0}^{\infty} d\lambda_2 \int_{\frac{1}{2}}^{1} dx
%\int_{0}^{1}dy \lambda_2^{4\epsilon-1}e^{-m\lambda_2}
%((1-x)^{-2\epsilon}-x^{-2\epsilon}) \notag\\
%&\quad
%\times((2x-1)^2 + i\varepsilon)^{\epsilon -1}y^{2\epsilon}P_{\epsilon}(y, \gamma_{12}) .
%\label{FMcalc3}
%\end{align}
%All Heaviside functions have now been used up, and the $\lambda_2$
%integral may be performed to give
\begin{align}
\mathcal{F}_a^{(2)} (\alpha_{12} , \epsilon)
&=\kappa^2\Gamma(4\epsilon)\frac{1}{\epsilon}\int_{\frac{1}{2}}^{1}dx\, (2x-1)^{2\epsilon -2}
((1-x)^{-2\epsilon}-x^{-2\epsilon}) \int_{0}^{1} dy\,
y^{2\epsilon}p_\epsilon(y, \alpha_{12}) \,,
\label{FMcalc4}
\end{align}
where the lower limit of the $x$ integral is implied by eq.~(\ref{theta31}).
% \Einan{Previous version has the statement namely:`
% Proceeding to evaluate this integral
% yields a power-like divergence from the limit $x=1/2$, which can be dealt with precisely as for the self-energy graph of section~\ref{sec:selfenergygraph}'. I believe this is incorrect.}
Proceeding to evaluate this integral, we note that in contrast to the self-energy graph of section~\ref{sec:selfenergygraph}, here there is no power divergence near $x\to \frac12$; instead, the factor $((1-x)^{-2\epsilon}-x^{-2\epsilon})$ suppresses the singularity in this limit, so that the integral is well-defined for $0<\epsilon<\frac12$.  Carrying out the integral one simply obtains:
\begin{align}
\label{FMcalc5}
\begin{split}
\mathcal{F}_a^{(2)}& (\alpha_{12}, \mu^2/m^2 , \epsilon) 
\,= \,\kappa^2\Gamma(4\epsilon)\frac{1}{\epsilon}\frac{1}{1-2\epsilon} \int_{0}^{1} dy y^{2\epsilon}p_\epsilon(y, \alpha_{12})
%\\=\,&\kappa^2\Gamma(4\epsilon)\frac{1}{\epsilon}\frac{1}{1-2\epsilon} \int_{0}^{1} dy p_0(y, \alpha_{12}) (1+2\epsilon\ln y) (1+\epsilon \ln q(y, \alpha_{12}) )\,
\\=\,&\left( \frac{g_s^2}{8\pi^2}\right)^2 \int_{0}^{1} dy
p_0(y, \alpha_{12}) \left[\frac{1}{4\epsilon^2} + \frac{1}{4\epsilon}\left(2+\ln q(y, \alpha_{12})+2\ln y
  +2\ln\left( \frac{\mu^2}{m^2} \right) \right) +\mathcal{O}(\epsilon^0)\right]\,,
\end{split}
\end{align}
where in the last step we expanded the expression in $\epsilon$, and switched from the scale $\bar \mu$ of eqs. (\ref{FRemit}) and (\ref{prefac}) to the $\overline{\text{MS}}$ renormalization scale, $\mu^2=\pi e^{-\gamma_E}\bar \mu^2$.
%\Einan{I removed the statement `` setting the renormalization scale as $\mu^2=m^2\frac{e^{\gamma_E}}{\pi}$, as explained following (\ref{subtracted_web_form}).}
% \Einan{I commented a part here: we should not attempt to use basis functions before constructing subtracted webs. Indeed it cannot be done for $\mathcal{F}_a^{(2, -1)} (\alpha_{12})$.
%Using the definition
%of the basis functions of eq.~(\ref{eq:Mbasis}), we then find that the
%mushroom graph has a double $\epsilon$ pole with coefficient
%\begin{equation}
%\mathcal{F}_a^{(2, -2)} (\alpha_{12})
%=\frac{1}{4} \left( \frac{g_s^2}{8\pi^2}\right)^2 \int_{0}^{1} dy p_0(y, \alpha_{12}) \\
%=\frac{1}{4} \left( \frac{g_s^2}{8\pi^2}\right)^2 r(\alpha_{12})M_{0,0,0}(\alpha_{12}) ,
%\label{mushroomeps2}
%\end{equation}
%as well as a single pole with coefficient
%\begin{equation}
%\mathcal{F}_a^{(2, -1)} (\alpha_{12})
%=\frac{1}{4} \left( \frac{g_s^2}{8\pi^2}\right)^2 \int_{0}^{1} dy p_0(y, \alpha_{12})[\mathrm{ln}q(y, \alpha_{12}) + \mathrm{ln}y^2 +2] .
%\label{mushroomeps1}
%\end{equation}
%}

\begin{figure}
\begin{center}
\scalebox{0.6}{\includegraphics{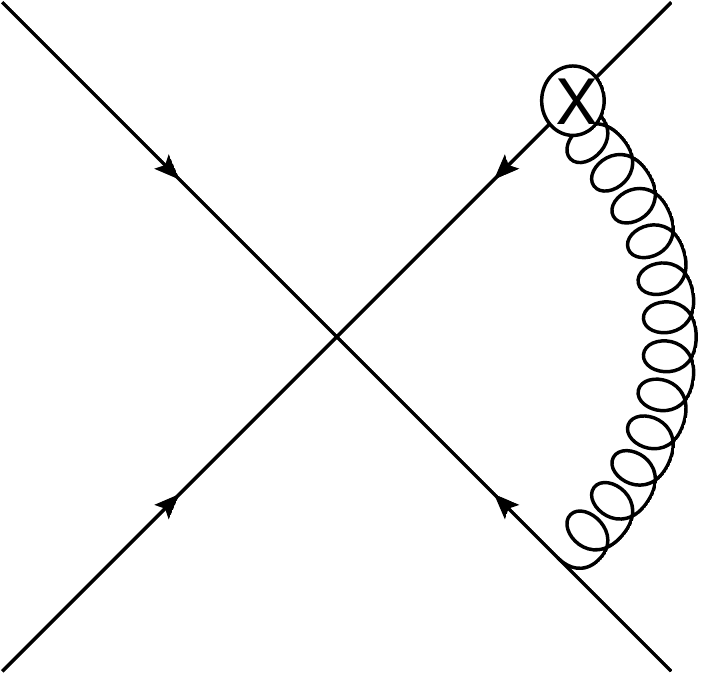}}
\caption{Counterterm graph for the mushroom diagram of
  figure~\ref{fig:13}$(a)$, where $\otimes$ denotes the counterterm for
  the gluon emission vertex from the Wilson line.}
\label{fig:counterterm}
\end{center}
\end{figure}

The appearance of a double pole at $\epsilon\to 0$ corroborates our above observation
that one expects a logarithmic singularity upon shrinking the boomerang gluon
to a point at the gluon emission vertex, in addition to the singularity associated with
renormalization of the cusp. Before renormalizing the cusp singularity as described in section~\ref{sec:review} we must renormalize the gluon emission vertex. To this end we add the counterterm graph of figure~\ref{fig:counterterm}, corresponding to the one-loop single gluon-exchange diagram, dressed by a gluon-emission vertex counterterm which we compute in appendix~\ref{app:counterterm}. The colour factor with which the counterterm graph enters is the same as the graph itself (eq.~(\ref{Cmushroom})), and its kinematic factor is given by
\begin{align}
\label{CT2}
\begin{split}
{\cal F}_a^{(2)\,{\rm CT}} (\alpha_{12},  &\mu^2/m^2, \epsilon)
\,\equiv \,Z_{v}^{(1)}(\epsilon)\,{\cal F}^{(1)}(\alpha_{12} , \epsilon)
\,=\,\frac{g_s^2}{8\pi^2\epsilon} \kappa \Gamma(2\epsilon) \int_{0}^{1} dy\, p_\epsilon(y, \alpha_{12})\,
\\  
 &=
 \left(\frac{g_s^2}{8\pi^2}\right)^2  \int_{0}^{1} dy p_0(y,
\alpha_{12})\left[-\frac{1}{2\epsilon^2}- \frac{1}{2\epsilon}\left(\ln q(y,
    \alpha_{12}) +\ln\left( \frac{\mu^2}{m^2} \right)\right) +\mathcal{O}(\epsilon^0)\right]\,,
    \end{split}
\end{align}
where $Z_{v}^{(1)}$ is the one-loop counterterm corresponding to the renormalization of the gluon emission vertex, and ${\cal F}^{(1)}$ is the kinematic part of the one-loop exchange graph. In the second step we inserted the result for ${\cal F}^{(1)}$ from eqs.~(\ref{F1res}) and the  counterterm from eq.~(\ref{Zgs1}), and in the third we expanded in $\epsilon$ and switched from $\bar{\mu}$ to $\mu$ as in eq.~(\ref{FMcalc5}).
%
%From eqs.~(\ref{F1res}, \ref{Zgs1}), eq.~(\ref{CT2}) then becomes
%\begin{equation}
%\mathcal{F}_{\rm CT}^{(2)}
%= -\frac{1}{\epsilon}\frac{g_s^2}{8\pi^2}\mathcal{F}^{(1)}
%= \kappa\Gamma(2\epsilon)\frac{1}{\epsilon}\frac{g_s^2}{8\pi^2}
%\int_{0}^{1} dy  p_\epsilon(y, \alpha_{12}) .
%\label{CT2b}
%\end{equation}
%\Einan{I believe that the right-hand-side of this equation is correct, even though the middle expression has a wrong sign. This is because the counterterm in (\ref{Zgs1}) was wrong (I corrected the sign there) and the expression for (\ref{F1res}) has a wrong sign (I have not yet corrected that one, but I inserted a comment there.}
%To renormalize the mushroom graph, we must add this contribution, and
%the kinematic parts will be additive given that the mushroom and
%counterterm graphs both have the same colour factor. This modifies
%both the double and single poles in $\epsilon$.
Summing up the results of the non-renormalized graph, eq.~(\ref{FMcalc5}), plus the counterterm graph, eq.~(\ref{CT2}), and using the basis functions of eq.~(\ref{eq:Mbasis}), one
finds for the coefficient of the double pole
\begin{align}
\mathcal{F}_{a,\,{\rm ren.}}^{(2, -2)}
&=\mathcal{F}_{a}^{(2, -2)} 
+ \mathcal{F}_{a}^{(2, -2)\, {\rm CT}} = -\frac{1}{4} \left( \frac{g_s^2}{8\pi^2}\right)^2r(\alpha_{12})
M_{0,0,0}(\alpha_{12}),
\label{mushroomren2}
\end{align}
and for the single pole
\begin{align}
\mathcal{F}_{a,\,{\rm ren.}}^{(2, -1)}
=\mathcal{F}_a^{(2, -1)} + \mathcal{F}_a^{(2, -1)\, {\rm CT}}
&=- \frac{1}{4} \left( \frac{g_s^2}{8\pi^2}\right)^2r(\alpha_{12})
\left[M_{1,0,0}(\alpha_{12})
-2M_{0,0,0}(\alpha_{12})\right]\, ,
\label{mushroomren1}
\end{align}
where the explicit expressions for $M_{0,0,0}$ and $M_{1, 0,0}$
can be found in appendix~\ref{app:functions}.
We stress that while the latter result can neatly be written in terms of basis
functions, the non-renormalized kinematic function $\mathcal{F}_a^{(2, -1)}$
cannot. This is a general feature\footnote{Generally, the additional stage of
  forming \emph{subtracted webs} will be required for the result to be
  expressible in terms of basis functions~\cite{Gardi:2013saa}.  We will encounter this in 
  section~\ref{sec:3lines}. }.  We also
point out that the dependence on $\ln(\mu^2/m^2)$ has cancelled in the
coefficient of the $1/\epsilon$ pole between eq.~(\ref{FMcalc5}) and
eq.~(\ref{CT2}), as it must do given that the infrared regulator $m$ cannot appear in the final result for the soft anomalous dimension.

We can now use these results to calculate the contribution of the
entire web of figure~\ref{fig:13} to the soft anomalous dimension. We
first need the web mixing matrix, that describes how to combine the
kinematic and colour parts of individual diagrams in the web. Using
the algorithm of ref.~\cite{Gardi:2010rn} (reviewed here in
appendix~\ref{app:Rcalc}) for the (3,1) web, we find that the combination of eq.~(\ref{Wnidef})
evaluates to
\begin{equation}
W_{(3,1)}=\frac12\left(\begin{array}{ccc}
{\cal F}(a) & {\cal F}(b) & {\cal F}(c)\end{array}\right)
\left(\begin{array}{rrr}2 & -1 & -1 \\
0 & 1 & -1 \\ 0 & -1 & 1\end{array}\right)
\left(\begin{array}{c} C(a) \\ C(b) \\ C(c)\end{array}\right).
\label{web13}
\end{equation}
We have already given the colour factor of diagram $(a)$ in
eq.~(\ref{Cmushroom}). The colour factors of the other two diagrams
are
\begin{align}
C(b)&=T_1^b\,T_1^b\,T_1^a\,T_2^a=C_{R_1} T_1^a T_2^a;\notag\\
C(c)&=T_1^a\,T_1^b\,T_1^b\,T_2^a=C(b),
\label{Cbc}
\end{align}
where as usual $C_{R_i}$ denotes a quadratic Casimir in the
representation of line $i$. The fact that the colour factors of
diagrams ($b$) and ($c$) are equal, and evaluate to the $C_{R_1}$-dependent part of diagram $(a)$, means that eq.~(\ref{web13}) simplifies
considerably to
\begin{equation}
W_{(3,1)}=\frac12{\cal F}(a)\left[2C(a)-C(b)-C(c)\right]
=-\frac12 N_c(T_1\cdot T_2){\cal F}(a)\,.
\label{web13b}
\end{equation}

The single pole of eq.~(\ref{web13b})
contributes to the two-loop soft anomalous dimension $\Gamma^{(2)}$,
as prescribed by eq.~(\ref{Gamres}). The commutator term that converts
the web into a subtracted web is zero, given that the only lower-order
subwebs in the (3,1) web are the self-energy bubble, and a single
gluon exchange between the two Wilson lines. As discussed in
section~\ref{sec:selfenergygraph}, the colour factor of the 
self-energy graph is a constant, and thus commutes with all other webs. We
can then immediately identify the contribution of (3,1) webs to the
two-loop soft anomalous dimension to be
\begin{equation}
\Gamma^{(2)}\Big|_{(3,1)}=-4w_{(3,1)}^{(2,-1)}=\sum_{i\neq j}
\frac{1}{2}N_c (T_i\cdot T_j)\left(\frac{1}{2\pi}\right)^2
r(\alpha_{ij})\Big[2M_{0,0,0}(\alpha_{ij})-M_{1,0,0}(\alpha_{ij})
\Big],
\label{Gam2_31}
\end{equation}
where we have used eq.~(\ref{mushroomren1}), and summed over all pairs
of Wilson lines $i$ and $j$ (n.b. each pair occurs twice in the sum,
given that the boomerang gluon can be on line $i$ or $j$).
The result in eq.~(\ref{Gam2_31}) agrees with previous calculations, in particular it can be checked that it reproduces the (non-Abelian part of the) coefficient of the single-logarithmic term in eq.~(42) of ref.~\cite{Korchemsky:1987wg} upon relating the kinematic variables according to $\mathrm{\gamma}=\ln \alpha$.

%\Einan{Note the change in formulation below. I avoid saying that one MUST do what we do. There may be other consistent regularizations. I also stress that the regularization of the non-boomerangs is sufficient.}
To summarise, we have shown in detail how to adapt the exponential
regulator of eq.~(\ref{Wilson2}) to the calculation of boomerang
webs. We do it by simply removing this regulator for boomerang gluons, so as to be able to cleanly isolate ultraviolet singularities
associated with the cusp, from those that have to do with the
renormalization of the coupling.
The regularization of the non-boomerang gluons at large distances is sufficient to render diagrams in which they are straddled by non-regularized boomerang gluons infrared-finite.
A simplification in the calculation
of the (3,1) web was that self-energy diagrams (i.e. diagrams $(b)$ and
$(c)$ in figure~\ref{fig:13}) do not contribute to the final expression
for the web, despite the fact that their individual colour factors and
kinematic parts are non-zero. In fact, this property persists at
higher perturbative orders, and thus greatly streamlines the
calculation of boomerang webs at three loops and beyond. We present a
proof of this result in section~\ref{sec:selfenergy}, so that we can
reliably use it throughout the remainder of the paper.

Considering the contribution of the (3,1) web to the soft anomalous dimension in eq.~(\ref{Gam2_31}), we note that the general structure is similar to that of non-boomerang MGEWs analysed in refs.~\cite{Gardi:2013saa,Falcioni:2014pka}, namely an overall rational function $r(\alpha_{ij})$ associated with the non-boomerang gluon, multiplying a \emph{pure} transcendental function. Furthermore, the latter may still be written in terms of the basis functions defined in eq.~(\ref{eq:Mbasis}). However, while non-boomerang MGEWs  are characterized by a uniform maximal weight (that is the contribution to the anomalous dimension at $n$ loops is of weight $2n-1$) the (3,1) web displays mixed (non-uniform) non-maximal weight: eq.~(\ref{Gam2_31}) features both weight 2 ($M_{1,0,0}$) and weight 1 ($M_{0,0,0}$) contributions. The origin of this weightdrop can be traced back to the integration over the boomerang gluon yielding the factor  $1/(1-2\epsilon)$ in eq.~(\ref{FMcalc5}) ({\it{cf.}} a similar factor appearing in the self-energy diagram of  eq.~(\ref{FSE3})). This weightdrop is a general characteristic of boomerang webs and is discussed further below in section~\ref{sec:kinboom} and in the context of the three-loop examples in section~\ref{sec:calculate}.

\subsection{Kinematic factors of boomerang webs}
\label{sec:kinboom}

In section~\ref{sec:MGEWkin}, we discussed the general procedure for
calculating MEGWs of
refs.~\cite{Gardi:2013saa,Falcioni:2014pka}, where an explicit
assumption of this method was that each gluon propagates between 
{\it different} Wilson lines. The latter is no longer true once
boomerang gluons are present, and the results of the previous two
sections can be used to guide us towards a suitable generalisation of
the MGEW integrand, which encompasses the new feature. Given a MGEW with $n$ 
gluon exchanges in total, out of which $b$ are boomerang gluons, 
we must modify eq.~(\ref{gendiag2}) as follows:
\begin{align}
\label{gendiagboom}
\begin{split}
  {\cal F}^{(n)} \left( D \right) & =
  \,  \left(\frac{1}{2} \, g_s^2 \bar{\mu}^{2\e} \,
  \frac{\Gamma(1 - \e)}{4 \pi^{2 - \e}}\right)^n \,
  \prod_{k = 1}^{n-b} \Bigg[ \int_0^\infty d \lambda_k \, \lambda_k^{- 1 + 2 \e} \,
  {\rm e}^{ - i (m - i \varepsilon) \lambda_k }
   \\
&  \qquad \times \, \,
  \int_0^1 d x_k \, \gamma_k\,
  \Big( - x_k^2 - (1 - x_k)^2 +  \gamma_k \, x_k (1 - x_k) +i\varepsilon\Big)^{- 1 + \e} \Bigg]
\\
&  \times \,\,
\prod_{l=n-b+1}^n \Bigg[\int_0^\infty d\lambda_l\, \lambda_l^{-1+2\epsilon}
\int_0^1 d x_l\, 2\left[-(2x_l-1)^2+i\varepsilon\right]^{\epsilon-1}\,
%\\& \times \,
  \, \Bigg]\Theta_D \big[ \left\{ x_i, \lambda_i \right\} \big]  \, .
\end{split}
\end{align}
%\Einan{A few changes here: Fixed the lower limit of summations in the last line from $b+1$ to~\hbox{$n-b+1$}.
%Replaced $\kappa^n$ in front with the relevant combination but without the $(-1)^n$. Inserted a minus into the boomerang propagator (it was already present in the non-boomerang one). Added $i\varepsilon$ prescriptions (they should be there so long as the $(-1)$ is). 
%}
The first two lines correspond to the $n-b$
non-boomerang gluon exchanges, 
and follow a similar format to
eq.~(\ref{gendiag2}), including the presence of the exponential regulator. The third line contains the integrations associated with the $b$ boomerang gluons, where the exponential regulator has been removed as discussed in the previous
section. Furthermore, the propagator function in each $x_l$ integral
has been replaced with its appropriate form for $\gamma_l\rightarrow
2$. 
Finally, the third line also contains the Heaviside functions
implementing the gluon orderings along the Wilson lines for a given
diagram, which may potentially involve both the boomerang, and
non-boomerang, gluons. 

While the convergence of the integrations over the distance parameters $\lambda_k$ for the non-boomerang gluons ($k\leq n-b$) is clearly guaranteed by the regulating exponentials, it is less obvious from eq.~(\ref{gendiagboom}) that also those for the boomerang gluons, that is, $\lambda_l$ for all $n-b+1\leq l\leq n$, are regulated. Closer inspection of these integrals reveals that they are in fact regulated in all cases of interest, namely so long as Wilson-line self-energy subdiagrams are excluded\footnote{As mentioned above, those which are excluded (see figure~\ref{fig:generalselfenergy}), will be shown to have a vanishing exponentiated colour factor in the next section.}. One way to see this is to observe that each boomerang gluon then necessarily straddles at least one other gluon emission, be it another boomerang gluon or a non-boomerang one. Furthermore, each boomerang cluster (a subdiagram involving one or more boomerang gluons) limits the upper integration limit over some non-boomerang gluon along the Wilson line, and it also limits the lower integration limit of some (possibly another) non-boomerang gluon along the same line. Upon performing the integration over all boomerang 
 $\lambda_l$ parameters first, one then necessarily hits both an upper and a lower limit of integration due to the Heaviside functions $\Theta_D \big[ \left\{ x_i, \lambda_i \right\} \big]$, linking the distance parameters $\lambda_l$ for the boomerang gluons to those of the non-boomerang ones, $\lambda_k$ for $k\leq n-b$, which are in turn regularised by the exponentials. This mechanism was seen already in the context of the (3,1) web above (see in particular eq. (\ref{theta31})); we now see that it is completely general, and we will give further examples at three loops in 
section~\ref{sec:calculate}.

%\Einan{I tried to explain why the boomerang integrals are regularised in the above paragraph. However, I have not linked it in detail to the change of variables below. Please see (1) whether you agree with the above argument (2) whether the relation with (3.24) can be better explained.}

It is convenient to rewrite eq.~(\ref{gendiagboom}) so as to expose the general properties of boomerang webs. To this end, we may introduce variable
transformations analogous to eq.~(\ref{defy}):
\begin{equation}
\lambda_k=(1-y_{k-1})\prod_{p=k}^{n-b} y_p,\quad k=1,\ldots,n-b,\quad
y_0=0,
\label{defyboom}
\end{equation}
where the product now includes the non-boomerang gluons only. One may
also decouple the distance parameters $\{\lambda_l\}$ of the boomerang
gluons from their non-boomerang counterparts by defining
\begin{equation}
\lambda_l=y_{n-b}\,\tilde{\lambda}_l,
\label{lamrescale}
\end{equation}
after which one may perform the $y_{n-b}$ integral in
eq.~(\ref{gendiagboom}) to obtain
\begin{align}
\label{gendiagboom2}
  {\cal F}^{(n)} \left( D \right) & =
 \,  \kappa^n\,\Gamma(2n\epsilon)
  \prod_{k = 1}^{n-b} \left[ \int_0^1 d x_k \, \gamma_k\,
  P_{\epsilon}(x_k,\gamma_k)\right]
 \prod_{l=n-b+1}^n\left[
\int_{\frac12}^1 d x_l\, [(2x_l-1)^2]^{\epsilon-1}\right]\, 
%\nonumber \\ &\times 
\phi_D^{(n)}(\{x_i\};\epsilon),
\end{align}
where the kernel is now defined by
\begin{align}
\phi_D^{(n)}(\{x_i\};\epsilon)&=
2^b
 \prod_{k=1}^{n-b-1} \left[\int_0^1 dy_k (1-y_k)^{-1+2\epsilon}
y_k^{-1+2k\epsilon}\right]
%\notag\\&\quad\times 
 \prod_{l=n-b+1}^n\left[\int_0^\infty d\tilde{\lambda}_l
\tilde{\lambda}_l^{-1+2\epsilon}\right]
\Theta_D \big[ \big\{ x_i, \tilde{\lambda}_i,y_i \big\} \big],
\label{phiboom}
\end{align}
where the $\tilde{\lambda}_l$ integrals are bounded by the Heaviside functions, as explained above.

The general representation of boomerang MGEWs in eq.~(\ref{gendiagboom2}) gives us an opportunity to recall some of the general properties of MGEWs~\cite{Henn:2013wfa,Gardi:2013saa,Falcioni:2014pka}, and then pinpoint the differences between those containing boomerang gluons and those which do not.  Equation~(\ref{gendiagboom2}) much like its boomerang-free analogue, eq.~(\ref{gendiag3}), represents at $n$-loop order an integration over the $2n$ positions of emission and absorption of the $n$ gluons along the Wilson lines.
As we have seen in the previous section,  these integrals ultimately lead to a result for the subtracted web, that is a contribution to the soft anomalous dimension, taking the form of  eq.~(\ref{subtracted_web_mge_kin}), with a rational factor consisting of a factor of $r(\alpha_{ij})$ for each gluon exchange between lines $i$ and $j$, multiplying a pure transcendental function of $\{\alpha_{ij}\}$ with polylogarithmic weight $2n-1$.  Equation~(\ref{gendiag3}) explains the origin of this pure, maximal weight structure: every integral over $y_k$ in eq.~(\ref{phi_D}) is an integral over a $d\log$ form, with endpoint singularities regularised by $\epsilon>0$. The resulting kernel is therefore a pure function of weight $n-1$, that is, the $\epsilon^0$ term in its Laurent expansion is of weight $n-1$, and upon assigning $\epsilon$ weight $-1$, all the terms in the Laurent expansion have the same weight.
A similar thing happens at the next stage, when  the kernel is integrated with respect to the propagators in eq.~(\ref{gendiag3}). 
At this stage the linear denominator is generated by the propagators (see eq.~(\ref{p0_part_fracs})), and again, each and every integral over $x_k$ results in an increase of one unit in the transcendental weight. 
This is true for each and every diagram contributing to the web, as well as the commutators entering the subtracted web.

Consider now the analogous structure of the integration in the case of boomerang webs. In eq.~(\ref{phiboom}) we see $n-b-1$ integrals over non-boomerang $y_k$  variables plus $b$ integrals over boomerang distance scales $\tilde{\lambda}_l$. Both are 
of $d\log$ form, regularized by $\epsilon>0$. Thus, again in total we have $n-1$ integrals each contributing to the weight of  $\phi_D$. The latter must therefore still be a pure function of weight $n-1$. The differences to non-boomerang webs occur at the next step, when integrating over the kernel in eq.~(\ref{gendiagboom2}). First, a factor of $r(\alpha_{ij})$ is only generated by 
the $n-b$ non-boomerang propagator integrals over $x_k$. Second, while each of the latter integrals is a d$\log$ form, regularised by $\epsilon>0$, which therefore increases the weight by one unit, 
the remaining~$b$ integrals over $x_l$ take a rather different form:
\begin{equation}
\label{x_l_integral}
\int_{\frac12}^1 dx_l \,(2x_l-1)^{2\epsilon-2}\left(\ldots\right),
\end{equation}
where the ellipsis denotes the remaining integrand, containing
Heaviside functions which may depend on the $\{x_l\}$. 
Such integrals contain a potential divergence associated with the lower limit
$x_l\rightarrow\frac12$. 
This lower limit of integration corresponds to a local  (instantaneous) 
emission and absorption of the boomerang gluon. 
Integration over $x_l$ in eq.~(\ref{x_l_integral}) produces a factor of
\begin{equation}
\label{x_l_integral_res}
\frac{1}{1-2\epsilon}
\end{equation}
in the final expression for the kinematic function of the corresponding boomerang web diagram. 

The pole at $\epsilon=\frac12$ represents a linear power divergence. 
We have already seen such a factor in the self-energy web in eq.~(\ref{FSE3}) as well as in the (3,1) 
web in eq.~(\ref{FMcalc5}). 
We now see that this is a general feature of boomerang webs. 
We further note however that there is a qualitative difference between the above two cases. In the self-energy web, the extra factor $\left(\ldots\right)$ in eq.~(\ref{x_l_integral}) is absent, so the integral only exists for $\epsilon>\frac12$, and one must first compute it there and then analytically continue the result towards $\epsilon=0$, where it is ultimately expanded. In contrast, in the case of the (3,1) 
web we can see from eq.~(\ref{FMcalc4}), as already discussed there, 
that a factor of the form $((1-x)^{-2\epsilon}-x^{-2\epsilon})$ regularises the endpoint singularity at $x=\frac12$. It is also clear that the occurrence of this regularising factor is rather general: it appears due to the difference of the two limits of integration over the boomerang gluon $\tilde{\lambda}_l$ (this is $\lambda_1$ in eq.~(\ref{theta31})), which must both coincide with the emission point of the other gluon when the boomerang gluon is contracted to a point. This would therefore be the precise form of the factor $\left(\ldots\right)$ in eq.~(\ref{x_l_integral}) whenever a boomerang gluon straddles a single gluon emission. 
The same considerations apply more generally: whenever a boomerang gluon straddles other emissions at some position $\lambda_0$ on the line, both the upper and lower limit of $\tilde{\lambda}_l$ coincide with $\lambda_0$ when the boomerang gluon is shrunk to a point, namely at $x_l \to \frac12$. We therefore expect that the factor $\left(\ldots\right)$ multiplying the singularity at $x_l=\frac12$, would always have a Taylor expansion that begins with \emph{a linear term}, $(x_l-\frac12)$. This factor regularises 
the double pole at $x_l=\frac12$ and renders eq.~(\ref{gendiagboom2}) well-defined for any $\epsilon>0$. 
Of course, poles at $\epsilon\to 0$ will be generated due to end-point singularities. In addition, despite the regularising factor in eq.~(\ref{x_l_integral}), the pole at $\epsilon=\frac12$ survives. 
These features are already present in the example of the (3,1) web in eq.~(\ref{FMcalc5}) and we shall illustrate them in more complex three-loop examples in section~\ref{sec:calculate}.

The implications the analysis above, and specifically the presence of the pole at $\epsilon=\frac12$, have on the transcendental structure of the kinematic function ${\cal F}^{(n)}$ in eq.~(\ref{gendiagboom2}) are clear: instead of increasing the weight by one, as the usual propagator integrals $x_k$ in  eq.~(\ref{gendiagboom2}) do, the $b$ boomerang integrals leave the weight unchanged in as far as the contributions arising from the leading term in the expansion of eq.~(\ref{x_l_integral_res}) are concerned, and decrease it further in contributions arising from higher-order terms in the $\epsilon$ expansion. This implies, first, that the maximal weight attained in the relevant subtracted web is $2n-1-b$, i.e. a weight drop of one unit for every boomerang web when compared to ordinary MGEWs of the same loop order, and second, that when higher-order terms in the $\epsilon$ expansion are relevant, the subtracted boomerang web would feature mixed (non-uniform) weight.  These phenomena were exemplified in the (3,1) web, which features basis functions with weights
2 and~1 (to be contrasted with the maximal weight of 3 of subtracted two-loop MGEWs, see~\cite{Gardi:2013saa,Falcioni:2014pka}). We will see further examples of this at three loops in section~\ref{sec:calculate} (see a summary in table~\ref{tab:weights_3loop} there).

In order to present explicit results, it is useful to generalise the definition of a {\it
  subtracted web kernel} from eq.~(\ref{subtracted_web_mge_kin}), to
the case in which boomerang gluons may be present:
\begin{equation}
  F^{(n)}_{W, \, j} \big( \alpha_i \big)  =
    \prod_{k = 1}^{n-b} \left[ \int_0^1\, d x_k \, p_0 (x_k, \alpha_k) \right]  \,
  {\cal G}^{(n)}_{W, \, j} \Big(x_i, q(x_i, \alpha_i) \Big).
\label{calGboom}
\end{equation}
Here, as above, $b$ is the number of boomerang gluons, and thus the
integration only includes propagator functions associated with
non-boomerang exchanges. Having discussed the general kinematic
properties of boomerang webs, let us now return to the decoupling property of self-energy-type diagrams discussed in the previous section, namely that web diagrams spanning two or more Wilson lines, which contain self-energy subdiagrams (such as those in figures~\ref{fig:13}$(b)$ and~\ref{fig:13}$(c)$) do not contribute to the soft anomalous dimension at any order.

\section{Decoupling of self-energy diagrams at all orders}
\label{sec:selfenergy}

Boomerang webs at arbitrary orders in perturbation theory will contain
many individual diagrams in which gluons form self-energy-like loops,
without straddling one or more gluon emissions that leave the Wilson
line. The aim of this section is to formally prove that such graphs do
not end up contributing to the web after combination with the web
mixing matrix of eq.~(\ref{Wnidef}). Put another way, if we define the
{\it exponentiated colour factors} of diagrams in a given web via
\begin{equation}
\tilde{C}(D)=\sum_{D'} R_{DD'} C(D'),
\label{ECFs}
\end{equation}
where $C(D)$ is the conventional colour factor of diagram $D$, then
the exponentiated colour factor of a diagram containing a self-energy loop is zero.

\begin{figure}[htb]
\begin{center}
\scalebox{0.9}{\includegraphics{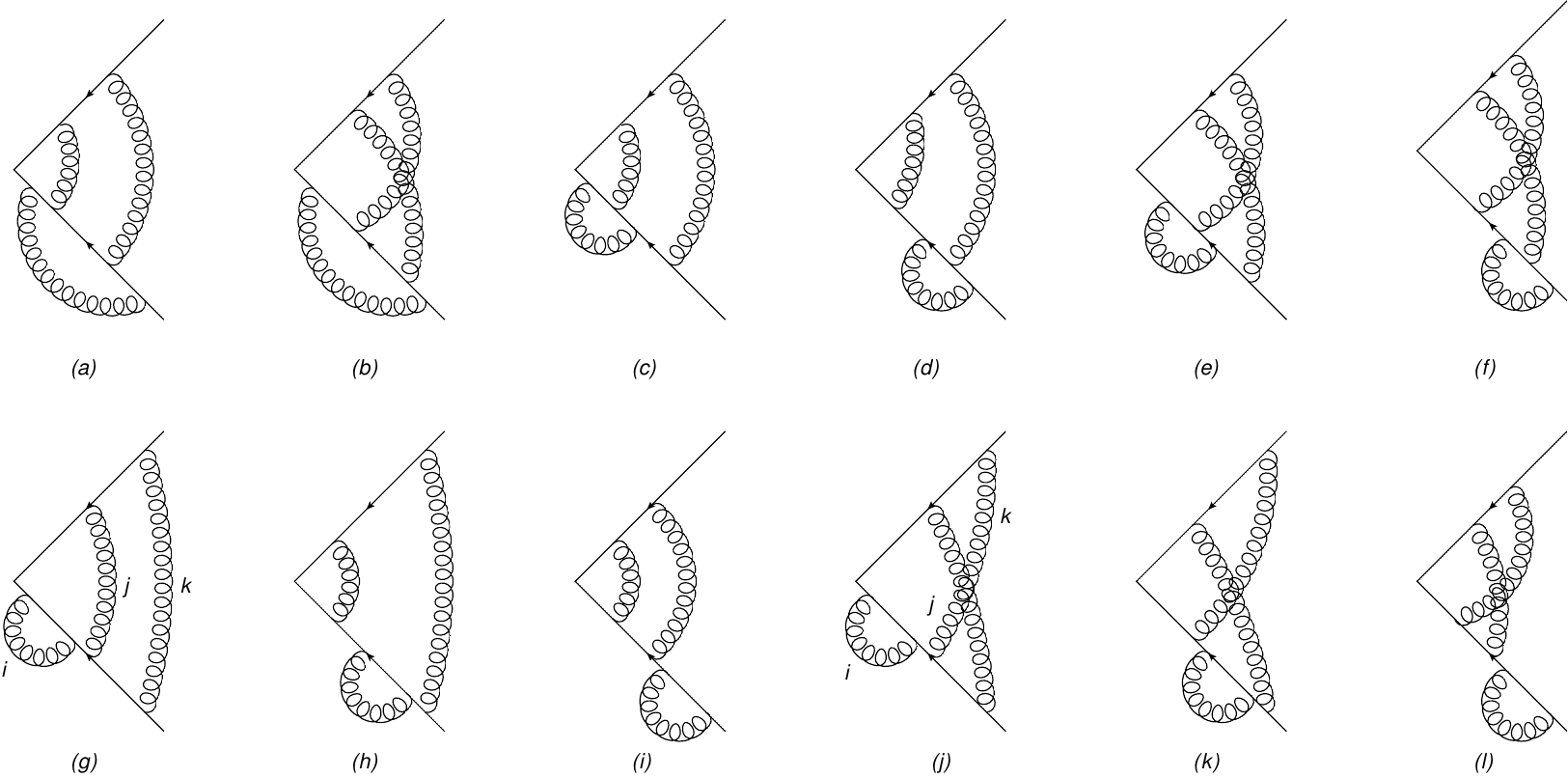}}
\caption{The (2,4) web, where we have assigned replica indices to the
  gluons in diagrams $(g)$ and~$(j)$. Although not drawn here, it is assumed that
  additional Wilson lines emanate from the Wilson-line cusp.}
\label{fig:24}
\end{center}
\end{figure}

To guide
the proof, let us first consider a non-trivial example, namely the
(2,4) web of figure~\ref{fig:24}.
This has 12 diagrams, 6 of which -- diagrams $(g)$ through $(l)$ -- involve self-energy bubbles. Let us take two of the graphs, namely $(g)$ and $(j)$, and
show that their exponentiated colour factors vanish. To this end, we can perform a replica
analysis as in ref.~\cite{Gardi:2010rn} or appendix~\ref{app:Rcalc},
and we have labelled the gluons in the figure with appropriate replica
indices $(i,j,k)$. In table~\ref{tab:24}, we show the possible
hierarchies $h$ of replica indices, together with their multiplicities
$M_N(h)$. We also show, for each diagram $D$, the colour factor of the
diagram obtained from ordering the replica indices along the Wilson
line (such that larger replica indices are closer to
the Wilson-line vertex), labelled by ${\cal R}[D|h]$. The exponentiated colour factors of the two diagrams
considered are given by eq.~(\ref{Ctildedef}), and we find
\begin{align}
\label{Ctilde24}
\begin{split}
\tilde{C}(g)&=\frac{C(g)}{2}-\frac{C(h)}{3}-\frac{C(i)}{6}; \\
\tilde{C}(j)&=-\frac{C(g)}{3}-\frac{C(h)}{3}+\frac{2C(i)}{3}
+\frac{C(j)}{2}-\frac{C(l)}{2}.
\end{split}
\end{align}
\begin{table}
\begin{center}
\begin{tabular}{c|c|c|c|c}
$h$ & ${\cal R}[g|h]$ & ${\cal R}[j|h]$ & $M_N(h)$ & ${\cal
  O}(N)$ part of $M_N(h)$ \\
\hline
$i=j=k$ & $C(g)$ & $C(j)$ & $N$ & 1\\
$i=j<k$ & $C(h)$ & $C(h)$ & $\frac12 N(N-1)$ & $-\frac12$ \\
$i=j>k$ & $C(g)$ & $C(g)$ & $\frac12 N(N-1)$ & $-\frac12$ \\
$i=k<j$ & $C(h)$ & $C(h)$ & $\frac12 N(N-1)$ & $-\frac12$ \\
$i=k>j$ & $C(g)$ & $C(g)$ & $\frac12 N(N-1)$ & $-\frac12$ \\
$j=k<i$ & $C(g)$ & $C(j)$ & $\frac12 N(N-1)$ & $-\frac12$ \\
$j=k>i$ & $C(i)$ & $C(l)$ & $\frac12 N(N-1)$ & $-\frac12$ \\
$i<j<k$ & $C(i)$ & $C(i)$ & $\frac16 N(N-1)(N-2)$ & $\frac13$ \\
$i<k<j$ & $C(i)$ & $C(i)$ & $\frac16 N(N-1)(N-2)$ & $\frac13$ \\
$j<i<k$ & $C(h)$ & $C(h)$ & $\frac16 N(N-1)(N-2)$ & $\frac13$ \\
$j<k<i$ & $C(g)$ & $C(g)$ & $\frac16 N(N-1)(N-2)$ & $\frac13$ \\
$k<i<j$ & $C(h)$ & $C(h)$ & $\frac16 N(N-1)(N-2)$ & $\frac13$ \\
$k<j<i$ & $C(g)$ & $C(g)$ & $\frac16 N(N-1)(N-2)$ & $\frac13$ \\
\end{tabular}
\caption{Replica analysis of the (2,4) web of figure~\ref{fig:24}.}
\label{tab:24}
\end{center}
\end{table}

We may now use the fact that a self-energy loop contributes a factor
$T_i^a T_i^a=C_{R_i}$ to the colour factor of any web diagram, which
is diagonal in colour space. Thus, graphs which differ only by the
placement of a self-energy loop on a given Wilson line have equal
colour factors. For the specific web of figure~\ref{fig:24}, this
implies
\begin{equation}
C(g)=C(h)=C(i),\qquad C(j)=C(k)=C(l).
\label{Cequal}
\end{equation}
Equation~(\ref{Ctilde24}) then immediately implies
\begin{equation}
\tilde{C}(g)=\tilde{C}(j)=0.
\label{Ctilde0}
\end{equation}

\begin{figure}[b]
\begin{center}
\scalebox{0.6}{\includegraphics{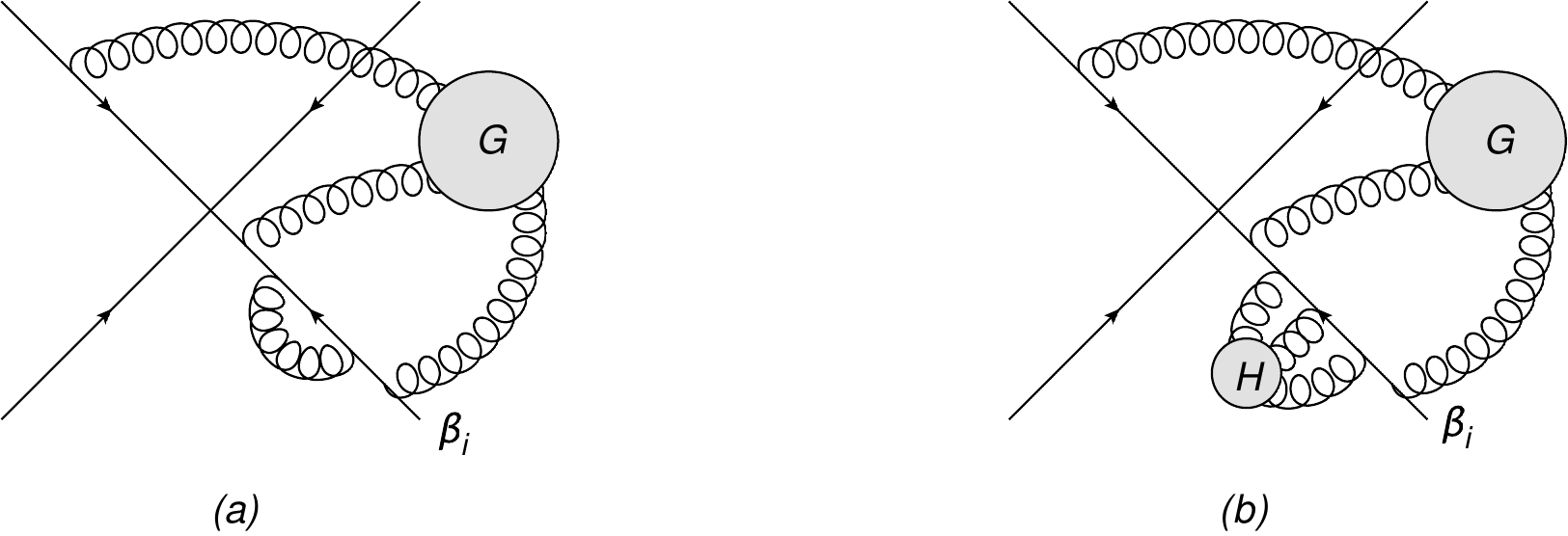}}
\caption{$(a)$ General web diagram containing a self-energy loop, where
  the rest of the diagram $G$ consists of a number of connected pieces;
  $(b)$ generalisation to include a non-trivial subdiagram $H$ in place of
  the self-energy loop.}
\label{fig:generalselfenergy}
\end{center}
\end{figure}
In the above replica analysis (and as noted in
ref.~\cite{Gardi:2010rn}), a hierarchy $h$ containing $r$ distinct
replica indices has multiplicity
\begin{equation}
{\cal M}_N[h(r)]=\left(\begin{array}{c}N\\ r\end{array}\right)=
\frac{1}{r!}N(N-1)\ldots (N-r+1),
\label{multr}
\end{equation}
and thus contributes the following to the ${\cal O}(N)$ part of the replicated
colour factor:
\begin{equation}
{\cal M}_N[h(r)]\Big|_{{\cal O}(N)}=\frac{(-1)^{r-1}}{r}.
\label{multrN}
\end{equation}
Now consider a general web diagram $D$ containing a self-energy loop
on line $i$, as shown in figure~\ref{fig:generalselfenergy}($a$). Here
$G$ is the rest of the graph, which must contain at least one gluon connecting to another Wilson line, and may potentially consist of a number of
connected pieces. 
Let us assume that a hierarchy of replica indices has already been assigned to $G$, and that this has $r$ distinct indices.
For a given hierarchy $h$ of replica indices for the entire diagram, we must reorder
those gluons on line $i$ whose replica numbers differ. However, this
reordering can never make the self-energy loop straddle another gluon
emission: both endpoints of the boomerang gluon have the same replica
number, and so cannot appear on opposite sides of another gluon
attachment whose replica number is different. The most that can happen
is that the self-energy loop as a whole is shifted along the line. According to eq.~(\ref{Ctildedef}), for each
hierarchy $h$, we must record the colour factor ${\cal R}[D|h]$
obtained after the reordering. The contribution to this colour factor
from the self-energy loop is $C_{R_i}$, which is diagonal in colour
space. Given that the hierarchy of replica indices for the subgraph $G$ have
already been fixed, it follows that for every hierarchy $h$ (including
the replica index for the self-energy loop), the reordered
colour factor ${\cal R}[D|h]$ is the same, and given by
\begin{equation}
{\cal R}[D|h]=C_{R_i} {\cal R}[G|h]
\label{CtildeSE1}
\end{equation}
There are then two possibilities when inserting replica index $i$ of the self-energy loop to give the full hierarchy $h$:
\begin{enumerate}
\item The index $i$ is the same as one of the $r$ indices already
  assigned in $G$. Here the $h$ contributes with a
  multiplicity factor $(-1)^{r-1}/r$, from eq.~(\ref{multrN}). There
  are $r$ choices for $i$,
  so that the contribution to the exponentiated colour factor from all such choices is 
  \begin{equation}
    r\frac{(-1)^{r-1}}{r}C_{R_i}{\cal R}[G|h]=(-1)^{r-1} C_{R_i}{\cal
      R}[G|h],
    \label{CtildeSE2}
  \end{equation}
  where we have used eq.~(\ref{CtildeSE1}) in eq.~(\ref{Ctildedef}).
\item The index $i$ is distinct from the $r$ indices already ordered
  in $G$. Now there are $r+1$ distinct replica numbers in total in $h$, so
  that $h$ has a multiplicity factor
  $(-1)^{r}/(r+1)$. There are $r+1$ possible placings for the replica
  index $i$ (i.e. it may be less than or greater than any of the
  existing $r$ replica numbers), and thus the total contribution to
  the exponentiated colour factor is
  \begin{equation}
    (r+1)\frac{(-1)^{r}}{r+1}C_{R_i}{\cal R}[G|h]=(-1)^{r} C_{R_i}{\cal
      R}[G|h].
    \label{CtildeSE3}
  \end{equation}
\end{enumerate}
%\Einan{Why is it correct to discard boundary cases where say  the largest replica number in G is actually $N$ or the smallest one is 1 (or both) in which case the number of assignments for the self energy loop replica is a $r$ (or $r-1$) rather than $r+1$. I suspect that the answer is that we cannot really assign the replica indices to G separately, but we must consider the entire graph when assigning the replica indices, in which case all (r+1) possibilities exist.}
Adding together eqs.~(\ref{CtildeSE2}) and~(\ref{CtildeSE3}), the
total contribution to $\tilde{C}(D)$ from the hierarchy $h$ for subdiagram $G$ is
\begin{equation}
\left((-1)^{r-1}+(-1)^r\right)C_{R_i}{\cal R}[G|h]=0.
\label{CtildeD0}
\end{equation}
The full calculation of $\tilde{C}(D)$ requires a sum over all hierarchies for
the \emph{full} diagram.  This is easily rewritten as a sum over all hierarchies, $h$,
of $G$ then a sum over all assignments of $i$ for the extra gluon.  Each of
these sub-sums is zero by eq.~(\ref{CtildeD0}), hence the required result that
indeed $\tilde{C}(D)=0$.

To clarify the above proof, we can revisit the replica
analysis of the (2,4) web in table~\ref{tab:24}. Considering first
diagram $(g)$, the subdiagram $G$ consists of a ladder of two gluon
exchanges between the two active Wilson lines. There are then three
possible assignments of replica indices $j$ and $k$ to $G$. If $j=k$
(corresponding to $r=1$ distinct indices), then in assigning a replica
index to the self-energy loop one may choose $i=j$ or $i\neq j$. When the index
$i$ is the same as $(j,k)$ one obtains the colour factor $C(g)$, with multiplicity factor
1. Alternatively, we may have $i<j$ or $i>j$, giving colour factors $C(i)$ or
$C(g)$ respectively, and each with a multiplicity factor of
$-1/2$. However, $C(g)=C(i)$, so that the total contribution to the
exponentiated colour factor is $C(g)\big(1-\frac12-\frac12\big)=0$. A similar analysis can be applied to the other
possible hierarchies $j<k$ and $j>k$ for the subdiagram $G$, and also
for the second diagram $(j)$ considered in table~\ref{tab:24}.

As well as the above result for self-energy loops, we can also prove a more general result. Consider replacing the self-energy loop in figure~\ref{fig:generalselfenergy}$(a)$ with a subdiagram $H$ consisting of more than one connected piece in general, but such that none of its gluon attachments on Wilson line~$i$ straddle any gluon emissions not in~$H$.
Let us again label the complete diagram by $D$,
containing the subdiagrams $G$ and $H$, such that $H$ is assumed to only attach to Wilson line $i$, while $G$ involves attachments to $i$ and other Wilson lines.
We shall now show that the exponentiated colour factors for such diagrams vanish (a
result we will use for three-loop boomerang webs in
section~\ref{sec:calculate}).
Let us assume that $r$ replica indices have been assigned to $G$, and $s$ indices to $H$, where the latter potentially
overlap with the indices in $G$.

As in the previous proof, we will split the sum over all replica index
hierarchies into a sum over sums, which between them cover all hierarchies.  We
will then show that each is zero.  To do this, we define a sub-hierarchy,
$\{\{r\},\{s\}\}$, of a replica hierarchy $h$ to be the separated ordering for $G$
and $H$.  This is equivalent to undoing a shuffle.  For example, $h=\{r_1 < s_1=r_2<s_2<r_3\}$, has sub-hierarchy
$\{\{r_1<r_2<r_3\},\{s_1 <s_2\}\}$.  Each $h$ has a unique
sub-hierarchy, but many different hierarchies may give the same sub-assignment
(e.g.~$r_1<s_1<r_2<s_2<r_3$ gives the same sub-hierarchy as above).  We then
split the sum in eq.~(\ref{ECFs}) according to sub-assignment:
\begin{equation}
\tilde{C}(D)=\sum_{{\rm all}\,h} {\cal M}_N[h(r,s)]\Big|_{{\cal O}(N)}
{\cal R}[D|h]
%C_h(D)
=
\sum_{\{\{r\},\{s\}\}} \,\, \sum_{{\rm all}\,h\to \{\{r\},\{s\}\}} {\cal
  M}_N[h(r,s)]\Big|_{{\cal O}(N)}\,
{\cal R}[D|h]\,,
%C_h(D),
\label{eq:splitsum}
\end{equation}
where ${\cal R}[D|h]$ is the colour factor of the diagram obtained from $D$ after
hierarchy $h$ is applied.

\begin{figure}
\begin{center}
\scalebox{0.7}{\includegraphics{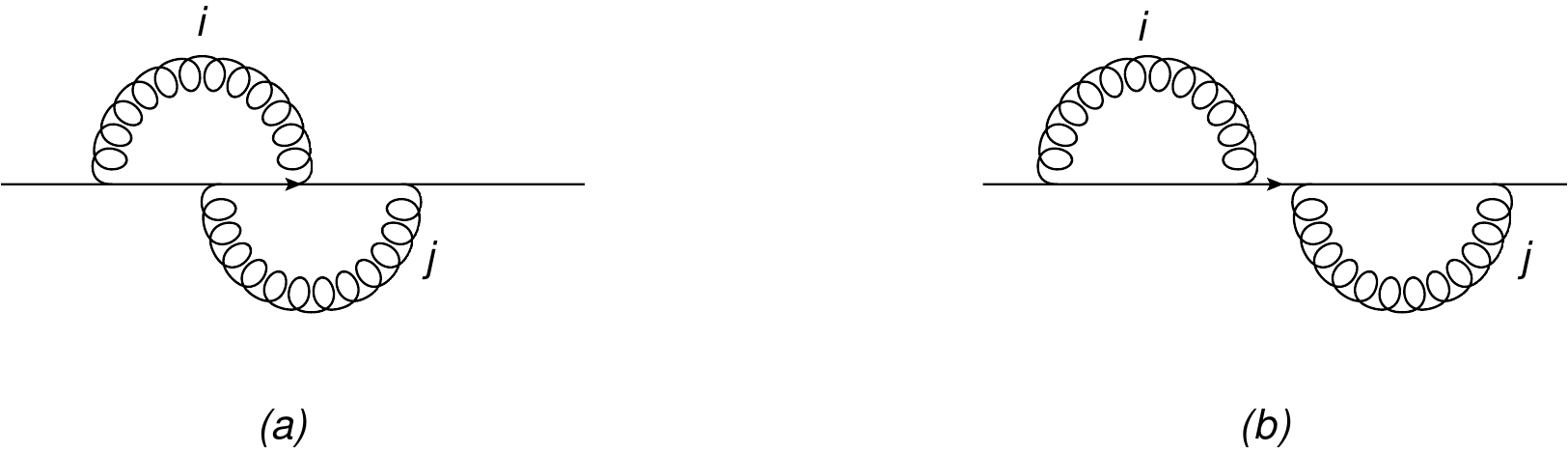}}
\caption{$(a)$ Generalised self-energy diagram on a single Wilson line, with replica indices $i$ and $j$; $(b)$ replica ordered version, if $i<j$.}
\label{fig:Hexample}
\end{center}
\end{figure}

For a
given assignment $\{s\}$ of replica indices to $H$, the subdiagram will split
into a number of pieces, each of which has a colour factor
proportional to the identity, as we are only considering diagrams where $H$ does not connect to any Wilson line
other than $i$ and furthermore does not straddle any gluon in $G$ on line $i$. An example for $H$ is given in
figure~\ref{fig:Hexample}$(a)$, which shows a subgraph on a single
Wilson line, consisting of two overlapping self-energy loops. If we
assign replica indices such that $i<j$, the graph will split into two
separate self-energy loops (figure~\ref{fig:Hexample}$(b)$), each with colour factor
$C_{R_i}\mathbb{1}$, which is different to the colour factor of
figure~\ref{fig:Hexample}$(a)$.  We use the organisation of the sum in
eq.~(\ref{eq:splitsum}), to treat the contribution for each hierarchy of $H$
separately, so in this example the hierarchies with $i=j$ are considered separately
to those with $i<j$.

%\Einan{Note that we must exclude the possibility of $G$ being empty, since webs where $H$ is the entire diagram are non-trivial (e.g. diagram $(a)$ in the (4) web has a non-vanishing exponentiated colour factor. This should be explained.} 

For a fixed assignment of the $r$ replica indices to $G$ and $s$ indices in $H$,
different hierarchies of the full set of replicas will potentially reorder the parts of $H$ along
the Wilson line, according to the mutual ordering between the indices
of $G$ and $H$. However, because both before and after this reordering none of
the subdiagrams in $H$ straddle any gluon emissions in $G$ \emph{and} the colour
factor of the subdiagram $H$ (and its subdiagrams if present) are proportional to the identity, it follows that
all choices with the same sub-hierarchy $\{\{r\},\{s\}\}$ will lead to a diagram with the
same colour factor.
We may express this common colour factor as the product
${\cal R}[G|\{r\}]\,{\cal R}[H|\{s\}]$,
%$C_{\{r\}}(G)C_{\{s\}}(H)$,
noting explicitly that it depends on the choice of the $r$ indices in
$G$ and $s$ indices in $H$, but not on how they are interleaved.

Combining $H$ and $G$, we will have a total number of indices
\begin{equation}
n=r+s-k,\quad k\in[0,\min(r,s)]
\label{ndef}
\end{equation}
where $k$ is the number of indices that overlap (i.e.~are set equal) between $G$ and
$H$. From eq.~(\ref{multrN}), the overall hierarchy $h$ will
contribute with a multiplicity factor
\begin{displaymath}
\frac{(-1)^{n-1}}{n}=\frac{(-1)^{r+s-k-1}}{r+s-k}.
\end{displaymath}
% The total contribution to the exponentiated colour factor from the sub-sum with fixed sub-hierarchy
% $\{\{r\},\{s\}\}$ will then be
Returning to the expression for the exponentiated colour factor of $D$ in eq.~(\ref{eq:splitsum}), we have
now found that the inner sum can be written as
\begin{equation}
\sum_{{\rm all}\,h\to \{\{r\},\{s\}\}} {\cal
  M}_N[h(r,s)]\Big|_{{\cal O}(N)}\, {\cal R}[D|h] \,=\,{\cal R}[G|\{r\}]\,{\cal R}[H|\{s\}]
 %C_{\{r\}}(G)C_{\{s\}}(H)
\sum_{k=0}^{\min(r,s)}N_{r,s,k}\frac{(-1)^{r+s-k-1}}{r+s-k},
\label{Ctildeproof1}
\end{equation}
where $N_{r,s,k}$ is the number of ways of assigning $r$ indices to
$G$ and $s$ indices to $H$, with $k$ overlaps. To find this,
first note that there are $r+s-k$ distinct indices in total. There are
\begin{displaymath}
\left( \begin{array}{c}r+s-k \\ k\end{array}\right)
\end{displaymath}
ways of choosing which indices correspond to the overlapping ones. Of
the $r+s-2k$ remaining indices, $(r-k)$ must be chosen to be the
remaining indices of $G$ (which has $r$ distinct indices in total), for which there are
\begin{displaymath}
\left(\begin{array}{c} r+s-2k \\ r-k\end{array}\right)
\end{displaymath}
possible choices. The remaining $s-k$ indices are then automatically
the remaining indices in $H$, and one thus finds
\begin{equation}
N_{r,s,k}=\left( \begin{array}{c}r+s-k \\ k\end{array}\right)
\left(\begin{array}{c} r+s-2k \\ r-k\end{array}\right).
\label{Nrsk}
\end{equation}
Note that $N_{r,s,k}$ is symmetric under $r\leftrightarrow s$ as it must be.
From eq.~(\ref{Ctildeproof1}), the total contribution to the
exponentiated colour factor of diagram $D$ is then
\begin{equation}
%C_{\{r\}}(G)C_{\{s\}}(H)
{\cal R}[G|\{r\}]\,{\cal R}[H|\{s\}]
\sum_{k=0}^{\min(r,s)}
\left( \begin{array}{c}r+s-k \\ k\end{array}\right)
\left(\begin{array}{c} r+s-2k \\ r-k\end{array}\right)
\frac{(-1)^{r+s-k-1}}{r+s-k}.
\label{Ctildeproof2}
\end{equation}
To carry out the sum, we may write it as an infinite sum, i.e.,
\begin{align}
&\sum_{k=0}^{\min(r,s)}
\left( \begin{array}{c}r+s-k \\ k\end{array}\right)
\left(\begin{array}{c} r+s-2k \\ r-k\end{array}\right)
\frac{(-1)^{r+s-k-1}}{r+s-k}\notag\\
&=\sum_{k=0}^{\min(r,s)}\frac{\Gamma(r+s-k)}{k!\Gamma(r-k+1)\Gamma(s-k+1)}
(-1)^{r+s-k-1}\notag\\
&=\sum_{k=0}^{\infty}\frac{\Gamma(r+s-k)}{k!\Gamma(r-k+1)\Gamma(s-k+1)}
(-1)^{r+s-k-1}
\end{align}
relying on the fact that for integer $r$ and $s$ the $\Gamma$ functions in the denominator render all terms with $k>\min(r,s)$ identically zero. Next we may consider generic values of $r$ and $s$, for which one may establish that
\begin{equation}
\sum_{k=0}^{\infty}\frac{\Gamma(r+s-k)}{k!\Gamma(r-k+1)\Gamma(s-k+1)}
(-1)^{r+s-k-1}= (-1)^{r+s-1} \frac{\sin(\pi r)\sin(\pi s) }{\pi r s \sin(\pi (r + s))}\,,
\end{equation}
which vanishes in the limit where $r$ and $s$ are positive integers.
Thus, each assignment of $r$ replica numbers to the subdiagram $G$  and $s$
replica numbers to the subdiagram $H$ in
figure~\ref{fig:generalselfenergy}$(b)$ leads, upon summing over all
hierarchies, to a vanishing contribution to the exponentiated colour factor of the whole diagram. Hence, exponentiated colour factors of diagrams which can be split as in figure~\ref{fig:generalselfenergy}$(b)$ are zero.

Note that a consistency check of the above proof is that for the case
$s=1$, the sum in eq.~(\ref{Ctildeproof2}) reduces to just two terms, $k=0$ and $k=1$, yielding
\begin{displaymath}
%C_{\{r\}}(G)C_{\{s\}}(H)
{\cal R}[G|\{r\}]\,{\cal R}[H|\{s\}]
\left[(r+1)\frac{(-1)^{r}}{r+1} + r\frac{(-1)^{r-1}}{r}\right]=0,
\label{sum02}
\end{displaymath}
as encountered in the previous proof ({\it{cf.}} eqs. (\ref{CtildeSE3}) and (\ref{CtildeD0})).

In summary, we have shown that graphs of the general form of
figure~\ref{fig:generalselfenergy}$(b)$, in which $G$ and~$H$ are
subdiagrams consisting (in general) of any number of connected pieces, such that $H$ connects to a single line and does not straddle any emission in $G$, have vanishing exponentiated colour factors. Thus, from
eq.~(\ref{Wnidef}), their kinematic parts do not contribute to the
logarithm of the soft function, and so do not have to be
calculated. For example, we will see in section~\ref{sec:calculate} that of the 15
diagrams in the (5,1) web, only 4 have non-zero exponentiated colour
factors. This greatly simplifies the calculation of boomerang webs
at three-loop order, which we proceed to do in the following section.

We note that in the above proof we required that $G$ contained at least one gluon connecting to other Wilson lines.  This means that the proof above does not apply to webs which consist of multiple boomerang gluons on a single line and nothing else.
Indeed these pure self-energy webs are not zero (as we saw with the one-loop self-energy graph in section 3.1) and they do contribute to the soft anomalous dimension. However, because they involve just a single line, these contributions must be entirely independent of kinematic variables and we do not consider them further in this paper.

\section{Boomerang webs at three-loop order}
\label{sec:calculate}

In the previous sections, we have prepared the necessary ingredients
for the calculation of boomerang MGEWs. We will now focus on the explicit calculation of these webs at the 
three-loop order. We have seen that at two loops there is a single boomerang web, the (3,1) web. At three loops there are five: two webs involving three Wilson lines, (1,1,4) and (1,2,3) and three webs involving two lines: (3,3), (5,1), and (2,4). As before we shall not assume colour conservation amongst the Wilson lines involved, allowing for a hard interaction vertex involving more  lines. The three-line webs are discussed first, in section~\ref{sec:3lines}, followed by the two-line ones in section~\ref{sec:2lines}.

\subsection{Boomerang webs connecting three Wilson lines}
\label{sec:3lines}

For three-loop webs connecting three lines, the colour basis of
eq.~(\ref{colbasis}) can be written in the explicit
form\footnote{Note that here we use the usual anticommutator notation $\{T_i^a,T_i^b\}=T_i^aT_i^b+T_i^bT_i^a$, rather than the notation of eq.~(\ref{Csym}) which includes an extra factor of $1/n!=1/2$.}~\cite{Gardi:2013ita}
\begin{align}
  \label{colbasis3}
  \begin{split}
    c_1^{[3,3]} &= - f^{a c e} f^{bde} T_1^{\{a,b\}} T_2^c T_3^d
= \{T_1^{a},T_1^{b}\}[T_2^{b},T_2^{c}][T_3^{a},T_3^{c}]\\
    c_2^{[3,3]} &= - f^{cae} f^{bde} T_1^a T_2^{\{b,c\}}
    T_3^d 
= [T_1^{a},T_1^{b}]\{T_2^{b},T_2^{c}\}[T_3^{a},T_3^{c}]\\
    c_3^{[3,3]} &= - f^{cbe} f^{ade} T_1^a T_2^b
    T_3^{\{c,d\}}
= [T_1^{a},T_1^{b}][T_2^{b},T_2^{c}]\{T_3^{a},T_3^{c}\}\\
    c_4^{[3,3]} &= -\frac12 i f^{acd} f^{bef} f^{def} T_1^a T_2^b
    T_3^c = \frac12 i N_c f^{abc} T_1^a T_2^bT_3^c
= [T_1^{a},T_1^{b}][T_2^{b},T_2^{c}][T_3^{a},T_3^{c}]\,.
  \end{split}
\end{align}
For each three-line web, the combinations of kinematic factors
accompanying each colour factor have already been derived in
ref.~\cite{Gardi:2013ita} using the corresponding mixing matrices.
There are two distinct boomerang webs
connecting three lines, namely the (1,1,4) web of
figure~\ref{fig:114}, and the (1,2,3) web of figure~\ref{fig:123}. Let
us consider each in turn and compute the relevant integrals.

\subsubsection{The jelly-fish (1,1,4) web}
\label{sec:114}

\begin{figure}[htb]
\begin{center}
\scalebox{0.8}{\includegraphics{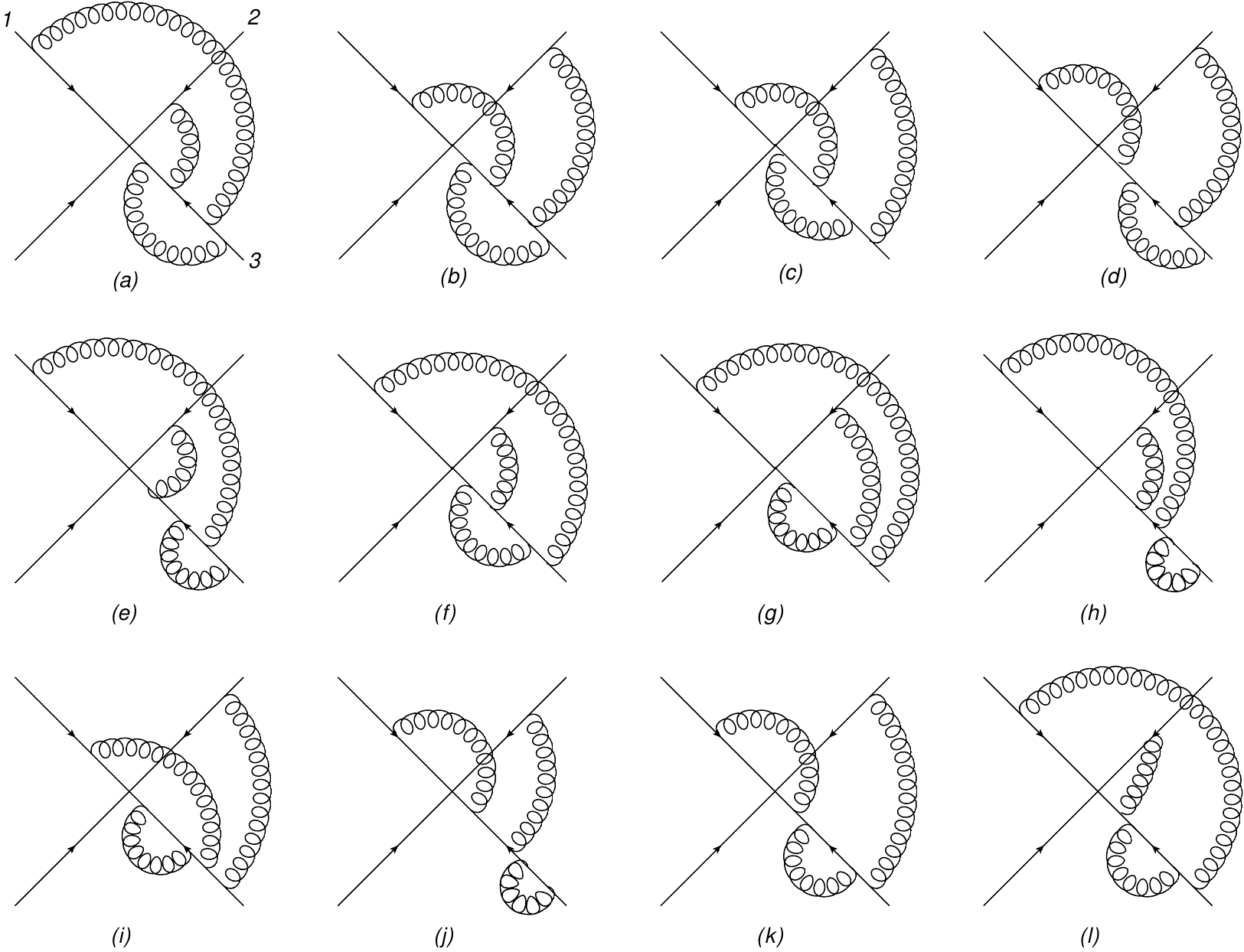}}
\caption{The (1,1,4) web.}
\label{fig:114}
\end{center}
\end{figure}

The (1,1,4) web consists of twelve distinct diagrams, where six of them
(diagrams $(g)$--$(l)$ in figure~\ref{fig:114}) contain self-energy loops,
and are thus irrelevant according to the results of
section~\ref{sec:selfenergy}.
Following the approach of
sections~\ref{sec:review} and~\ref{sec:mushroom}, the kinematic
parts of the first six diagrams can be written as
\begin{equation}
\mathcal{F}_{D} (\alpha_{13} ,\alpha_{23} ,  \epsilon)
=\kappa^3 \Gamma(6\epsilon) \int_{\frac{1}{2}}^{1}\frac{dx}{[(2x-1)^2]^{1-\epsilon}} \int_{0}^{1} dy dz \,p_\epsilon(y, \alpha_{23})\, p_\epsilon(z, \alpha_{13}) \phi_D (x,y,z;\epsilon)\,,
\label{kerneldef}
\end{equation}
where $D\in \{a,b,c,d,e,f\}$, and we present the calculation and results for kernels $\phi_D$ in
appendix~\ref{app:114calc}. The results~$\phi_D (x,y,z;\epsilon)$ are presented in eqs.~(\ref{eq:phia114}) through (\ref{kernels114f}), both as hypergeometric functions depending on $x$, $y$ and $z$, with $\epsilon$ shifting the parameters away from integer values, as well as through an expansion in powers of $\epsilon$. These expressions are useful to illustrate the general discussion around eq.~(\ref{x_l_integral}). Indeed, we observe that each diagram kernel consists of differences between pairs of hypergeometric functions, which vanish at $x=\frac12$, providing a linear suppression near the lower endpoint of the $x$ integration in eq.~(\ref{kerneldef}), such that the double pole associated with the boomerang gluon propagator is regularised, rendering this integral well-defined for small positive $\epsilon$. 

We further point out that the suppression of the enpoint singularity at $x=\frac12$ can also be easily seen after the $\epsilon$ expansion of the kernel. For diagrams $D\in \{c,d,e,f\}$, where the boomerang gluon straddles a single emission, the $\epsilon$-expanded kernel~$\phi_D (x,y,z;\epsilon)$ begins with $\sim \frac{1}{\epsilon}\ln((1-x)/x)$ providing linear suppression of the $x=\frac12$ singularity. In turn, for diagrams $D\in \{a,b\}$, where the boomerang gluon straddles two emissions, the $\epsilon$-expanded kernel~$\phi_D (x,y,z;\epsilon)$ begins with $\ln^2((1-x)/x)$ providing a quadratic suppression of this singularity.

The exponentiated colour factors for the six diagrams are~\cite{Gardi:2013ita}
\begin{align}
  \label{eq:Ctilde114}
  &\tilde{C}(a)=\frac12 (c_3^{[3,3]}-c_4^{[3,3]});\ \qquad \tilde{C}(b)=\frac12
  (c_3^{[3,3]}+c_4^{[3,3]}); \\ \notag
  &\tilde{C}(c)=\tilde{C}(d)=\frac12 c_4^{[3,3]};\   \qquad
  \tilde{C}(e)=\tilde{C}(f)=-\frac12 c_4^{[3,3]}\,,
\end{align}
where we used the basis of colour factors of eq.~(\ref{colbasis3}).
The complete renormalized web can then be written in this basis as
\begin{align}
\label{W114}
\begin{split}
W_{(1,1,4)} =&\, c_3^{[3,3]}  {\cal F}_{(1,1,4);3}  + c_4^{[3,3]} \Big[ {\cal F}_{(1,1,4);4} + {\cal F}_{(1,1,4);4}^{\rm
  CT} \Big],  \\
  =& \, c_3^{[3,3]} \Big[\frac12 \big({\cal F}_a+{\cal F}_b\big)  \Big]  + c_4^{[3,3]} \Big[ \frac12\big(  -{\cal F}_a +{\cal F}_b  +{\cal F}_{c}
    +{\cal F}_{d}  -{\cal F}_{e} -{\cal F}_{f} \big)
 \\
  & \qquad\qquad\qquad\qquad\qquad\qquad \qquad +  \frac12\big( {\cal
    F}_{c}^{\rm CT}   +{\cal F}_{d}^{\rm CT}  -{\cal F}_{e}^{\rm CT} -{\cal
    F}_{f}^{\rm CT} \big) \Big],
\end{split}
\end{align}
% \Einan{It is quite clear at this point that we need a notation for the unrenormalised ${\cal F}_{(1,1,4);j}$ which is computed in teh appendix. We still need to decide whether we want any notation for the unrenormalized $W$. }
% with
% \begin{align}
% {\cal F}_{(1,1,4);1}&={\cal F}_{(1,1,4);2}=0;\notag\\
% {\cal F}_{(1,1,4);3}&=\frac12 \Big[{\cal F}_a+{\cal F}_b\Big];\notag\\
% {\cal F}_{(1,1,4);4,\,{\rm ren.}}&=\frac12\Big[
% -{\cal F}_a
% +{\cal F}_b
% +{\cal F}_{c,\,{\rm ren.}}
% +{\cal F}_{d,\,{\rm ren.}}
% -{\cal F}_{e,\,{\rm ren.}}
% -{\cal F}_{f,\,{\rm ren.}}
% \Big],
% \label{F114}
% \end{align}
% and for diagrams~$(c)$, $(d)$, $(e)$ and $(f)$ we have used the notation
% \begin{equation}
%   {\cal F}_{i,\,{\rm ren.}}={\cal F}_i+{\cal F}_{i}^{\text{CT}},
% \label{tildeFdef}
% \end{equation}
where ${\cal F}_{D}^{\text{CT}}$ is the counterterm contribution associated with the renormalization of the gluon emission vertex
in diagram $D$, analogous to eq.~(\ref{CT2}) for the (3,1)
web. Note that the counterterm contributions enter with the same coefficient as the diagram they renormalize, thus removing any singularities associated with the shrinking of boomerang gluon loops to a point.
% That is, ${\cal F}_{i,\,{\rm ren.}}$ is such that all additional
% singularities associated with the shrinking of boomerang gluon loops to zero have been removed.
As is implied from eq.~(\ref{W114}),
counterterm graphs are not required for diagrams $(a)$ and $(b)$ in
figure~\ref{fig:114}: such a counterterm would correspond to the
renormalization of a two-gluon emission vertex coupling to the Wilson
line, which is not required.

Using the integrals of eq.~(\ref{kerneldef}) we immediately find the results for the
kinematic functions ${\cal  F}_{(1,1,4);3}$ and ${\cal F}_{(1,1,4);4}$ defined by the combinations of contributions from individual diagrams in eq.~(\ref{W114}); these two are given respectively by  eqs.~(\ref{F1143calc2}) and~(\ref{eq:postBintegration}).
In turn, each of the counterterm contributions entering ${\cal F}_{(1,1,4);4}^{\text{CT}}$
consists of the factor $Z_{v}^{(1)}$ of eq.~(\ref{Zgs1}) multiplying
a lower order graph, obtained by shrinking the boomerang gluon to a
point. For each of the diagrams $(c)$--$(f)$ in figure~\ref{fig:114}, the
lower-order diagram will be one of the members of the (1,1,2) web of
figure~\ref{fig:121}, after relabelling of the Wilson lines. Indeed,
the combination of graphs appearing in eq.~(\ref{W114}) is precisely
such as to construct the combination of kinematic factors found in the
(1,1,2) web (see ref.~\cite{Gardi:2010rn}) and we may thus write
\begin{align}
\label{eq:114CT}
\begin{split}
\cFi{1,1,4}{4}{\text{CT}} &(\alpha_{13} ,\alpha_{23} ,  \mu^2/m^2 , \epsilon)
= -Z_{v}^{(1)}{\cal
  F}_{(1,1,2)}^{(2)}(\alpha_{13},\alpha_{23})  \\
 =& -\frac12\left( \frac{g_s^2}{8\pi^2} \right)^3 \int_{0}^{1} dy dz\,  p_0(y,
   \alpha_{23}) p_0(z, \alpha_{12}) \Bigg \{ \frac{1}{\epsilon^2}\ln \left( \frac{z}{y}\right)
    + \frac{1}{\epsilon}\bigg[ 2\text{Li}_2 \left( -\frac{z}{y} \right) \\
 & \quad -2 \text{Li}_2 \left( -\frac{y}{z} \right)+ \ln \left(
   \frac{z}{y}\right) \left(\ln \left(q(y,\alpha_{23}) q(z,\alpha_{13})\right)
   +2\ln\left( \frac{\mu^2}{m^2}\right) \right) \bigg]  +
 \mathcal{O}(\epsilon^0) \Bigg\},
\end{split}
\end{align}
using eqs.~(\ref{F121form}) and (\ref{Zgs1}).

To obtain the renormalized kinematic function multiplying $c_4^{[3,3]}$ in (\ref{W114}) we now sum up
 the unrenormalised  function of eq.~(\ref{eq:postBintegration})  and the counterterm contribution of eq.~(\ref{eq:114CT}),
obtaining
\begin{align}
\label{eq:114total}
\begin{split}
&{\cal F}_{(1,1,4);4}(\alpha_{13} ,\alpha_{23} ,  \mu^2/m^2 , \epsilon) + {\cal
  F}_{(1,1,4);4}^{\rm  CT} (\alpha_{13} ,\alpha_{23} ,  \mu^2/m^2 , \epsilon) \\
 =&\left( \frac{g_s^2}{8\pi^2} \right)^3 \frac18 \int_{0}^{1} dy dz\,
p_0(y, \alpha_{23}) p_0(z, \alpha_{12}) \,\ln \left( \frac{z}{y}\right)
\\
&
\quad \Bigg\{ - \frac{1}{\epsilon^2}    + \frac{1}{\epsilon}
\left[  6 + 3 \ln \left(y z\right)- \ln \left(q(y,\alpha_{23})q(z,\alpha_{13})\right)
  + \ln\left( \frac{\mu^2}{m^2}\right) \right]  + \mathcal{O}(\epsilon^0) \Bigg\},
\end{split}
\end{align}
where we note the cancellation of all dilogarithms at ${\cal O}(1/\epsilon)$.  With this we have completed the computation of all ingredients in the renormalized $(1,1,4)$ web of eq. (\ref{W114}).

For the contribution of this web to the soft anomalous dimension, we
must combine eq.~(\ref{W114}) with the commutator contributions
appearing in eq.~(\ref{Gamres}), to form the subtracted web
$\overline{w}_{(1,1,4)}$ as follows:
\begin{align}
  \label{eq:Wbar114}
  \overline{w}_{(1,1,4)}^{(3,-1)} = w_{(1,1,4)}^{(3,-1)} + \frac12 \left[w^{(2,-2)},
  w^{(1,1)} \right] -\frac12 \left[ w^{(1,0)},
  w^{(2,-1)} \right] -\frac12 \left[ w^{(2,0)},
  w^{(1,-1)} \right].
\end{align}
Note that, as always, all $w^{(n,k)}$ entering this expression are defined including the counterterms associated with the renormalization of the gluon emission of the Wilson line.

The two lower-order webs which occur in the (1,1,4) web are the single gluon
exchange web and the (3,1) web, where the latter has been calculated in
section~\ref{sec:mushroom}.  The commutators of these do not yield anything
proportional to the colour factor $c_3^{[3,3]}$.
They do give a contribution to
the colour factor $c_4^{[3,3]}$, with the kinematic function
\begin{align}
\label{eq:114comm}
  \begin{split}
    &{\cal F}_{(1,1,4);4}^{\text{Comm}} (\alpha_{13} ,\alpha_{23} ,  \mu^2/m^2 , \epsilon)
\\
&=\frac12 \left[ {\cal
        F}^{(2,-2)}_{(3,1)}(\alpha_{23}){\cal F}^{(1,1)}(\alpha_{13}) -  {\cal
        F}^{(2,-2)}_{(3,1)}(\alpha_{13}){\cal F}^{(1,1)}(\alpha_{23}) + {\cal
        F}^{(2,-1)}_{(3,1)}(\alpha_{23}){\cal F}^{(1,0)}(\alpha_{13}) \right. \\
    & \qquad \left. -  {\cal
        F}^{(2,-1)}_{(3,1)}(\alpha_{13}){\cal F}^{(1,0)}(\alpha_{23}) - {\cal
        F}^{(2,0)}_{(3,1)}(\alpha_{23}){\cal F}^{(1,-1)}(\alpha_{13}) + {\cal
        F}^{(2,0)}_{(3,1)}(\alpha_{13}){\cal F}^{(1,-1)}(\alpha_{23}) \right]
\\ &= \frac{1}{8\epsilon} \left( \frac{g_s^2}{8\pi^2} \right)^3 \int_{0}^{1} dy dz\,  p_0(y,
   \alpha_{23}) p_0(z, \alpha_{13}) \Bigg \{
\frac12\ln^2 q(z,\alpha_{13})- \frac12 \ln^2
   q(y,\alpha_{23}) \\
   & \qquad \qquad +(2+\ln(yz))\left(\ln q(y,\alpha_{23})  -\ln
     q(z,\alpha_{13}) - \ln\left(\frac{z}{y}\right)\right) -\ln\left(\frac{z}{y}\right) \ln\left( \frac{\mu^2}{m^2}\right) \Bigg\}.  \end{split}
\end{align}
Note that similarly to eq.~(\ref{eq:114total}) this result is manifestly antisymmetric under the interchange of $\alpha_{23}$ and
$\alpha_{13}$. This permutation symmetry is of course consistent with Bose symmetry and the fact that the colour factor $c_4^{[3,3]}$ in (\ref{colbasis3}) is antisymmetric.

Similarly to eqs.~(\ref{subtracted_web_form},
\ref{subtracted_web_mge_kin}) we can write the final result for the
(1,1,4) subtracted web
in terms of integrals over
subtracted web kernels. Specifically, we define
\begin{equation}
  \label{eq:F114defs}
  F^{(3)}_{(1,1,4);i}(\alpha_{13} ,\alpha_{23} ,  \epsilon)=\int_0^1 dx_1 dx_2 \, p_0(x_1,\alpha_{13})p_0(x_2,\alpha_{23})\,
  {\cal G}_{(1,1,4);i}^{(3)}(x_1,x_2,\alpha_{13},\alpha_{23})
\end{equation}
for $i=3,4$.
After adjusting for normalisation, we find ${\cal G}_{(1,1,4);3}^{(3)}$ from eq.~(\ref{F1143calc2}) and
${\cal G}_{(1,1,4);4}^{(3)}$ from the sum of the $\epsilon^{-1}$ pole in eq.~(\ref{eq:114total}) and eq.~(\ref{eq:114comm}):
\begin{align}
  {\cal G}_{(1,1,4);3}^{(3)}(x_1,x_2,\alpha_{13},\alpha_{23})&=-\frac{4\pi^2}{9};\notag\\
  {\cal G}_{(1,1,4);4}^{(3)}(x_1,x_2,\alpha_{13},\alpha_{23}) &= \frac12\left[4\ln\left(\frac{q(x_2,\alpha_{23})}{x_2^2}\right)
                                                                -4\ln\left(\frac{q(x_1,\alpha_{13})}{x_1^2}\right)
                                                                \right.  \label{G114res}
  \\
                                                             & \qquad  \qquad
                                                                      \left.
                                                                      -\ln^2\left(\frac{q(x_2,\alpha_{23})}{x_2^2}\right)+\ln^2\left(\frac{q(x_1,\alpha_{13})}{x_1^2}\right)\right]. \notag
\end{align}
Again, the antisymmetry under the interchange of $\alpha_{23}$ and
$\alpha_{13}$ is manifest. The $\ln( \mu^2/m^2)$ terms have cancelled in the
$1/\epsilon$ pole, as they must do to ensure that the soft anomalous dimension does
not depend on
the regulator $m$. Carrying out the remaining integrals, we obtain the kinematic factors for the
subtracted web (defined as in eq.~(\ref{subtracted_web_mge_kin})):
\begin{align}
F^{(3)}_{(1,1,4);3}(\alpha_{13},\alpha_{23})&=-\frac{4\pi^2}{9}r(\alpha_{13})
r(\alpha_{23})M_{0,0,0}(\alpha_{13})M_{0,0,0}
(\alpha_{23});\notag\\
F^{(3)}_{(1,1,4);4}(\alpha_{13},\alpha_{23})&=\frac12r(\alpha_{13})
r(\alpha_{23})\Big[ 4 M_{1,0,0}(\alpha_{23})M_{0,0,0}(\alpha_{13})-4 M_{1,0,0}(\alpha_{13})M_{0,0,0}(\alpha_{23}) \notag\\
&\quad- M_{2,0,0}(\alpha_{23})M_{0,0,0}(\alpha_{13}) +M_{2,0,0}(\alpha_{13})M_{0,0,0}(\alpha_{23})\Big]\,,
\label{F114res}
\end{align}
where the $M_{k,l,n}(\alpha_{ij})$ are defined in eq.~(\ref{eq:Mbasis}). Explicit expressions for these functions and their symbols are summarised in appendix~\ref{app:functions}. Note that the first result in eq.~(\ref{F114res}) contains an overall factor of $\pi^2$, which itself has a non-zero transcendental weight. It is then natural to ask whether one can rewrite this result to be purely in terms of (products of) our basis functions with purely {\it rational} coefficients. That this is indeed the case can be seen by noting that (see appendix~\ref{app:functions})
\begin{equation}
\frac{4\pi^2}{3}M_{0,0,0}(\alpha)=4M_{0,2,0}(\alpha)-M_{0,0,2}(\alpha).
\label{M000rel}
\end{equation}
Thus, one may write
\begin{align}
    F_{(1,1,4);3}^{(3)}&=-\frac16r(\alpha_{13})r(\alpha_{23})
    \Big[\Big(4M_{0,2,0}(\alpha_{13})-M_{0,0,2}(\alpha_{13})\Big)M_{0,0,0}(\alpha_{23})\notag\\
    &\quad+M_{0,0,0}(\alpha_{13})
    \Big(4M_{0,2,0}(\alpha_{23})-M_{0,0,2}(\alpha_{23})\Big)\Big],
    \label{F114res2}
\end{align}
where we have made the symmetry of the web under the interchange of lines 1 and 2 manifest. We see once again the same general pattern previously seen in MGEWs and in the (3,1) web in section~\ref{sec:13}:
the subtracted web kinematic function takes the form of a rational function consisting of one factor of $r(\alpha_{ij})$ for each gluon which connects distinct Wilson lines $i$ and $j$, 
multiplied by a \emph{pure} transcendental function. The latter consists of a sum of products of polylogarithmic functions of individual $\alpha_{ij}$. The latter are again drawn from the basis of $M_{k,l,n}(\alpha_{ij})$ proposed in ref.~\cite{Falcioni:2014pka}.

We further note that as in the case of the (3,1) web -- and in contrast to non-boomerang MGEWs -- the polylogarithmic function in eq.~(\ref{F114res}) is of mixed, non-maximal weight, here weight 3 and weight 4, while the soft anomalous dimension at three loops receives contributions starting at weight 5. We will see a similar mixed, non-maximal weight structure across all boomerang webs.

\subsubsection{The (1,2,3) web}
\label{sec:123}

\begin{figure}[htb]
\begin{center}
\scalebox{0.7}{\includegraphics{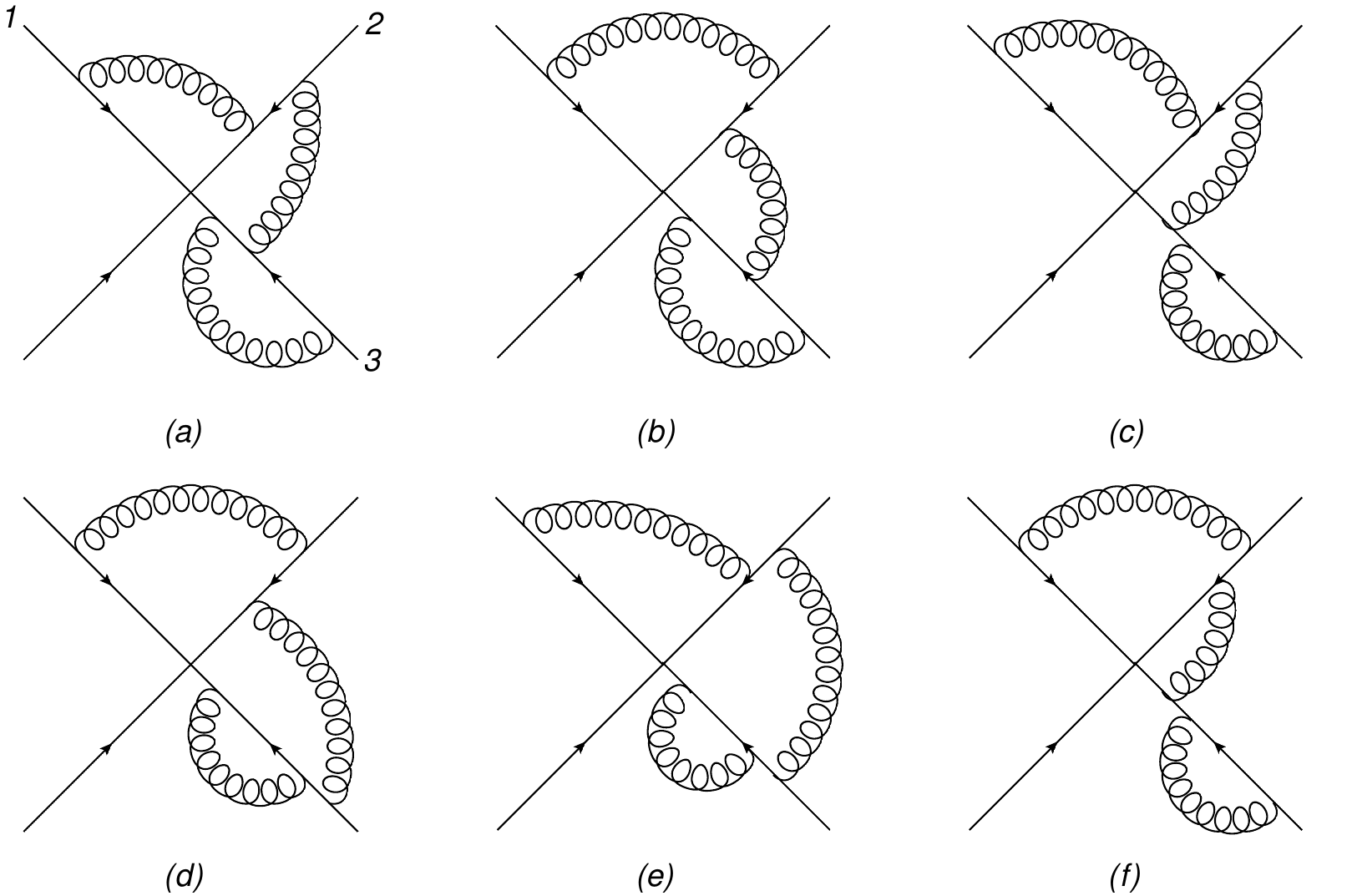}}
\caption{The (1,2,3) web.}
\label{fig:123}
\end{center}
\end{figure}

Next, we consider the (1,2,3) web of figure~\ref{fig:123}, consisting
of six diagrams. However, four of these (diagrams $(c)$--$(f)$) contain self-energy loops,
and thus do not contribute to the logarithm of the soft function,
using the results of section~\ref{sec:selfenergy}.
 From
ref.~\cite{Gardi:2013ita}, the exponentiated colour factors of diagrams $(a)$ and $(b)$ are
\begin{equation}
  \label{eq:Ctilde123}
  \tilde{C}(a) = -\frac12 c_4^{[3,3]}\quad {\rm and} \quad \tilde{C}(b) = \frac12 c_4^{[3,3]}.
\end{equation}
Therefore, the renormalized web is given by% the only nonzero contribution to the
% logarithm of the soft function is (in the notation of eq.~(\ref{W114}))
\begin{align}
\label{F123res}
\begin{split}
W_{(1,2,3)} &=\, c_4^{[3,3]} \Big[ {\cal F}_{(1,2,3);4} + {\cal F}_{(1,2,3);4}^{\rm
  CT} \Big] \\
  &=\, c_4^{[3,3]} \Big[-\frac12\big({\cal F}_a-{\cal F}_b\big)
    -\frac12\big({\cal F}_a^{\rm CT}-{\cal F}_b^{\rm CT} \big) \Big].
\end{split}
\end{align}
Using similar
methods to the (1,1,4) web, we find that the
kinematic parts of these diagrams are given by
\begin{equation}
\mathcal{F}_D (\alpha_{12},\alpha_{23},\epsilon) =
\kappa^3 \Gamma(6\epsilon)\frac{1}{\epsilon}\frac{1}{1-2\epsilon}
\int_{0}^{1} dy dz \, y^{2\epsilon}p_\epsilon(y,
\alpha_{23})p_\epsilon(z, \alpha_{12}) \phi_D (y,z),
\label{kernel123}
\end{equation}
where 
\begin{align}
\begin{split}
\phi_a (y,z)&=\frac{1}{2\epsilon} \left(\frac{1-y}{z} \right)^{2\epsilon}
                 {_2}F_1\left(6\epsilon,2\epsilon;1+2\epsilon;-\frac{1-y}{z}\right) \\
&=  \frac{1}{2\epsilon} + \ln \left( \frac{1-y}{z}\right) + \left\{\ln ^2 \left( \frac{1-y}{z}\right) + 6\text{Li}_2 \left( -\frac{1-y}{z}\right) \right\}\epsilon + \mathcal{O}(\epsilon^2);\\
\phi_b (y,z)&= \frac{1}{4\epsilon} \left( \frac{z}{1-y} \right)^{4\epsilon}
                 {_2}F_1\left(6\epsilon, 4\epsilon; 1+4\epsilon;-\frac{z}{1-y}\right) \\
&=  \frac{1}{4\epsilon} + \ln \left( \frac{z}{1-y}\right) + \left\{2\ln ^2
\left( \frac{z}{1-y}\right) + 6\text{Li}_2 \left( -\frac{z}{1-y}\right) \right\}\epsilon + \mathcal{O}(\epsilon^2)\,,
\end{split}
\label{phiab123}
\end{align}
and where  ${_2}F_1(a,b;c;z)$ is the Gauss hypergeometric function.
The integrals corresponding to the boomerang gluon\footnote{We point out that prior to performing this integral we observed the same regularisation of the double pole at the $x=\frac12$ endpoint as in eq.~(\ref{FMcalc4}).} have already been performed here, generating the pole at $\epsilon=\frac12$ in eq.~(\ref{kernel123}). To these, we must add the ultraviolet counterterm graphs, ${\cal F}_a^{\rm CT}$ and ${\cal F}_b^{\rm CT}$, associated with
renormalization of the gluon emission vertex in graphs $(a)$ and $(b)$ respectively. As for the (1,1,4) web,
the counterterms construct the (1,1,2) web of eq.~(\ref{F121form}), where line $2$ in figure~\ref{fig:123} is the one having two gluon attachments. The resulting counterterm contribution in eq.~(\ref{F123res}) is
\begin{align}
\begin{split}
{\cal F}_{(1,2,3);4}^{\rm CT}&(\alpha_{12},\alpha_{23},\mu^2/m^2,\epsilon) =-\frac12 Z_{v}^{(1)}{\cal F}^{(2)}_{(1,1,2)}(\alpha_{12},\alpha_{23},\epsilon)\\
=&\ \frac{1}{4} \left( \frac{g_s^2}{8\pi^2} \right)^3 \int_{0}^{1} dy dz\,  p_0(y,
  \alpha_{23}) p_0(z, \alpha_{12}) \Bigg \{ \frac{1}{\epsilon^2}\ln \left( \frac{z}{y}\right)
   + \frac{1}{\epsilon}\Bigg[ 2\text{Li}_2 \left( -\frac{z}{y} \right) \\
& -2\text{Li}_2 \left( -\frac{y}{z} \right)  + \ln \left(
  \frac{z}{y}\right) \left( \ln \left( q(y,\alpha_{23}) q(z,\alpha_{12}) \right) +2\ln\left( \frac{\mu^2}{m^2}\right) \right) \Bigg ]  + \mathcal{O}(\epsilon^0) \Bigg
\},
\end{split}
\label{counterterm123}
\end{align}
using eq.~(\ref{Zgs1}).  Combining this with the results from
eq.~\eqref{phiab123} gives the final result for the kinematic function multiplying $c_4^{[3,3]}$
in the renormalized (1,2,3) web in eq.~(\ref{F123res}) to be
\begin{align}
  \label{eq:renorm123}
  \begin{split}
    &{\cal F}_{(1,2,3);4}(\alpha_{12},\alpha_{23},\mu^2/m^2,\epsilon)  +{\cal
      F}_{(1,2,3);4}^{\rm CT}(\alpha_{12},\alpha_{23},\mu^2/m^2,\epsilon) \\
    =& \ \frac{1}{48} \left( \frac{g_s^2}{8\pi^2} \right)^3 \int_{0}^{1} dy dz\,  p_0(y,
  \alpha_{23}) p_0(z, \alpha_{12}) \, \Bigg \{ \frac{1}{\epsilon^3}  \\ & \quad +
  \frac{1}{\epsilon^2} \left[ 2 + 3\ln\left( \frac{\mu^2}{m^2}\right) + \ln
    \left( q(y,\alpha_{23}) q(z,\alpha_{12}) \right] -2\ln(y) + 4\ln(z) \right]
  \\ & \quad  + \frac{1}{2\epsilon} \left[ 8 + \frac{13\pi^2}{2}  + 40 \ln(y)-32\ln(z) + \ln^2\left(
      q(y,\alpha_{23}) q(z,\alpha_{12}) \right)
    -4\ln^2(y) \right. \\
& \qquad +32\ln(y)\ln(1-y)-16\ln(y)\ln(z)-8\ln^2(z) \\
& \qquad + 4 \ln\left(
      q(y,\alpha_{23}) q(z,\alpha_{12}) \right) \left(1 - \ln(y)+2\ln(z) \right)
    \\
& \qquad \left.
    + 3 \ln\left( \frac{\mu^2}{m^2}\right) \left(4 + 2 \ln\left(
      q(y,\alpha_{23}) q(z,\alpha_{12}) \right) + 4\ln(y) +3 \ln\left(
      \frac{\mu^2}{m^2}\right) \right) \right] + {\cal O}(\epsilon^0).
\end{split}
\end{align}
Once again the dilogarithms have cancelled to this order in $\epsilon$ in the renormalised result.

In order to obtain the corresponding contribution of this web to the soft anomalous dimension, the $(1,2,3)$ subtracted web, we
must now add to eq.~(\ref{eq:renorm123}) the commutator contributions involving lower-order webs.
As is clear from figure~\ref{fig:123}, the relevant lower-order webs include the
single gluon exchange (one-loop) diagram, and the (3,1) web of figure~\ref{fig:13}
and we find
\begin{align}
  \label{eq:commterms123}
  \overline{w}_{(1,2,3)}^{(3,-1)} = w_{(1,2,3)}^{(3,-1)} \,+\, c_4^{[3,3]} {\cal F}^{\rm Comm}_{(1,2,3);4}(\alpha_{12},\alpha_{23},\mu^2/m^2,\epsilon),
\end{align}
where
\begin{align}
  \label{eq:comms123}
  \begin{split}
    {\cal F}^{\rm Comm}_{(1,2,3);4}&(\alpha_{12},\alpha_{23},\mu^2/m^2,\epsilon)
   \\
=\,& \frac{1}{2\epsilon} \left( - {\cal F}^{(1,1)}(\alpha_{12}){\cal F}^{(2,-2)}_{(1,3)}(\alpha_{23}) -
      {\cal F}^{(1,0)}(\alpha_{12}) {\cal F}^{(2,-1)}_{(1,3)}(\alpha_{23}) + {\cal
        F}^{(2,0)}_{(1,3)}(\alpha_{23}){\cal F}^{(1,-1)}(\alpha_{12}) \right) \\
    =\, & -\frac{1}{32\epsilon} \left( \frac{g_s^2}{8\pi^2} \right)^3 \int_{0}^{1} dy dz\,  p_0(y,
  \alpha_{23}) p_0(z, \alpha_{12}) \\ & \quad \times \bigg[ 8 + \frac{13 \pi^2}{6} +
    4\ln(q(y,\alpha_{23})) - 4\ln(q(z,\alpha_{12})) - \ln^2(q(y,\alpha_{23}))
    \\ & \qquad \qquad
    + \ln^2(q(z,\alpha_{12}))+ 2 \ln(q(y,\alpha_{23}))\ln(q(z,\alpha_{12}))
    + 8\ln(y) +  4\ln^2(y) \\
    & \qquad \qquad + 4 \ln(q(y,\alpha_{23}))\ln(y)- 4\ln(q(z,\alpha_{12})) \ln(y)  \\
    & \qquad \qquad + \ln\left( \frac{\mu^2}{m^2} \right) \left( 4 +2
      \ln\left(q(y,\alpha_{23}) q(z,\alpha_{12})\right) + 4\ln(y) + 3\ln\left( \frac{\mu^2}{m^2} \right) \right)  \bigg].
\end{split}
\end{align}
Upon adding this to the $1/\epsilon$ term of eq.~(\ref{eq:renorm123}),  as
expected, all dependence on the regulator $m$ cancels and the integrated subtracted web
is given by
\begin{equation}
  \label{eq:F114result}
  F_{(1,2,3);4}^{(3)}(\alpha_{12},\alpha_{23})=\int_0^1 dx_1 dx_2 \, p_0(x_1,\alpha_{12})p_0(x_2,\alpha_{23})\,
  {\cal G}_{(1,2,3);4}^{(3)}(x_1,x_2,\alpha_{12},\alpha_{23})
\end{equation}
where the subtracted web kernel is given by
\begin{align}
{\cal
  G}^{(3)}_{(1,2,3);4}(x_1,x_2,\alpha_{12},\alpha_{23}) =\,& \frac{1}{6}\left[-8
  -4\ln\left(\frac{q(x_2,\alpha_{23})}{x_2^2}\right)
  +8\ln\left(\frac{q(x_1,\alpha_{12})}{x_1^2}\right) \right. \notag \\
  & \qquad  +2\ln^2\left(\frac{q(x_2,\alpha_{23})}{x_2^2}\right)-
  \ln^2\left(\frac{q(x_1,\alpha_{12})}{x_1^2}\right) \notag
    -16\ln^2\left(\frac{x_2}{1-x_2}\right) \\
 & \qquad \left. -
  2\ln\left(\frac{q(x_1,\alpha_{12})}{x_1^2}\right) \ln\left(\frac{q(x_2,\alpha_{23})}{x_2^2}\right)\right],
\label{G123;4}
\end{align}
so that the integrated contribution to the soft anomalous dimension is
\begin{align}
F^{(3)}_{(1,2,3);4}(\alpha_{12},\alpha_{23}) &=   \frac{1}{6}
r(\alpha_{12})r(\alpha_{23})\Big \{ -8M_{0,0,0} (\alpha_{12})M_{0,0,0} (\alpha_{23}) \notag\\
&-4M_{0,0,0} (\alpha_{12})M_{1,0,0}(\alpha_{23}) +8M_{1,0,0} (\alpha_{12})M_{0,0,0} (\alpha_{23})\notag\\
& + 2M_{0,0,0} (\alpha_{12})M_{2,0,0}(\alpha_{23}) - M_{2,0,0} (\alpha_{12})M_{0,0,0}(\alpha_{23})\notag\\
& -16M_{0,0,0} (\alpha_{12})M_{0,2,0}(\alpha_{23}) -2M_{1,0,0} (\alpha_{12})M_{1,0,0}(\alpha_{23})\Big \}\,,
\label{F123;4}
\end{align}
where explicit expressions for the functions $M_{k,l,n}(\alpha_{ij})$ are summarised in appendix~\ref{app:functions}.  
We observe again the same pattern described following eq.~(\ref{F114res}).
Also the (1,2,3) web can be expressed in terms of the previously-defined basis functions
 (see eq.~(\ref{eq:Mbasis})), and similarly to the (1,1,4) web we see mixed, non-maximal weight. We note that in eq.~(\ref{F123;4}) the weight ranges from 4 all the way down to~2.

\subsection{Boomerang webs connecting two Wilson lines}
\label{sec:2lines}

Having calculated all boomerang MGEWs connecting three Wilson lines, we
now turn our attention to those connecting two lines. Without loss of
generality, we will take the labels of the lines to be 1 and 2. As
explained in ref.~\cite{Falcioni:2014pka}, if the Wilson lines are not
in a colour singlet state (as will be the case in general if the lines
1 and 2 are chosen out of total of $L>2$ Wilson lines), 
the effective vertex formalism
generates two independent colour structures, which we label using our
standard notation as
\begin{align}
\label{colbasis2}
\begin{split}
c_{1}^{[3,2]}&=\frac{N_c^2}{4}T_1\cdot T_2\\
c_{2}^{[3,2]}&=-\frac14 f^{cbd}f^{ace}\{T_1^b,T_1^a\}\,
\{T_2^d,T_2^e\}.
\end{split}
\end{align}
Of these two colour factors, the second contains symmetrised
combinations of colour generators, whereas the first does not. These
observations will be useful when considering collinear reduction later
on. 

A further simplification occurs in the two-line webs considered here compared to the three-line ones described in the previous section. This is the fact that no commutators of lower-order webs can arise, owing to the fact that all the relevant one- and two-loop subdiagrams form webs with colour factors proportional to $T_1\cdot T_2$, which are hence mutually commuting.
Thus, the contribution of each web to the anomalous dimension in eq.~(\ref{Gamres}) -- the subtracted web -- simply corresponds to the single pole of that web. 

In the following sections we consider in turn the three boomerang MGEWs connecting two lines at three loops, the (3,3), (5,1) and (2,4) webs, compute their contribution to the soft anomalous dimension, and comment on how they manifest the general properties discussed above.

\subsubsection{The (3,3) web}
\label{sec:33}

\begin{figure}[h]
\begin{center}
\scalebox{.9}{\includegraphics{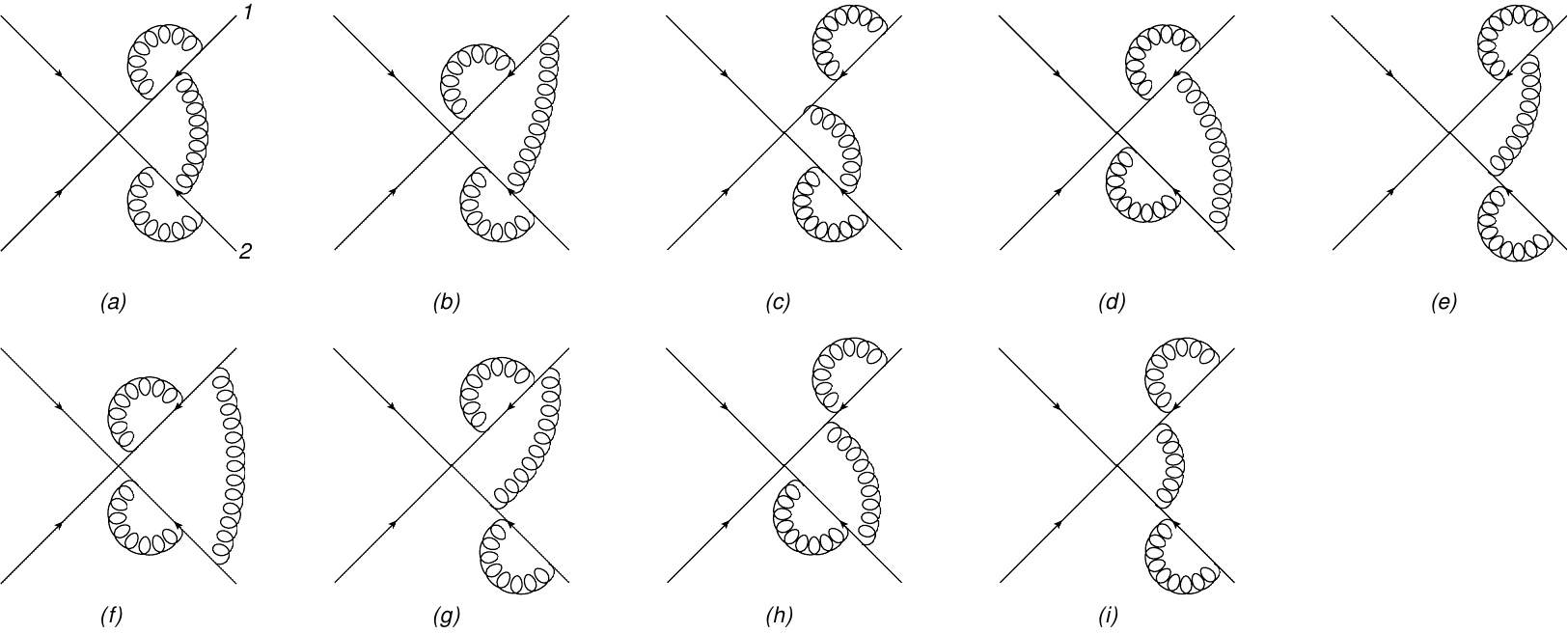}}
\caption{The (3,3) web.}
\label{fig:33}
\end{center}
\end{figure}

The (3,3) web is shown in figure~\ref{fig:33}, and contains nine diagrams. However, eight of these contain self-energy loops, and thus do not
contribute to the soft anomalous dimension, using the results of
section~\ref{sec:selfenergy}.

%\Einan{Note below I changed from $w_{(3,3)}$ to $W_{(3,3)}$ and added eq. (\ref{calF33}). }
The contribution of the entire web is
then simply
\begin{equation}
W_{(3,3)}=\tilde{C}(a)\, \left({\cal F}_a+{\cal F}_a^{\rm CT}\right),
\label{w33}
\end{equation}
where $\tilde{C}(a)$ is the exponentiated colour factor of
diagram~$(a)$,  ${\cal F}_{a}$ its kinematic part before renormalization, while ${\cal F}_a^{\rm CT}$ is the corresponding counterterm
contribution.
%
% that is:
%\begin{equation}
%{\cal F}_{a,\,{\rm ren.}}= {\cal F}_a+{\cal F}_a^{\rm CT}\,.
%\label{calF33}
%\end{equation}
 The exponentiated colour factor is found to~be
\begin{align}
\label{Ctildea33}
\begin{split}
\tilde{C}(a)&=C(a)-\frac12\Big(C(b)+C(c)+C(d)+C(e)+C(f)+C(i)\Big)\\
&\qquad\quad\,\,+\frac13\Big(C(g)+C(h)+2C(i)+2C(f)\Big)\\
&=c_1^{[3,2]},
\end{split}
\end{align}
where the second line follows after performing the appropriate colour algebra, with $c_1^{[3,2]}$ as defined in eq.~(\ref{colbasis2}).

Applying the Feynman rules similarly to the previous webs, one finds
\begin{equation}
  {\cal F}_a(\alpha_{12})=\frac{\kappa^3\Gamma(6\epsilon)}
  {\epsilon^2(1-2\epsilon)^2}\int_0^1dy[y(1-y)]^{2\epsilon}
  p_\epsilon(y,\alpha_{12}).
  \label{Fa33}
\end{equation}
Note that we see explicitly the presence of a double pole at $\epsilon=\frac12$ associated with the power divergence of the boomerang loop. The double pole is to be expected given that there
are two independent boomerang gluons, one on each Wilson
line. 
Expanding the result yields
\begin{align}
\label{dumbbell}
\begin{split}
{\cal F}_a\left(\alpha_{12} \right) &= -\left(
\frac{g_s^2}{8\pi^2}\right)^3 \int_0^1 dy p_0(y, \alpha_{12}) \Bigg \{
\frac{1}{6\epsilon^3}+\frac{1}{6\epsilon^2} \bigg[ 4 + \ln q(y,
  \alpha_{12}) +4 \ln y +3\ln\Big(\frac{\mu^2}{m^2}\Big)\bigg]\\ & +
\frac{1}{12\epsilon}\bigg[ 24 + 32 \ln y + 8 \ln q(y, \alpha_{12})  +\frac{13}{2} \pi^2+ \ln ^2 q(y,
  \alpha_{12}) +8 \ln q(y, \alpha_{12}) \ln y  + 8 \ln ^2 y\\ &  +
  8 \ln (1-y) \ln y+\ln\Big(\frac{\mu^2}{m^2}\Big)\Big(6\ln
  q(y,\alpha_{12})+24\ln(y)+24 \Big)+9\ln^2\Big(\frac{\mu^2}{m^2}\Big)
  \bigg] + \mathcal{O}(\epsilon^0) \Bigg \}.
\end{split}
\end{align}
One must combine this with counterterm contributions, and as is evident from figure~\ref{fig:33}$(a)$ there are
three possibilities. One can shrink both boomerang gluon loops to a point, in which case one obtains the
one-loop single gluon exchange web dressed by two counterterms. Or,
one can shrink only the upper or lower boomerang gluon loops, recovering the (3,1) web dressed by one counterterm.
The full counterterm contribution to the web is then
\begin{align}
\label{F33CT}
\begin{split}
{\cal F}_{a}^{\rm CT}(\alpha_{12})&=
(Z_{v}^{(1)})^2 \mathcal{F}^{(1)}_{(1,1)}(\alpha_{12})
+ 2 Z_v^{(1)}
{\cal F}^{(3)}_{(3,1)}(\alpha_{12})\\
= & \left( \frac{g_s^2}{8\pi^2}\right)^3 \int_0^1 dy p_0(y, \alpha_{12}) \Bigg
    \{\frac{1}{\epsilon^2} \bigg[1 +\ln y +\frac12\ln\Big(\frac{\mu^2}{m^2}\Big)\bigg]  \\
&+\frac{1}{\epsilon} \bigg[2+2 \ln y+ \ln q(y, \alpha_{12})+\frac{13}{24} \pi^2+ \ln q(y, \alpha_{12}) \ln y+\ln ^2 y 
\\
&+\ln\Big(\frac{\mu^2}{m^2}\Big)\Big(\frac12\ln q(y,\alpha_{12})+2\ln y +2\Big)
+\frac34\ln^2\Big(\frac{\mu^2}{m^2}\Big)
\bigg] + \mathcal{O}(\epsilon^0) \Bigg \}\,.
\end{split}
\end{align}
%\Einan{Note that I changed from ${\cal F}_{(3,3)}^{\rm CT}$ to ${\cal F}_{a}^{\rm CT}(\alpha_{12})$ above.}
There are no lower-order webs one can use to form commutator
structures, and thus summing eqs.~(\ref{dumbbell}, \ref{F33CT}) leads
directly to the contribution of this web to the soft anomalous dimension:
\begin{equation}
F^{(3)}_{(3,3);1}(\alpha_{12})=\int_0^1 dx_1 p_0(x_1,\alpha_{12}){\cal G}_{(3,3);1}
(x_1,\alpha_{12}),
\label{FtoG33}
\end{equation}
where we have defined the subtracted web kernel
\begin{align}
{\cal G}_{(3,3);1}(x_1,\alpha_{12})&=-\frac23\left[
\ln^2\left(\frac{q(x_1,\alpha_{12})}{x_1^2}\right)
-4\ln\left(\frac{q(x_1,\alpha_{12})}{x_1^2}\right)
-4\ln^2\left(\frac{x_1}{1-x_1}\right)\right]\,.
\label{G33}
\end{align}
The integrated subtracted web can then be written in terms of the basis
functions of eq.~(\ref{eq:Mbasis})~as
\begin{align}
F^{(3)}_{(3,3);1}(\alpha_{12})&=-\frac23 r(\alpha_{12})
\Big[M_{2,0,0}(\alpha_{12})-4M_{1,0,0}(\alpha_{12})-4M_{0,2,0}(\alpha_{12})\Big]\,,
\label{F33res}
\end{align}
where explicit expressions for the functions $M_{k,l,n}(\alpha_{ij})$ are summarised in appendix~\ref{app:functions}. We see again the same pattern described following eq.~(\ref{F114res}): the (3,3) web can also be expressed in terms of the previously-defined basis functions, and similarly to other boomerang webs it features mixed, non-maximal weights, in this case, weights 3 and 2.

\subsubsection{The (5,1) web}
\label{sec:15}

\begin{figure}[htb]
\begin{center}
\scalebox{0.9}{\includegraphics{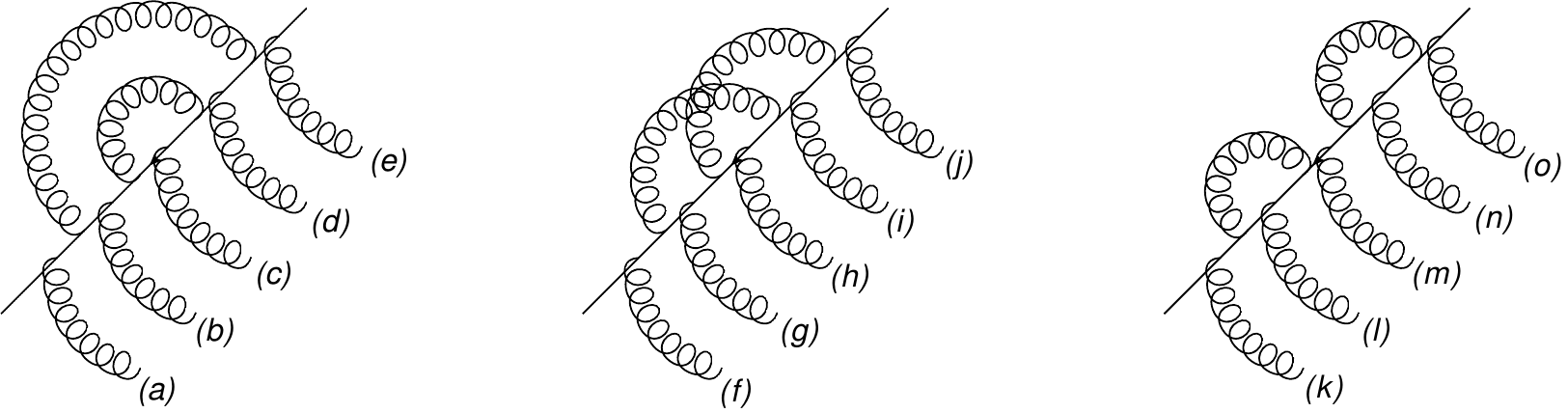}}
\caption{The (5,1) web in a compact representation. Note that
in any given diagram in this web there is just one non-boomerang emission.
Labels attached to the gluons correspond to the diagram obtained by placing the gluon as shown, and connecting
  it with a second Wilson line.  }
\label{fig:15}
\end{center}
\end{figure}

Next, we consider the (5,1) web, consisting of two boomerang gluons on
one Wilson line, and a single gluon emitted from the same line that
lands on another. We represent this web compactly as in
figure~\ref{fig:15}.
There are three configurations of the boomerang
gluons, and the non-boomerang gluon can be in any of the places shown,
where each of these constitutes a different diagram in the web. Thus,
there are fifteen diagrams in total.

By the results of section~\ref{sec:selfenergy}, the only diagrams
which end up contributing to the web (i.e. which do not contain
self-energy loops) are $(c)$, $(g)$, $(h)$ and $(i)$. Furthermore, we find the
exponentiated colour factors to be
\begin{align}
\tilde{C}(c)=\tilde{C}(g)=\tilde{C}(i)=c_1^{[3,2]},\quad\qquad
\tilde{C}(h)=2c_1^{[3,2]}.
\label{ECF15}
\end{align}
Thus, the renormalised web including counterterm contributions may be expressed as
\begin{align}
\label{1-5i}
W_{(5,1)} =& \, c_1^{[3,2]}\Big[ \mathcal{F}_{(5,1);1} +
  \mathcal{F}_{(5,1);1}^{\rm CT} \Big] \\ \notag
=& \, c_1^{[3,2]}\Big[{\cal F}_c+{\cal F}_g
+2{\cal F}_h+{\cal F}_i+{\cal F}_c^{\rm CT}+{\cal F}_g^{\rm CT}
+2{\cal F}_h^{\rm CT}+{\cal F}_i^{\rm CT} \Big].
\end{align}
Using eq.~(\ref{FDform}) we may write the kinematic function of each of the four surviving diagrams as
\begin{align}
{\cal F}_D(\alpha_{12})&=g_s^2\bar{\mu}^{2\epsilon}{\cal N}
\beta_{1}\cdot \beta_{2}\int_0^\infty du dv \left[-(\beta_{1}u-\beta_{2}v)^2\right]^{-1+\epsilon}  \!
\exp\left[-im \left(u\sqrt{\beta_{1}^2}+ v\sqrt{\beta_{2}^2}\right)\right]  \mathcal{B}_{D}^{[0,\infty]}(u)
\label{FDform_one_gluon}
\end{align}
where the function $\mathcal{B}_{D}^{[0,\infty]}(u)$ represents the two-loop subdiagram consisting of boomerang gluons with their emission and absorption positions -- whose order depends on the specific diagram~$D$ considered -- integrated along the $\beta_1$ Wilson line over the entire range $[0,\infty]$ without any regulators.  Applying similar methods to above, the kinematic functions may be brought to the following form:
\begin{equation}
\label{51calFD}
\mathcal{F}_D(\alpha_{12}) = \kappa^3 \frac{\Gamma(6\epsilon)}{2\epsilon^2}\int^{1}_{0}dx\,dy\,dz\,p_\epsilon(z,\alpha_{12})z^{4\epsilon}(2x-1)^{2\epsilon-2}(2y-1)^{2\epsilon-2} \phi_D(x,y) \theta\Big(x>\frac12\Big)
\theta\Big(y>\frac{1}{2}\Big),
\end{equation}
where
%\Einan{Previous version:
%\begin{align}
%\phi_c (x,y)=\,& \Big\{(1-x)^{-2\epsilon}\big[(1-y)^{-2\epsilon}-y^{-4\epsilon}(1-y)^{2\epsilon}\big] + x^{-2\epsilon}\big[y^{-2\epsilon}-y^{2\epsilon}(1-y)^{-4\epsilon}\big]\Big\}\\
%&\theta(x>y), \notag\\
%\phi_g (x,y)=\,& \Big\{(1-x)^{-2\epsilon}\big[y^{-2\epsilon}+y^{2\epsilon}(1-y)^{-4\epsilon}\big] - %x^{-2\epsilon}y^{2\epsilon}(1-y)^{-4\epsilon} -y^{-2\epsilon}x^{2\epsilon}(1-x)^{-4\epsilon}\Big\}\\
%&\theta(x<y) +
%  \Big\{(1-x)^{-2\epsilon}\big[y^{2\epsilon}(1-y)^{-4\epsilon}+y^{-2\epsilon}-2(1-y)^{-2\epsilon}\big]\Big\}\,\theta(x>y),
%  \notag \\
%\phi_h (x,y)=\,& \Big\{y^{-2\epsilon}\big[x^{-2\epsilon}-2(1-x)^{-2\epsilon}
%+ x^{2\epsilon}(1-x)^{-4\epsilon}\big]\Big\}\,\theta(x<y)\\
%&+ \Big\{(1-x)^{-2\epsilon}\big[(1-y)^{-2\epsilon} - 2y^{-2\epsilon} +
%  (1-y)^{2\epsilon}y^{-4\epsilon}\big]\Big\}\,\theta(x>y), \notag \\
%\phi_i (x,y)=\,& \Big\{(1-x)^{-2\epsilon}y^{-2\epsilon} -
%          2x^{-2\epsilon}y^{-2\epsilon}+y^{-2\epsilon}x^{-4\epsilon}(1-x)^{2\epsilon}\Big\}\,\theta(x<y)
%          + \Big\{(1-x)^{-2\epsilon}\\ \notag
%&\big[y^{-2\epsilon}-y^{-4\epsilon}(1-y)^{2\epsilon}\big] +
%  y^{-2\epsilon}x^{-4\epsilon}(1-x)^{2\epsilon} -
%  (1-y)^{-2\epsilon}x^{-4\epsilon}(1-x)^{2\epsilon}\Big\}\,\theta(x>y).
%\end{align}
%}
\begin{align}
\label{phicghi}
\begin{split}
\phi_c (x,y)
%=\,& \Bigg\{(1-x)^{-2\epsilon}(1-y)^{-2\epsilon}
%\left[1-\left(\frac{1-y}{y}\right)^{4\epsilon}\right] + x^{-2\epsilon}y^{-2\epsilon}
%\left[1-\left(\frac{y}{1-y}\right)^{4\epsilon}\right]\Bigg\}\,\theta(x>y),\\
=\,& x^{-2\epsilon}y^{-2\epsilon}\left[1-\left(\frac{y}{1-y}\right)^{4\epsilon}\right]\left[1-\left(\frac{1-y}{y}\right)^{2\epsilon}\left(\frac{x}{1-x}\right)^{2\epsilon}\right]\,\theta(x>y),\\
\phi_g (x,y)
%=\,& (1-x)^{-2\epsilon} \Bigg\{
%y^{-2\epsilon}
%\left[1-\left(\frac{x}{1-x}\right)^{2\epsilon}\right]
%+y^{2\epsilon}(1-y)^{-4\epsilon}  \left[1-\left(\frac{1-x}{x}\right)^{2\epsilon}\right] \Bigg\}\,\theta(x<y) \\&
%+
%  (1-x)^{-2\epsilon}
%  y^{-2\epsilon}\left[1-\left(\frac{y}{1-y}\right)^{2\epsilon}
%  \right]^2\,\theta(x>y),
%  \\
  =\,& (1-x)^{-2\epsilon}y^{-2\epsilon} \left[1-\left(\frac{x}{1-x}\right)^{2\epsilon}\right] \left[1-\left(\frac{y}{1-y}\right)^{4\epsilon}\left(\frac{1-x}{x}\right)^{2\epsilon}\right]
\theta(x<y) \\&
+
  (1-x)^{-2\epsilon}y^{-2\epsilon}\left[1-\left(\frac{y}{1-y}\right)^{2\epsilon}
  \right]^2\,\theta(x>y), 
  \\
\phi_h (x,y)=\,& 
x^{-2\epsilon}y^{-2\epsilon}\left[1-\left(\frac{x}{1-x}\right)^{2\epsilon}\right]^2\,\theta(x<y)\\
&+ (1-x)^{-2\epsilon}(1-y)^{-2\epsilon}
\left[1 - \left(\frac{1-y}{y}\right)^{2\epsilon}\right]^2\,\theta(x>y),  \\
\phi_i (x,y)
%=\,& y^{-2\epsilon}(1-x)^{-2\epsilon}
%\left[1 -
%\left(\frac{1-x}{x}\right)^{2\epsilon}\right]^2
%\,\theta(x<y)\\ 
%         & + y^{-2\epsilon}
%         \Bigg\{(1-x)^{-2\epsilon}
%\left[1-\left(\frac{1-y}{y}\right)^{2\epsilon}\right] +
%  x^{-4\epsilon}(1-x)^{2\epsilon}\left[1 -
%  \left(\frac{y}{1-y}\right)^{2\epsilon}\right]\Bigg\}\,\theta(x>y)\\
  =\,& (1-x)^{-2\epsilon}y^{-2\epsilon}
\left[1 -
\left(\frac{1-x}{x}\right)^{2\epsilon}\right]^2
\,\theta(x<y)\\ 
         & + (1-x)^{-2\epsilon}y^{-2\epsilon}
\left[1-\left(\frac{1-y}{y}\right)^{2\epsilon}\right] \left[
  1-\left(\frac{1-x}{x}\right)^{4\epsilon}
  \left(\frac{y}{1-y}\right)^{2\epsilon}\right]\,\theta(x>y).
  \end{split}
\end{align}

%\Einan{I find that this gives:
%\begin{align}
%\begin{split}
%\phi_{(5,1);1}(x,y)=\,
%x^{-2\epsilon} y^{-2\epsilon}\Bigg\{&(1-\xi) \left(1-\eta ^2+\frac{1}{\xi }-\xi \right)\theta(x<y)\\
%&+
%\frac{1}{\xi}(1-\eta) (1-\xi ) \Big(1+\xi (2+\eta)\Big)\theta(x>y)\Bigg\}
%\end{split}
%\end{align}
%where 
%\[
%\xi\equiv \left(\frac{x}{1-x}\right)^{2\epsilon}\qquad 
%\eta\equiv \left(\frac{y}{1-y}\right)^{2\epsilon}
%\]
%\Bigg\{\left(1-\left(\frac{x}{1-x}\right)^{2\epsilon}\right)
%\left(1-\left(\frac{y}{1-y}\right)^{2\epsilon}+\left(\frac{x}{1-x}\right)^{-2\epsilon}-\left(\frac{x}{1-x}\right)^{2\epsilon}\right)\Bigg\}
%\theta(x<y)\\
%&+
%\left(\frac{x}{1-x}\right)^{-2\epsilon}
%\left(1-\left(\frac{x}{1-x}\right)^{2\epsilon}\right)
%\left(1-\left(\frac{y}{1-y}\right)^{2\epsilon}\right)
%\left(1+\left(\frac{x}{1-x}\right)^{2\epsilon}\left(2+
%\left(\frac{y}{1-y}\right)^{2\epsilon}\right)\right)\theta(x>y)
%\end{align}
%}
Importantly, as discussed on general grounds in section~\ref{sec:kinboom}, the integral in eq.~(\ref {51calFD}) exists for any $\epsilon>0$. Specifically, we observe that the functions $\phi_D(x,y)$ provide linear suppression in $(x-\frac12)$ whenever $x=\frac12$ is a limit of integration and similarly for $y$, so the double poles in $\mathcal{F}_D(\alpha_{12})$ associated with the boomerang gluon propagators are always accompanied by a regularising factor, such that the integral is well defined for small positive values of $\epsilon$. 

After adding together all contributions in the combination of
eq.~(\ref{1-5i}), one may carry out the $x$ and $y$ integrals and
expand in $\epsilon$, obtaining
\begin{align}
\begin{split}
  \mathcal{F}_{(5,1);1}(\alpha_{12})&=
  \kappa^3\frac{\Gamma(6\epsilon)}{2\epsilon^2}
  \frac{1}{2\epsilon-1}
  \left[\frac{1}{4\epsilon-1}+\frac{\Gamma(2\epsilon-1)\Gamma(2\epsilon+1)}
    {\Gamma(4\epsilon)}\right]\int^{1}_{0}dz\,p_\epsilon(z,\alpha_{12})z^{4\epsilon}\\  
&=-\left(\frac{g_s^2}{8\pi^2}\right)^3\frac{1}{12\epsilon^3}
\Bigg[3+\Big(14+9\ln\Big(\frac{\mu^2}{m^2}\Big)\Big)\epsilon\\
&\quad +\Big(52+\frac{23\pi^2}{3}+42\ln\Big(\frac{\mu^2}{m^2}\Big)
+\frac{27}{2}\ln^2\Big(\frac{\mu^2}{m^2}\Big)\Big)\epsilon^2+\ldots \Bigg]\int^{1}_{0}dz\,p_\epsilon(z,\alpha_{12})z^{4\epsilon}.
\end{split}
\end{align}

\begin{figure}[tb]
\begin{center}
\scalebox{0.8}{\includegraphics{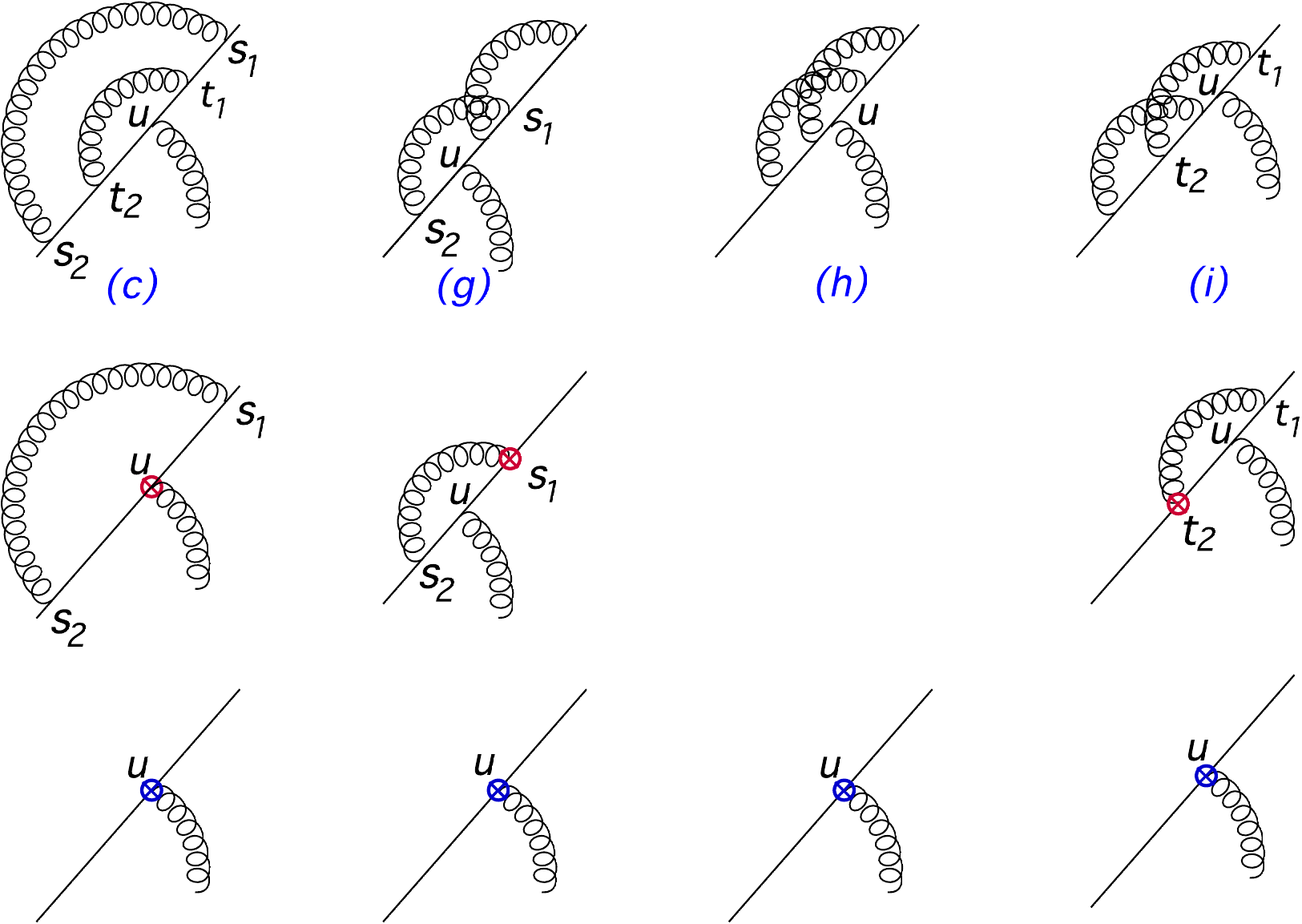}}
\caption{Renormalization of the (5,1) web. Upper row: the boomerang gluon subdiagram in diagrams $(c)$, $(g)$, $(h)$ and~$(i)$; Second row: one-loop counterterm contributions to the above diagrams associated with shrinking the innermost boomerang gluon to a point; Third row: two-loop counterterm contributions to the above diagrams associated with simultaneously shrinking both boomerang gluons to a point.}
\label{fig:15web_counterterms}
\end{center}
\end{figure}

According to eq.~(\ref{1-5i}), we must now combine $\mathcal{F}_{(5,1);1}$ with counterterms for the ultraviolet subdivergences
associated with shrinking the boomerang gluons to a point.  To remove all divergences, we may write a renormalization factor $Z_{v}$ for the gluon emission vertex from the Wilson line to two loops,
%\begin{equation}
%Z_{v}=1+Z_{v}^{(1)}+Z_{v}^{(2)}+\ldots,
%\label{Zgsdef}
%\end{equation}
%where $Z_{v}^{(n)}\sim{\cal O}(\alpha_s^n)$, and 
where the one-loop $Z_{v}^{(1)}$ was computed in appendix~\ref{app:counterterm} (see eq.~(\ref{Zgs1})) and the two-loop one,
$Z_{v}^{(2)}$, will be determined below. The renormalization of this web is described in  fig.~\ref{fig:15web_counterterms}, which shows the four contributing diagrams $(c)$, $(g)$,$(h)$ and~$(i)$ along with the counterterm contributions arising upon contracting one or both boomerang gluons to a point. 
%\begin{figure}
%\begin{center}
%\scalebox{0.6}{\includegraphics{2loopcounterterm.pdf}}
%\caption{$(A)$~Combination of gluons occuring in diagram $(c)$ of the
%  (5,1) web of figure~\ref{fig:15}; $(B)$~Counterterm graph associated
%  with shrinking the innermost boomerang gluon to a point.}
%\label{fig:2loopcounterterm}
%\end{center}
%\end{figure}
In diagram $(c)$, a one-loop counterterm is required to compensate for the subdivergence associated with shrinking the innermost boomerang gluon to the emission vertex of the non-boomerang gluon at point $u$, going to the second line in fig.~\ref{fig:15web_counterterms}.  The remaining diagram after this contraction is the mushroom graph of figure~\ref{fig:13}$(a)$, which was computed in section~\ref{sec:13}. 
An additional renormalization is required to compensate for the divergence associated with shrinking both gluons to a point, going to the third row in fig.~\ref{fig:15web_counterterms}. The remaining diagram is simply the one gluon exchange diagram.
Similarly, one- and two-loop counterterms are required for diagrams~$(g)$ and~$(i)$, with the only difference to the above being that now the first boomerang gluon being shrunk renormalizes one of the vertices of the second, rather than the non-boomerang emission vertex. Finally, diagram $(h)$ has no one-loop subdivergences, as each of the boomerang gluons straddles two emission vertices, but it does require a two-loop counterterm corresponding to shrinking both boomerang gluons simultaneously to a point.

As implied by the above description, the required counterterm contributions can be all identified considering the boomerang subdiagrams in fig.~\ref{fig:15web_counterterms}, without reference to the integration over~$u$ nor the other Wilson line. We therefore proceed to compute these considering the pole terms in the integrand of eq.~(\ref{FDform_one_gluon}). To this end we introduce an upper limit on the furthest boomerang gluon attachment  along the Wilson line, which we denote by $u_{\max}$, and consider the poles arising in $\mathcal{B}_{i}^{[0,u_{\max}]}(u)$ for each of the contributing diagrams.
For diagram $(c)$ we find
\begin{equation}
\begin{split}
  \mathcal{B}_{c}^{[0,u_{\max}]}(u)=\,& \bar{\mu}^{4\epsilon}
  (-\beta_1^2)^{2\epsilon}
g_s^4\mathcal{N}^2\int^{u}_{0}ds_2
\int_{s_2}^u dt_2\int^{u_{\max}}_{u}ds_1\int_{u}^{s_1}dt_1\,((s_1-s_2)^2)^{\epsilon-1}((t_1-t_2)^2)^{\epsilon-1}
%\theta(s_1>t_1)\theta(t_2>s_2) 
\\
=\,&\Big(\frac{g_s^2}{4\pi^2}\Big)^2\bigg\{\frac{1}{8\,\epsilon^2}+\frac{1}{4\,\epsilon} \Big[1+2\gamma_E+2\,\text{ln}\Big(\frac{\sqrt{-\beta_1^2}\mu(u_{\max}-u)u}{u_{\max}}\Big)\Big]+ \mathcal{O}(\epsilon^0)\bigg\},
\end{split}
\end{equation}
where in the second line we performed the four integrations and expanded in $\epsilon$, discarding finite terms.
Similarly, for the remaining diagrams we find
\begin{align}
 \mathcal{B}_{g}^{[0,u_{\max}]}(u)=\,&\Big(\frac{g_s^2}{4\pi^2}\Big)^2\bigg\{\frac{1}{8\,\epsilon^2}+\frac{1}{2\,\epsilon} \Big[1+\gamma_E+\text{ln}\Big(\frac{\sqrt{-\beta_1^2}\mu(u_{\max}-u)u}{u_{\max}}\Big)\Big]+ \mathcal{O}(\epsilon^0)\bigg\};\notag\\
 \mathcal{B}_{h}^{[0,u_{\max}]}=\,&\Big(\frac{g_s^2}{4\pi^2}\Big)^2\bigg\{\frac{1}{4\,\epsilon} + \mathcal{O}(\epsilon^0)\bigg\};\notag\\
 \mathcal{B}_{i}^{[0,u_{\max}]}(u)=\,&\Big(\frac{g_s^2}{4\pi^2}\Big)^2\bigg\{\frac{1}{8\,\epsilon^2}+\frac{1}{2\,\epsilon} \Big[1+\gamma_E+\text{ln}\Big(\frac{\sqrt{-\beta_1^2}\mu(u_{\max}-u)u}{u_{\max}}\Big)\Big]+ \mathcal{O}(\epsilon^0)\bigg\}.
\label{FghiCT}
\end{align}
Note that diagram $(h)$, being free of one-loop subdivergences, does not have a double pole, nor does its single-pole carry any dependence on the position of the non-boomerang attachment $u$ and the cutoff $u_{\max}$.   

Next, consider the counterterm contributions removing the one-loop subdivergences in diagrams $(c)$, $(g)$ and~$(i)$ described by the second row in fig.~\ref{fig:15web_counterterms}. These three diagrams are all the same: they simply correspond to the one-loop counterterm $Z_{v}^{(1)}$ of eq.~(\ref{Zgs1}) times 
\begin{equation}
\label{B_Mushroom}
\begin{split}
  {\cal B}_M^{[0,u_{\max}]}(u) =\,& -\bar{\mu}^{2\epsilon}
(-\beta_1^2)^{\epsilon}
  g_s^2\mathcal{N}\int^{u_{\max}}_{u}ds_1\int^{u}_{0}ds_2 ((s_1-s_2)^2)^{\epsilon-1}\\
  =\,& \,
  \bar{\mu}^{2\epsilon} (-\beta_1^2)^{\epsilon} g_s^2\mathcal{N}
  \frac{u_{\max}^{2\epsilon}-(u_{\max}-u)^{2\epsilon}-u^{2\epsilon}}{2\epsilon(1-2\epsilon)},
\end{split}
\end{equation}
where the subscript $M$ indicates that this function forms the integrand of the mushroom diagram, that is inserting 
${\cal B}_M^{[0,u_{\max}]}(u)$ into eq.~(\ref{FDform_one_gluon}) and sending $u_{\max}\to \infty$, one recovers the mushroom diagram of section~\ref{sec:13}.
Upon expanding eq.~(\ref{B_Mushroom}) in $\epsilon$ and multiplying by $Z_{v}^{(1)}$ of eq.~(\ref{Zgs1})  we obtain:
\begin{equation}
  Z_{v}^{(1)} {\cal B}_M^{[0,u_{\max}]}(u) = \Big(\frac{g_s^2}{4\pi^2}\Big)^2\bigg\{-\frac{1}{4\epsilon^2} - \frac{1}{2\epsilon} \Big[1+\gamma_E+\text{ln}\Big(\frac{\sqrt{-\beta_1^2}
      \mu(u_{\max}-u)u}{u_{\max}}\Big)\Big]+ \mathcal{O}(\epsilon^0)\bigg\}.
\label{x1}
\end{equation}

To complete the renormalization of the web we proceed to determine the relevant two-loop counterterm by
requiring that the sum of all contributions to the web, weighted by the appropriate exponentiated colour factors displayed in the second line of eq. (\ref{1-5i}), is ultraviolet finite:
\begin{equation}
\label{15web_Two_loop_UV_finiteness}
\begin{split}
&\Big(\mathcal{B}_{c}^{[0,u_{\max}]}(u) + Z_{v}^{(1)}{\cal B}_M^{[0,u_{\max}]}(u) +Z_{v(c)}^{(2)} \Big) 
+ \Big(\mathcal{B}_{g}^{[0,u_{\max}]}(u) +Z_{v}^{(1)} {\cal B}_M^{[0,u_{\max}]}(u)
+Z_{v(g)}^{(2)}
\Big) \\
& + 2\Big(\mathcal{B}_{h}^{[0,u_{\max}]}(u)+Z_{v(h)}^{(2)}\Big)
+\Big(\mathcal{B}_{i}^{[0,u_{\max}]}(u)+ Z_{v}^{(1)}{\cal B}_M^{[0,u_{\max}]}(u) +Z_{v(i)}^{(2)} \Big) = \mathcal{O}(\epsilon^0),
\end{split}
\end{equation}
It is straightforward to verify that for each of the three diagrams $(c)$, $(g)$ and~$(i)$ the logarithmic dependence on the position of the non-boomerang attachment $u$ and the cutoff $u_{\max}$ cancels with the corresponding $Z_{v}^{(1)} {\cal B}_M^{[0,u_{\max}]}(u)$ counterterm. Ultraviolet finiteness of the sum of diagrams in eq.~(\ref{15web_Two_loop_UV_finiteness}) fixes the two-loop vertex renormalization factor associated with multiple gluon exchange graphs to be
\begin{equation}
\begin{split}
Z_{v({\rm MGE})}^{(2)}  
\equiv
Z_{v(c)}^{(2)}+Z_{v(g)}^{(2)}+2Z_{v(h)}^{(2)}+Z_{v(i)}^{(2)}
%=& -\Big(\mathcal{F}^{\rm CT}_{c}+\mathcal{F}^{\rm CT}_{g}+2\,\mathcal{F}^{\rm CT}_{h}+\mathcal{F}^{\rm CT}_{i}+3\otimes_1\mathcal{M}\Big)\Big|_{\text{poles}}\\
=&\Big(\frac{g_s^2}{4\pi^2}\Big)^2\bigg\{\frac{3}{8\,\epsilon^2} - \frac{1}{4\,\epsilon}\bigg\}.
\end{split}
\end{equation}
The total contribution to the web from the second-order counterterm
may be written as $Z_{v({\rm MGE})}^{(2)}$ multiplying the one-gluon exchange web:
\begin{equation}
\begin{split}
 Z_{v({\rm MGE})}^{(2)} \mathcal{F}^{(1)}_{(1,1)}&=\left(\frac{g_s^2}{8\pi^2}\right)^3\int_0^1 dz p_\epsilon(z,\alpha_{12})
\left\{
\frac{1}{\epsilon^2}\left[2+\frac{3}{2}\ln z
+\frac34\ln\left(\frac{\mu^2}{m^2}\right)\right]
+\frac{1}{\epsilon}\left[3+\frac{3\pi^2}{4}
\right.\right.\\
&\left.\left. +3\ln z+\frac32\ln^2 z+\ln\left(\frac{\mu^2}{m^2}\right)
\left(\frac72+3\ln z
\right)+\frac98\ln^2\left(\frac{\mu^2}{m^2}\right)\right]
\right\}.
\end{split}
\end{equation}

The total renormalized web is now given by
\begin{equation}
W_{(5,1)}=c_1^{[3,2]}\Big[{\cal F}_c+{\cal F}_g+2{\cal F}_h+{\cal F}_i
+3Z_{v}^{(1)}{\cal F}_{M}+ Z_{v({\rm MGE})}^{(2)} \mathcal{F}^{(1)}_{(1,1)}\Big],
\label{w15ren}
\end{equation}
where ${\cal F}_{M}$ denotes the mushroom graph of figure~\ref{fig:13}$(a)$. 
As in other two-line webs no additional subtraction of commutators is necessary and thus, the contribution of the (5,1) web to the soft anomalous dimension is directly given by the single-pole contribution to eq.~(\ref{w15ren}).
Combining all contributions leads to 
\begin{equation}
F^{(3)}_{(5,1);1}(\alpha_{12})=\int_0^1 dx_1 p_0(x_1,\alpha_{12}){\cal G}^{(3)}_{(5,1);1}
(x_1,\alpha_{12}),
\label{FtoG15}
\end{equation}
where the appropriate web kernel is
\begin{align}
{\cal G}^{(3)}_{(5,1);1}(x_1,\alpha_{12})&=-8\left[\frac43-\frac{\pi^2}{9}
-\frac56 \ln\left(\frac{q(x_1,\alpha_{12})}{x_1^2}\right)
+\frac18 \ln^2\left(\frac{q(x_1,\alpha_{12})}{x_1^2}\right)\right].
\label{G15}
\end{align}
Finally, performing the integration over $x_1$ we readily obtain the result in terms of the basis functions $M_{k,l,n}(\alpha_{ij})$ of eq.~(\ref{eq:Mbasis}):
\begin{align}
F^{(3)}_{(5,1);1}(\alpha_{12})=-8r(\alpha_{12})&\left[\frac43
M_{0,0,0}(\alpha_{12})-\frac13M_{0,2,0}(\alpha_{12})+\frac{1}{12}M_{0,0,2}(\alpha_{12})-\frac56M_{1,0,0}(\alpha_{12})\right.\notag\\
&\left.\quad+\frac18M_{2,0,0}(\alpha_{12})\right].
\label{F15}
\end{align}
We confirm once more the pattern described following eq.~(\ref{F114res}). Indeed, the (5,1) web can also be expressed in terms of these  basis functions, and similarly to other boomerang webs it features mixed, non-maximal weights, here weights 3, 2 and 1.

\subsubsection{The (2,4) web}
\label{sec:24}

The last remaining boomerang web is the (2,4) web of
figure~\ref{fig:24} where, as already discussed, only diagrams
$(a)$ through $(f)$ will contribute to the final result. Applying a replica
analysis to obtain the web mixing matrix, we find (after applying
colour algebra) that the corresponding exponentiated colour factors are given by
\begin{align}
\tilde{C}(a)&=-\frac12 c_1^{[3,2]}+c_2^{[3,2]};\notag\\
\tilde{C}(b)&=-\frac32 c_1^{[3,2]}+c_2^{[3,2]};\notag\\
\tilde{C}(c)&=\tilde{C}(d)=0\notag\\
\tilde{C}(e)&=\tilde{C}(f)=-c_1^{[3,2]},
\label{ECF24}
\end{align}
so that the contribution of the entire renormalised web is given by
\begin{align}
\label{comb}
W_{(2,4)} =&\, c^{[3,2]}_1 \Big[ {\cal F}_{(2,4);1} + {\cal F}_{(2,4);1}^{\rm
             CT} \Big] + c^{[3,2]}_2 \Big[ {\cal F}_{(2,4);2} \Big] \\ \notag
=& - c^{[3,2]}_1\left(\frac{1}{2}
{\cal F}_a + \frac{3}{2}{\cal F}_b + {\cal F}_e
+ {\cal F}_f +{\cal F}_e^{\rm CT}
+ {\cal F}_f^{\rm CT} \right)+c^{[3,2]}_2\left({\cal F}_a
+ {\cal F}_b\right)\,,
\end{align}
As indicated in the second line, $ {\cal F}_{(2,4);1}^{\rm CT}$ is included because
 diagrams $(e)$ and $(f)$ must be supplemented by graphs in
which the one-loop vertex counterterm of
appendix~\ref{app:counterterm} dresses the two-loop crossed gluon web
of figure~\ref{fig:cross}, whose kinematic factor can be found in
eqs.~(\ref{F2-1}, \ref{F20}). This gives
\begin{equation}
  \label{eq:F24CT}
   {\cal F}_{(2,4);1}^{\rm CT} = -2Z_v^{(1)}{\cal F}_X.
\end{equation}
Using the calculational approach adopted
for the other boomerang webs, we find that the kinematic part of each 
diagram $D\in \{a, b, e, f\}$ in figure~\ref{fig:24} can be written as
\begin{equation}
{\cal F}_D(\alpha_{12})=\kappa^3\frac{\Gamma(6\epsilon)}{\epsilon(1-2\epsilon)}
\int_0^1 dy\,dz\,p_\epsilon(y,\alpha_{12})p_\epsilon(z,\alpha_{12})
\phi_D (y,z;\alpha_{12})\,,
\label{Fi24}
\end{equation}
where we have already performed the integration over the boomerang gluon.
Considering first diagram $(a)$ we find
\begin{equation}
\begin{split}
\phi_a (y,z;\alpha_{12})
= (1-y)^{2\epsilon}\int_{0}^{1}dw\, w^{4\epsilon-1}(1-w)^{2\epsilon-1}&\bigg[1-\bigg(1-\frac{1-w}{w}\frac{1-z}{1-y}\bigg)^{2\epsilon} \bigg]\\
&\times \theta\bigg(\frac{1-w}{w}<\frac{y}{z}\bigg)\theta\bigg(\frac{1-w}{w}<\frac{1-y}{1-z}\bigg).
\label{phia24}
\end{split}
\end{equation}
Making the change of variable to:
\begin{equation}
q = \frac{1-w}{w},
\end{equation}
eq.~(\ref{phia24}) becomes:
\begin{align}
\label{phia24b}
\phi_a (y,z;\alpha_{12}) = (1-y)^{2\epsilon}\int_{0}^{\infty}dq \hspace{0.1cm} q^{2\epsilon-1}(1+q)^{-6\epsilon} \bigg[1-\bigg(1-\frac{1-z}{1-y}q\bigg)^{2\epsilon}\bigg]\theta\bigg(q<\frac{y}{z}\bigg)\theta\bigg(q<\frac{1-y}{1-z}\bigg)\,,
\end{align}
which can be integrated to give
\begin{equation}
\begin{split}
\phi_a&(y,z;\alpha_{12}) =\frac{(1-y)^{2\epsilon}}{2\epsilon} \times \\&\Bigg\{ 
\Big(\frac{y}{z}\Big)^{2\epsilon} \theta(z>y) \Bigg[
{}_2F_1\left(2\epsilon,6\epsilon;1+2\epsilon;-\frac{y}{z}\right)
- F_1\left(2\epsilon,6\epsilon,-2\epsilon,1+2\epsilon;-\frac{y}{z},\frac{y(1-z)}{z(1-y)}\right)\Bigg]\,+ \\
&
\Big(\frac{1-y}{1-z}\Big)^{2\epsilon} \theta(y>z) \Bigg[{}_2F_1\left(2\epsilon,6\epsilon;1+2\epsilon;-\frac{1-y}{1-z}\right)- \frac{\Gamma^2(1+2\epsilon)}{\Gamma(1+4\epsilon)} {{}_2}F_1\left(2\epsilon,6\epsilon;1+4\epsilon;-\frac{1-y}{1-z}\right)\Bigg]\Bigg\} .
\label{phia24c}
\end{split}
\end{equation}
Here ${_2}F_1(a,b;c;z)$ is the Gauss hypergeometric function, and
$F_1(a,b,c,d;x,y)$ the Appell $F_1$ function. We may expand the former in
$\epsilon$ using the \texttt{HypExp} package in
Mathematica~\cite{Huber:2005yg,Huber:2007dx}. We explain how to expand
the Appell function in appendix~\ref{app:Appell}, such that after
expansion in $\epsilon$ eq.~(\ref{phia24c}) assumes the form
\begin{equation}
    \phi_a (y,z;\alpha_{12})  
    =  2\epsilon\, \,\text{Li}_2\Big(\frac{z}{y}\frac{1-y}{1-z}\Big) \,\theta(y>z)
    +2\epsilon\,\zeta_2 \theta(z>y)
    +{\cal O}(\epsilon^2)\,.
\label{phi24d_first_version}
\end{equation}
Next we note that owing to the symmetry of the propagator functions in (\ref{Fi24}), we are free to transform the integration parameters according to $y\rightarrow 1-y$ and $z\rightarrow 1-z$, in any of the terms. Using this freedom we eliminate the $\theta(z>y)$ component, shifting the $\zeta_2$ term to the $\theta(y>z)$ component, obtaining 
\begin{equation}
\phi_a (y,z;\alpha_{12})  = \epsilon\, \bigg\{2\,\zeta_2 + 2 \,\text{Li}_2\Big(\frac{z}{y}\frac{1-y}{1-z}\Big) \bigg\}\,\theta(y>z) +{\cal O}(\epsilon^2).
\label{phi24d}
\end{equation}

Similarly, the kinematic parts
of diagrams $(b)$, $(e)$ and $(f)$ in figure~\ref{fig:24} can be written in the form of
eq.~(\ref{Fi24}) with the kernels
\begin{align}
\phi_D (y,z;\alpha_{12}) &= (1-y)^{2\epsilon}\int_{0}^{\infty}dq \, q^{2\epsilon-1}(1+q)^{-6\epsilon} \theta\bigg(q>\frac{y}{z}\bigg)\theta\left(q<\frac{1-y}{1-z}\right)\,\, \psi_D\left(\displaystyle{\frac{1-z}{1-y}q}\right)
\end{align}
where 
\begin{align}
\begin{split}
\psi_b(Q)&=1-\psi_f(Q);\qquad
\psi_e(Q)=Q^{2\epsilon}-\psi_b(Q);\qquad
\psi_f(Q)=\left(1-Q\right)^{2\epsilon}\,.
\end{split}
\end{align}
%\begin{align}
%\begin{split}
%\psi_b(q,y,z)&=1-\bigg(1-\frac{1-z}{1-y}q\bigg)^{2\epsilon}\qquad %\psi_f(q,y,z)= 1-\psi_b(q,y,z)
%\\
%\psi_e(q,y,z)&=\bigg(1-\frac{1-z}{1-y}q\bigg)^{2\epsilon}-1+\bigg(\frac{1-z}{1-y}\bigg)^{2\epsilon}q^{2\epsilon}
%\\
%\psi_f(q,y,z)&=\bigg(1-\frac{1-z}{1-y}q\bigg)^{2\epsilon}
%\end{split}
%\end{align}
The corresponding $\epsilon$-expanded kernels are given by
\begin{align}
\label{phi24kernels}
\phi_b (y,z;\alpha_{12}) = &\, \epsilon\, \bigg\{2\,\zeta_2 - 2 \,\text{Li}_2\Big(\frac{z}{y}\frac{1-y}{1-z}\Big) \bigg\}\,\theta(y>z)+{\cal O}(\epsilon^2);\\
\phi_e(y,z;\alpha_{12})  = &\, \bigg\{\text{ln}\Big(\frac{y}{z}\frac{1-z}{1-y}\Big) + \Big[\frac{1}{2} \text{ln}^2\Big(\frac{1-z}{1-y}\Big) - \frac{1}{2} \text{ln}^2\Big(\frac{z^2}{y^2}\frac{1-y}{1-z}\Big) + 2\,\text{ln}\Big(\frac{y}{z}\frac{1-z}{1-y}\Big)\text{ln}(1-z) \notag\\
&- 2\,\zeta_2 + 6 \,\text{Li}_2\Big(-\frac{1-z}{1-y}\Big) - 6 \,\text{Li}_2\Big(-\frac{z}{y}\Big) + 2 \,\text{Li}_2\Big(\frac{z}{y}\frac{1-y}{1-z}\Big)\Big]\epsilon\, \bigg\}\,\theta(y>z)+{\cal O}(\epsilon^2);\notag\\
\phi_f (y,z;\alpha_{12}) =&\,  \bigg\{\text{ln}\Big(\frac{y}{z}\frac{1-z}{1-y}\Big) + \Big[\text{ln}^2\Big(\frac{1-z}{1-y}\Big) - \text{ln}^2\Big(\frac{z}{y}\Big) + 2\,\text{ln}\Big(\frac{y}{z}\frac{1-z}{1-y}\Big)\text{ln}(1-z) - 2\,\zeta_2 \notag\\
&+ 6 \,\text{Li}_2\Big(-\frac{1-z}{1-y}\Big) - 6 \,\text{Li}_2\Big(-\frac{z}{y}\Big) + 2 \,\text{Li}_2\Big(\frac{z}{y}\frac{1-y}{1-z}\Big)\Big]\epsilon\,  \bigg\}\,\theta(y>z)+{\cal O}(\epsilon^2). \notag
\end{align}
%\Chris{I added the following two equations with discussion, so that one can explicitly verify the cancellation of scale logs.} 

The combination of kinematic factors appearing in eq.~(\ref{comb}) for the
colour factor $c_1^{[3,2]}$ then evaluates to
\begin{align}
\label{F24combination}
\begin{split}
&{\cal F}_{(2,4);1}(\alpha_{12}) 
=-\left(\frac{g_s^2}{8\pi^2}\right)^3\int_0^1 dy\,dz\, p_0(y,\alpha_{12})
p_0(z,\alpha_{12})\theta(y>z)\\
&\times\left\{\frac16\left[-12{\rm Li}_2
\left(-\frac{1-z}{1-y}\right)+12{\rm Li}_2\left(-\frac{z}{y}\right)-2
{\rm Li}_2\left(\frac{z(1-y)}{y(1-z)}\right)+2\ln\left(\frac{z(1-y)}{y(1-z)}
\right)
\ln\left(\frac{q(y,\alpha_{12})}{y^2}\right)\right.\right.\\
&\left.\left.
\quad -2\ln\left(\frac{y}{1-y}\right)\ln\left(\frac{q(z,\alpha_{12})}{z^2}\right)
-2\ln\left(\frac{1-z}{z}\right)\log\left(\frac{q(z,\alpha_{12})}{(1-z)^2}
\right)-6\ln\left(\frac{y}{1-y}\right)\ln\left(\frac{1-z}{z}\right)\right.\right.\\
&\left.\left.
\quad-4\ln\left(\frac{y(1-z)}{z(1-y)}\right)-12\ln z\ln\left(
\frac{y(1-z)}{z(1-y)}\right)-\ln^2\left(\frac{y}{1-y}\right)
-9\ln^2\left(\frac{1-z}{z}\right)\right]\right.\\
&\left.\quad-\ln\left(\frac{\mu^2}{m^2}\right)\ln\left(
\frac{y(1-z)}{z(1-y)}\right)\right\}\frac{1}{\epsilon}+{\cal O}(\epsilon^0).
\end{split}
\end{align}
Furthermore, the counterterm contribution amounts to
\begin{align}
\label{24counterterm}
\begin{split}
{\cal F}_{(2,4);1}^{\rm CT}(\alpha_{12}) &=-\frac12\left(\frac{g_s^2}{8\pi^2}\right)^3
\int_0^1 dy\,dz\, p_0(y,\alpha_{12})
p_0(z,\alpha_{12})\theta(y>z)\left\{
\ln\left(\frac{y(1-z)}{z(1-y)}\right)\left[2\ln\left(\frac{\mu^2}{m^2}\right)
\right.\right.\\
&\left.\left.
+\ln\left(\frac{q(y,\alpha_{12})}{y^2}\right)+\ln\left(\frac{q(z,\alpha_{12})}
{z^2}\right)+2\ln (y z)\right]
+4{\rm Li}_2\left(-\frac{1-z}{1-y}\right)-4{\rm Li}_2\left(-\frac{z}{y}\right)
\right.\\
&\left.+\ln^2\left(\frac{1-z}{1-y}\right)-\ln^2\left(\frac{z}{y}\right)
\right\}\frac{1}{\epsilon}+{\cal O}(\epsilon^0).
\end{split}
\end{align}

Using the fact that there are no lower-order contributions that form non-zero commutators in eq.~(\ref{Gamres}), the ${\cal O}(\epsilon^{-1})$ coefficient of the renormalized (2,4) web directly determine its contribution to the soft anomalous dimension.
Putting things together in the combination of eq.~(\ref{comb}), we find the integrals multiplying the two colour structures $c_i^{[3,2]}$ for $i=1,2$  to be
\begin{equation}
F_{(2,4);i}^{(3)}(\alpha_{12})=\int_0^1 dy\int_0^1 dz\, p_0(y,\alpha_{12})\,
p_0(z,\alpha_{12})
\,\theta(y>z)\,{\cal G}_{(2,4);i}^{(3)}(y,z,\alpha_{12})\,,
\label{FtoG24}
\end{equation}
where we displayed the overall Heaviside function restricting the integration range, and where
\begin{align}
{\cal G}^{(3)}_{(2,4);1}(y,z,\alpha_{12})&= \frac43
\bigg[ 2\,\text{Li}_2\Big(\frac{z}{y}\frac{1-y}
{1-z}\Big) + 4\,\text{ln}\Big(\frac{y}{1-y}\Big) - \text{ln}\Big(
\frac{y}{1-y}\Big)\text{ln}\Big(\frac{q(y,\alpha_{12})}{y^2}\Big)
- 2\,\text{ln}^2\Big(\frac{y}{1-y}\Big)\notag\\
&+ 4\,\text{ln}\Big(\frac{1-z}{z}\Big) - \text{ln}\Big(\frac{1-z}{z}\Big)
\text{ln}\Big(\frac{q(z,\alpha_{12})}{(1-z)^2}\Big) - \text{ln}\Big(
\frac{y}{1-y}\Big) \text{ln}\Big(\frac{q(z,\alpha_{12})}{z^2}\Big) \notag\\
&+ \text{ln}\Big(\frac{z}{1-z}\Big) \text{ln}\Big(\frac{q(y,\alpha_{12})}
{y^2}\Big) \bigg];\notag\\
{\cal G}^{(3)}_{(2,4);2}(y,z,\alpha_{12})&=-\frac{16}{3}\zeta_2\,.
\label{G24}
\end{align}
The second kernel can be straightforwardly integrated over $y$ and $z$ and written in terms of the basis functions of
eq.~(\ref{eq:Mbasis}) as
%\begin{equation}
%F^{(3)}_{(2,4);2}(\alpha_{12})= %-\frac13r^2(\alpha_{12})
%\Big(4M_{0,2,0}(\alpha_{12})-M_{0,0,2}(\alpha_{12})\Big)M_{0,0,0}(\alpha_{12}).
%\label{F242}
%\end{equation}
\begin{equation}
F^{(3)}_{(2,4);2}(\alpha_{12})= -\frac49r^2(\alpha_{12})
\pi^2 M_{0,0,0}^2(\alpha_{12})
=-\frac49r^2(\alpha_{12})
\Big(6 M_{0,2,1}(\alpha_{12}) - M_{0,0,3}(\alpha_{12})\Big)\,.
\label{F242}
\end{equation}

The integration of the first kernel in eq.~(\ref{G24}), however, is not immediately
interpretable in terms of basis functions. Firstly, it contains
dilogarithms involving the parameters $y$ and $z$, which are not part
of the integrand of eq.~(\ref{eq:Mbasis}), consisting exclusively of powers of logarithms. 
The webs previously calculated in this paper and in
refs.~\cite{Gardi:2013saa,Falcioni:2014pka} contained dilogarithms at
intermediate stages, but these completely cancelled at the level of
the subtracted web integrand. Here this is not the case, and we have
furthermore found no variable transformation (or dilogarithm identity)
that removes the dilogarithms from eq.~(\ref{G24}). Secondly, there is
a remaining Heaviside function in eq.~(\ref{G24}), which also does not
appear in the definition of eq.~(\ref{eq:Mbasis}). It thus appears
that our previously conjectured basis of functions is
incomplete. However, this conclusion is premature and
incorrect. Remarkably, the kernel of eq.~(\ref{G24}) may be integrated
fully analytically and found to respect the basis of
eq.~(\ref{eq:Mbasis}) after all. 
In appendix~\ref{app:Polylog24} we show how this works in detail.
The final result reads
\begin{align}
\label{F241_main_text}
\begin{split}
F^{(3)}_{(2,4);1}(\alpha_{12})&=\frac43r^2(\alpha_{12})\Big[8M_{0,1,1}(\alpha_{12})
- 4M_{1,1,1}(\alpha_{12}) - 3M_{0,2,1}(\alpha_{12}) - \frac{1}{6}
\,M_{0,0,3}(\alpha_{12})  \Big]\,,
\end{split}
\end{align}
%\begin{align}
%\label{F241_main_text}
%\begin{split}
%F^{(3)}_{(2,4);1}(\alpha_{12})&=\frac43r^2(\alpha_{12})\Big[8M_{0,1,1}(\alpha_{12})
%- 4M_{1,1,1}(\alpha_{12}) - 3M_{0,2,1}(\alpha_{12}) - \frac{1}{24}
%\,M_{0,0,0}^4(\alpha_{12})  \Big]\,,
%\end{split}
%\end{align}
%\begin{align}
%\label{F241_main_text}
%\begin{split}
%F^{(3)}_{(2,4);1}(\alpha_{12})&=\frac43r^2(\alpha_{12})\Big[8M_{0,1,1}(\alpha_{12})
%- 4M_{1,1,1}(\alpha_{12}) - 2M_{0,2,1}(\alpha_{12}) - \frac{1}{24}
%\,M_{0,0,0}^4(\alpha_{12}) \\
%&\qquad \qquad \qquad - \frac{1}{2}\,M_{0,0,0}(\alpha_{12})M_{0,2,0}(\alpha_{12})\Big]\,,
%\end{split}
%\end{align}
featuring weight 4 as well as weight 3 contributions.

We have now calculated all multiple-gluon exchange boomerang webs up to three-loop order, and shown that they conform with the expected functional form: an overall rational function consisting of powers of $r({\alpha_{ij}})$ for any non-boomerang exchange between lines $i$ and $j$, multiplying a pure transcendental function. The latter may be expressed in terms of sums of products of our $M_{k,l,n}(\alpha_{ij})$ basis functions, each dependent on a single $\alpha_{ij}$ and having a restricted symbol alphabet, eq.~(\ref{alphabet}). We found one salient difference compared to non-boomerang MGEWs: while MGEWs without boomerang gluons have uniform, maximal weight (weight 5 for subtracted webs at three loops), boomerang ones have mixed, non-maximal weight. Table~\ref{tab:weights_3loop} presents the weights occurring in each case:
\begin{table}[h]
\begin{center}
\begin{tabular}{|c|c|l|}
\hline
web & boomerangs & weights\\
\hline
(1,1,4)\,\,$c^{[3,3]}_3$ &1&4\\
(1,1,4)\,\,$c^{[3,3]}_4$ &1&4,3\\
(1,2,3)\,\,$c^{[3,3]}_4$ &1& 4,3,2\\
  \hline
(3,3)\,\,$c^{[3,2]}_1$ &2& 3,2\\
(5,1)\,\,$c^{[3,2]}_1$ &2& 3,2,1\\
(2,4)\,\,$c^{[3,2]}_2$ &1& 4  \\
(2,4)\,\,$c^{[3,2]}_1$ &1& 4,3  \\
\hline
\end{tabular}
\caption{\label{tab:weights_3loop}
Transcendental weights in the kinematic functions of three-loop webs. The left column specifies the web and the colour factor component using the bases in eqs.~(\ref{colbasis3}) and (\ref{colbasis2}), the middle column presents the number of boomerang gluons, and the right one the weights of the transcendental functions entering the anomalous dimension.}
\end{center}
\end{table}
the fact that the maximal weight occurring reduces by one unit with each additional boomerang gluon, is expected based on the observation in section~\ref{sec:kinboom}, namely that for each boomerang gluon, there is one integration, over $x_l$ in eq.~(\ref{gendiagboom2}), which instead of increasing the weight by one unit as in non-boomerang webs, generates a factor of~$1/(1-2\epsilon)$, which reflects a power divergence due to an instantaneous interaction. At leading order in the $\epsilon$ expansion this readily translates into a weight drop compared to the non-boomerang case, and furthermore, when subleading powers in $\epsilon$ hit higher-order pole terms in the web kernel, further lower weight terms emerge.

As discussed in section~\ref{sec:bases}, a further consistency check
on higher-loop webs is that a subset of the information in a given web
can be obtained from webs connecting a greater number of Wilson lines,
through the process of collinear reduction. We discuss how this
applies to the boomerang webs calculated in this paper in the following section.

\section{Collinear reduction for boomerang webs}
\label{sec:collinear}

In section~\ref{sec:review}, we reviewed how kinematic factors have
been previously obtained for boomerang-free MGEWs~\cite{Gardi:2010rn,Gardi:2011yz,Gardi:2013ita,Gardi:2013saa,Falcioni:2014pka}. In
such cases, it was possible to perform an additional (albeit partial)
cross-check of the final results for web kinematic factors, using the
process of {\it collinear reduction}~\cite{Falcioni:2014pka}. Roughly speaking, this states
that one may take a web connecting $n$ Wilson lines, and obtain
kinematic results pertaining to webs connecting $m<n$ lines, by
identifying the 4-velocities of two or more Wilson lines in the
original web. In carrying out such a procedure, gluon emissions from
different Wilson lines may end up on the same line, and we must then
reinterpret the colour indices of such emissions appropriately. To
this end, the effective vertex formalism reviewed here in
section~\ref{sec:bases} becomes useful: eq.~(\ref{Csym}) implies that
there is no natural ordering of effective vertices if more than one of
them occurs on a given Wilson line. Thus, as far as the ordering of
vertices is concerned, it is similar to the case where they occur on
different Wilson lines. To explore the converse of this, imagine a web
in which two Wilson lines $l_1$ and $l_2$ contain a single vertex
each, which we may write as $V_{K_1}^{(l_1)}$ and $V_{K_2}^{(l_2)}$
respectively. The collinear reduction process then consists of the
following steps:
\begin{enumerate}
\item One sets the 4-velocities $\beta_1$ and $\beta_2$ to be equal
 (hence the term ``collinear'').
\item One identifies the partonic colour indices of the generators in
  the two vertices as living in the same (single) colour space.
\item One symmetrises over the colour factors of the two vertices, as
  in eq.~(\ref{Csym}).
\item One must include symmetry factors present in the obtained web,
  that are missing in the original web.
\end{enumerate}
This procedure generalises straightforwardly to any number of
effective vertices, and was already used in
ref.~\cite{Falcioni:2014pka}, where it provided a highly non-trivial
check of non-boomerang MGEWs at three-loop order. Note that not all information in
the fewer-line webs can be obtained by collinear reduction. Due to the
symmetrisation over multiple vertices, the fully antisymmetric colour
part of a given web cannot be generated. This is precisely the
contribution in which there is a fully connected colour factor on each
individual Wilson line, namely that arising from having at most a
single effective vertex on each line. For example, for three-line webs collinear reduction may yield the components involving $c_i^{[3,3]}$ for $i=1$ through $3$ in eq.~(\ref{colbasis3}), 
which contain two $C_{2,1}$ effective vertices (defined in eq.~(\ref{V2colour})) on line $i$,
but not $c_4^{[3,3]}$, which is fully antisymmetric. Similarly for two-line webs collinear reduction may yield the $c_2^{[3,2]}$ component defined in eq.~(\ref{colbasis2}), but not the $c_1^{[3,2]}$ one.

It is possible to use a similar collinear reduction procedure to check
some of the results of this paper. However, the presence of boomerang gluons
creates additional complications, which were not necessary to consider
in ref.~\cite{Falcioni:2014pka}. At the outset, in computing non-boomerang MGEWs using collinear reduction, the collinear limit $\beta_{i}=\beta_j$ simply amounts to identifying $\alpha_{ik}=\alpha_{jk}$ in the kinematic function of the original web. In turn, in computing boomerang webs, the collinear limit also involves taking $\alpha_{ij}\to -1$ for the boomerang gluon itself. In this limit $r(\alpha_{ij})$ of eq.~(\ref{r_def}) diverges and the basis functions $M_{k,l,n}(\alpha_{ij})$ become complex, so it is not a priori clear that the collinear limit exists. To better understand the problem we first consider the relation between the single-gluon exchange (1,1) web diagram and the self-energy one upon taking the collinear limit. We will subsequently explain how the problem is resolved in boomerang webs in which boomerang gluons straddle one or more emissions along the Wilson line. We will demonstrate the application of the collinear reduction procedure in the rather non-trivial example of the (1,1,4) web in section~\ref{coll_reduction_boomerang}.

\subsection{The collinear limit of the (1,1) web}
\label{sec:11reduce}

\begin{figure}[b]
\begin{center}
\scalebox{0.8}{\includegraphics{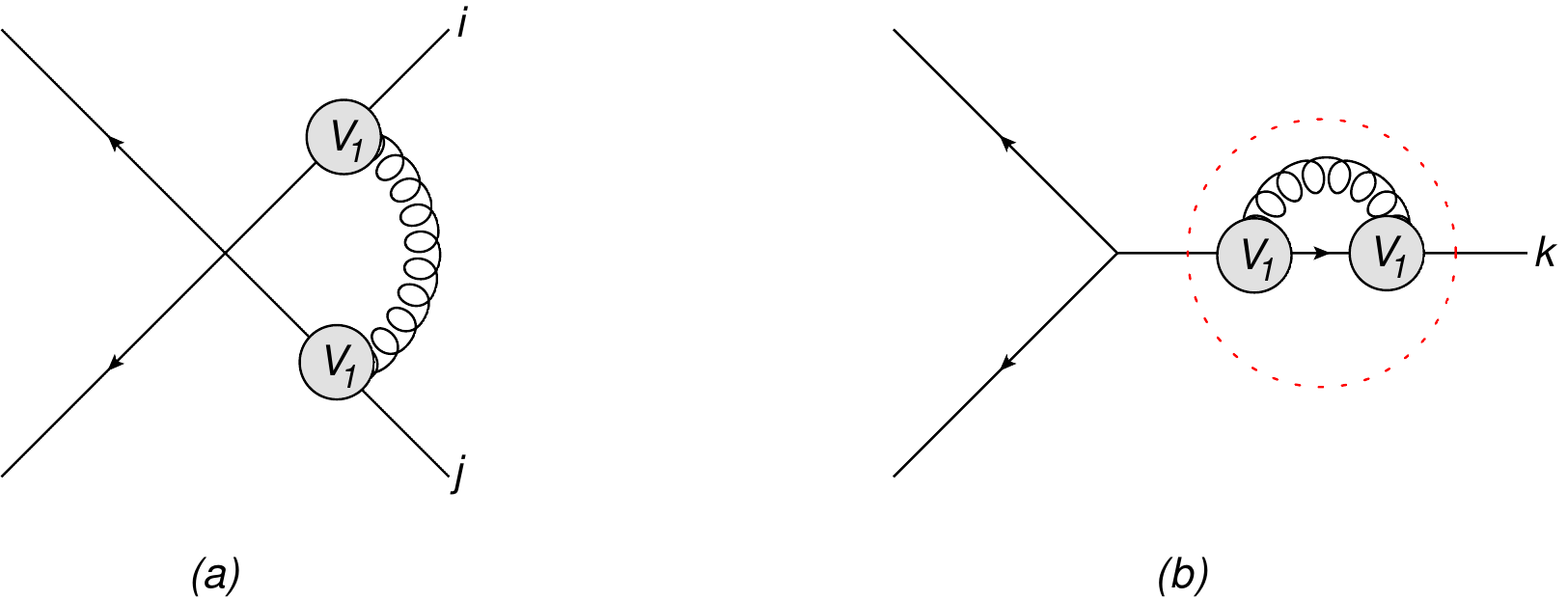}}
\caption{($a$) The (1,1) web, which features a 1-gluon emission vertex
  on each line; ($b$) collinear reduction of Wilson lines $i$ and $j$ to
  make a self-energy web on the single line $i$.}
\label{fig:11reduce}
\end{center}
\end{figure}

In this section, we consider the (1,1) web of
figure~\ref{fig:11reduce}($a$) and use it to analyse the collinear limit.
Upon identifying the two Wilson lines $i$ and $j$, we obtain the self-energy diagram of
figure~\ref{fig:11reduce}($b$), whose integral was
computed in section~\ref{sec:selfenergygraph}.
The kinematic factor for the (1,1) web is given in eq.~(\ref{F1res}),
\begin{equation}
\mathcal{F}^{(1)} (\alpha_{ij} , \epsilon)
\,=\,  \kappa \Gamma(2\epsilon) \int_{0}^{1} dx p_\epsilon(x, \alpha_{ij})=  
\frac{\kappa}{\epsilon}  \,  \frac{1 + \alpha_{ij}^2}{1 - \alpha_{ij}^2} \ln( \alpha_{ij})+\,{\cal O}(\epsilon^0)\,,
\label{F11reduce1}
\end{equation}
where in the second equality we substituted for $r (\alpha_{ij})$ using eq.~(\ref{r_def}) and expanded in $\epsilon$, keeping only the leading-order term.
Taking the collinear limit to obtain
figure~\ref{fig:11reduce}($b$), one identifies $\beta_i=\beta_j$, which
from eq.~(\ref{gammaij}) implies
\begin{equation}
\gamma_{ij}\rightarrow 2,\quad \alpha_{ij}\rightarrow -1.
\label{gammalim}
\end{equation}
One could therefore expect that the self-energy web would be given by the $\alpha_{ij}\rightarrow -1$ limit of eq.~(\ref{F11reduce1}), that is
\begin{equation}
\frac12\lim_{\alpha_{ij}\rightarrow -1} 
\mathcal{F}^{(1)} (\alpha_{ij} , \epsilon)
\,=\, \frac12 \frac{\kappa}{\epsilon}\lim_{\alpha_{ij}\rightarrow -1} 
  \,  \frac{1 + \alpha_{ij}^2}{1 - \alpha_{ij}^2} \ln( \alpha_{ij})+\cdots
\,=\, \frac12 \frac{\kappa}{\epsilon} \lim_{\alpha_{ij}\rightarrow -1} 
 \,  \frac{i\pi}{1 + \alpha_{ij}} +\cdots,
 \label{Coll_red_11_LO_fail}
\end{equation}
where the factor of $1/2$ originates from the
symmetrisation of the colour generators. 
Perhaps surprisingly, we see that the
kinematic limit required for the collinear reduction is ill-defined,
whereas the result we are expecting to reproduce -- the self-energy
factor of eq.~(\ref{FSE4}) -- is perfectly well-behaved. 

To see what has gone wrong, we may examine the kinematic factor for the (1,1) web
in more detail. Following the procedures outlined in
section~\ref{sec:review}, one finds (\emph{cf.} eq.~(\ref{FSE1}))
\begin{align}
{\cal F}_{(1,1)}^{(1)}(\alpha_{ij})&=g_s^2\bar{\mu}^{2\epsilon}{\cal N}\beta_i\cdot \beta_j
\int_0^\infty ds \int_0^\infty
dt\left[-(s\beta_i-t\beta_j)^2+i\varepsilon
\right]^{\epsilon-1}\,e^{-im\left(s\sqrt{\beta_i^2}
+t\sqrt{\beta_j^2-i\varepsilon}\right)},\,
\label{11kin}
\end{align}
where $s$ and $t$ are distance variables for the gluon attachments on
lines $i$ and $j$ respectively. Transforming according to
eqs.~(\ref{strescale}, \ref{sigmatau}), eq.~(\ref{11kin}) can be
rewritten as
\begin{align}
{\cal F}_{(1,1)}^{(1)}(\alpha_{ij})&=\kappa \gamma_{ij}
\int_0^\infty d\lambda \lambda^{2\epsilon-1} e^{-\lambda}
\int_0^1 dx
\left[x^2+(1-x)^2-\gamma_{ij}x(1-x)\right]^{\epsilon-1}\,
\notag\\
&=-\kappa\Gamma(2\epsilon) \left(\alpha_{ij}+\frac{1}{\alpha_{ij}}\right)
\int_0^1 dx
\left[x^2+(1-x)^2+\left(\alpha_{ij}+\frac{1}{\alpha_{ij}}\right)x(1-x)\right]^{\epsilon-1},
\label{11kin2}
\end{align}
where the $\lambda$ integral was carried out as
previously. In general kinematics one may subsequently perform the integration over $x$ using the factorization property of the expression in the square brackets ($q(x,\alpha)$ in eq.~(\ref{propafu2})) yielding the hypergeometric functions in eq.~(\ref{F1res}) corresponding to the two poles at 
\begin{equation}
\label{x_pole}
x_{\rm poles}=\left\{\frac{1}{1-\alpha_{ij}},\frac{\alpha_{ij}}{\alpha_{ij}-1}\right\}.
\end{equation}
Since we are dealing with an analytic function of $\alpha_{ij}$ we may start with real $0<\alpha_{ij}<1$, corresponding to space-like kinematics (say $\beta_i$ incoming and $\beta_j$ outgoing) to ensure that both poles in (\ref{x_pole}) are outwith the integration domain $x\in[0,1]$, and subsequently analytically continue to the required kinematic point (for time-like kinematics  $\alpha_{ij}$ is located just above the cut on the negative real axis). Note that this computation, and eq.~(\ref{F11reduce1}) in particular, is consistent with expanding under the integral about $\epsilon=0$, and integrating term by term.

In performing the collinear reduction of eq.~(\ref{11kin2}), one
must take the limit $\alpha_{ij}\rightarrow -1$, and include a factor
of 1/2 from symmetrisation of the gluon vertices. The relevant limit of the (1,1) web is then
\begin{align}
\label{collinear_limit_11}
\begin{split}
&\frac12 \lim_{\alpha_{ij}\to -1+i\varepsilon} {\cal F}_{(1,1)}^{(1)}(\alpha_{ij})\\
&=
-\kappa\Gamma(2\epsilon) \frac12 \lim_{\alpha_{ij}\to -1+i\varepsilon} 
\left(\alpha_{ij}+\frac{1}{\alpha_{ij}}\right)
\int_0^1 dx
\left[x^2+(1-x)^2+\left(\alpha_{ij}+\frac{1}{\alpha_{ij}}\right)x(1-x)\right]^{\epsilon-1}\,.
\end{split}
\end{align}
This limit may be compared with the direct computation of the self-energy web integral in eq.~(\ref{FSE3}), which yields
\begin{equation}
\label{self_enregy_target_result}
2\kappa\Gamma(2\epsilon)\int_{1/2}^1 \frac{ dx}{\left[(2x-1)^2\right] ^{1-\epsilon}}=2\kappa\Gamma(2\epsilon) \frac12\frac{1}{2\epsilon-1}\,.
\end{equation}
We have already seen in eq.~(\ref{Coll_red_11_LO_fail}) that the two fail to agree at leading order in $\epsilon$. Having discussed the pole structure of eq.~(\ref{11kin2}) we clearly see the origin of the problem: while for generic kinematics we may expand in $\epsilon$ under the integral near $\epsilon=0$, upon considering the special point $\alpha_{ij}=-1$
the two poles in eq.~(\ref{x_pole}) \emph{coincide} leading to a double pole at $x=\frac12$, which is not integrable near $\epsilon=0$; it requires instead $\epsilon>\frac12$, followed by analytical continuation in $\epsilon$, before an expansion can be performed. It is therefore not surprising that the result we obtained for general kinematic in (\ref{F11reduce1}) is incompatible with the special case of $\alpha_{ij}=-1$. 

The issue we encountered in eq.~(\ref{Coll_red_11_LO_fail}) boils down to an obstruction in performing analytic continuation of the general kinematic result in $\alpha_{ij}$, computed as an expansion in $\epsilon$, to the strict collinear limit where $\alpha_{ij}=-1$, where the function has a branch point. Indeed, adhering to the $i\varepsilon$ prescription, the collinear limit in eq.~(\ref{collinear_limit_11}) is itself 
well-defined, and furthermore, is equal to the self-energy web in eq.~(\ref{collinear_limit_11}), as we now show.
Taking the limit under the integral in eq.~(\ref{collinear_limit_11}) for $\epsilon>\frac12$ we arrive at
\begin{align}
\label{collinear_limit_11_calc}
\begin{split}
&\frac12 \lim_{\alpha_{ij}\to -1+i\varepsilon} 
{\cal F}_{(1,1)}^{(1)}(\alpha_{ij})
=\kappa\Gamma(2\epsilon) 
\int_0^1 \frac{dx}{\left[(2x-1)^2-i\delta\right]^{1-\epsilon}}\\
&=
2\kappa\Gamma(2\epsilon) \left\{\frac14 (-i \delta)^{\epsilon-\frac12}
    \frac{\sqrt{\pi} \Gamma(\frac12 - \epsilon)}{\Gamma(1 - \epsilon)} 
   + \frac12 \frac{1}{2\epsilon-1}\,\,
 {}_2F_1\left(\frac12 - \epsilon, 1 - \epsilon, \frac32 - \epsilon, i \delta\right)\right\}
 \\&
 \xrightarrow[\delta \to 0^+]{} 2\kappa\Gamma(2\epsilon) \frac12 \frac{1}{2\epsilon-1}\,,
\end{split}
\end{align}
where at the first step we conveniently account for the prescription using a new small parameter $\delta>0$, in the second we evaluate the integral exactly as a function of $\delta$ and in the final stage consider the limit $\delta\to 0^+$. Of course this limit is taken with $\epsilon>\frac12$, where the first term vanishes while the hypergeometric function in the second reduces to 1.
We thus observe that the collinear limit of the (1,1) web does indeed reproduce the self-energy web result of eq.~(\ref{self_enregy_target_result}) as a function of $\epsilon$.

With this example we have reassured ourselves that the collinear limit of webs can be consistently taken also when it gives rise to boomerang webs. We have seen that a subtle situation arises in the collinear limit $\alpha_{ij}\to -1$, in which the two singularities in eq.~(\ref{x_pole}) coincide, potentially prohibiting an expansion about $\epsilon=0$ prior to integration over $x$. In eq.~(\ref{collinear_limit_11_calc}) we have overcome this by keeping the $i\delta$ prescription prior to integrating over $x$, in which case the result of the direct computation (eq.~(\ref{self_enregy_target_result})) is exactly recovered from the collinear limit.

Next we turn our attention to the question of how to obtain subtracted web results for boomerang webs from final results for (subtracted) webs with a larger number of lines by taking collinear limits. Importantly, in this case the expansion in $\epsilon$ for the originally-computed web has already been done (recall that the subtracted web is defined in eq.~(\ref{w_calF}) 
by considering the coefficient of~$\epsilon^{-1}$), violating the proper order of limits we adhered to above. We have already seen in eq.~(\ref{Coll_red_11_LO_fail}) that the self-energy diagram \emph{cannot} be recovered in this way from the (1,1) web.
Nevertheless, in the following section we will show that the limit  $\alpha_{ij}\to -1$ can actually be taken to determine boomerang webs in which (as we have seen) all boomerang gluons necessarily straddle at least one extra emission along the Wilson line. To this end one must not consider the collinear limit of individual subtracted webs, but instead identify the combination of subtracted webs forming together the boomerang web of interest in the collinear limit. Upon considering this combination the limit exists and is bound to reproduce the result of the direct computation.

%%%%%%%%%%%

\subsection{Collinear reduction into boomerang webs}
\label{coll_reduction_boomerang}

To begin, we would like to explain the qualitative difference there is between the hopeless attempt to recover the self-energy web from the $\epsilon$-expanded (1,1) web, along the lines of eq.~(\ref{Coll_red_11_LO_fail}), and the well-defined collinear-reduction procedure into boomerang webs where boomerang gluons necessarily straddle other emissions along the Wilson line. As shown above, the self-energy web requires setting $\epsilon>\frac12$ when the integration is performed, owing to the double pole in eq.~(\ref{self_enregy_target_result}), which explains why the $\alpha_{ij}\to - 1$ limit of the (1,1) web does not commute with the $\epsilon$ expansion. In contrast, as discussed in section~\ref{sec:kinboom}, and demonstrated in several examples in section~\ref{sec:calculate} (see specifically the discussion following eq.~(\ref{kerneldef}) regarding the (1,1,4) web and following eq.~(\ref{phicghi}) regarding the (5,1) web) non-self-energy boomerang webs, where each boomerang gluon straddles other emissions, have the key property described following eq.~(\ref{x_l_integral}) where the integrand features an extra suppression factor $\sim(x_l-\frac12)$, regularising the double pole at $x_l=\frac12$, and rendering the integral in eq.~(\ref{gendiagboom2}) well-defined for small positive values of $\epsilon$. In such webs then, an expansion in $\epsilon$ will be valid (provided of course logarithmic end-point singularities are properly regularised by $\epsilon>0$). It is therefore expected that such boomerang webs could be reproduced order-by-order in the $\epsilon$-expansion by considering collinear limits of non-boomerang webs, by taking the $\alpha_{ij}\to -1$ limit. Specifically, this can be done directly for the subtracted web.

Having cleared the conceptual issue, let us now show how to apply the collinear reduction procedure in practice, by considering three-line boomerang webs. 
For a given three-line web $W$, we must find webs connecting four
Wilson lines that can produce the diagrams contained in $W$ upon identifying two of the Wilson lines. Furthermore, as shown in ref.~\cite{Falcioni:2014pka}, in the language of effective connected vertices of ref.~\cite{Gardi:2013ita}, the process of merging the two lines involves symmetrisation of the order of the effective vertices which are now placed on a single line.
This implies that the specific colour components of a given web which may be recovered upon applying collinear reduction are those where at least one of the lines features two or more effective vertices.

Our first and central example is the (1,1,4) web of
figure~\ref{fig:114}. We will show that its colour component $c^{[3,3]}_3$, whose corresponding kinematic function was determined through a direct calculation in the previous section (see eq.~(\ref{F114res})), may be derived\footnote{Note that the other colour component of the (1,1,4) web, involving $c^{[3,3]}_4$, does not have two effective colour structures on any of the lines, and therefore cannot be obtained through collinear reduction.} through the collinear reduction of non-boomerang four-line MGEWs computed in ref.~\cite{Gardi:2013saa}. 
The basic observation is that the colour structure $c^{[3,3]}_3$ defined in eq.~(\ref{colbasis3}) may be expressed in two different ways using connected multiple-gluon-emission effective colour matrices as follows:
\begin{subequations}
\label{two_ways_of_writing_c3}
\begin{align}
c^{[3,3]}_3 &= -T_1^a T_2^b \left\{T_3^c,\, C_{3,2}^{ ac,b}(3)\right\}\,,
\label{two_ways_of_writing_c3_1113}\\
c^{[3,3]}_3 &= T_1^a T_2^b \left\{C_{2,1}^{be}(3) , \,C_{2,1}^{ ea}(3)\right\}\,,
\label{two_ways_of_writing_c3_1122}
\end{align}
\end{subequations}
where the double- and triple-gluon-emission matrices, defined in eqs. (\ref{V2colour}) and (\ref{V3colour}), are always placed on Wilson line $3$, inside an anticommutator.  Therefore, in total four gluons are emitted from this line, out of which one pair is contracted to form the boomerang gluon,
as relevant to the (1,1,4) web of figure~\ref{fig:114}. 

\begin{figure}
  \centering
  \scalebox{0.6}{\includegraphics{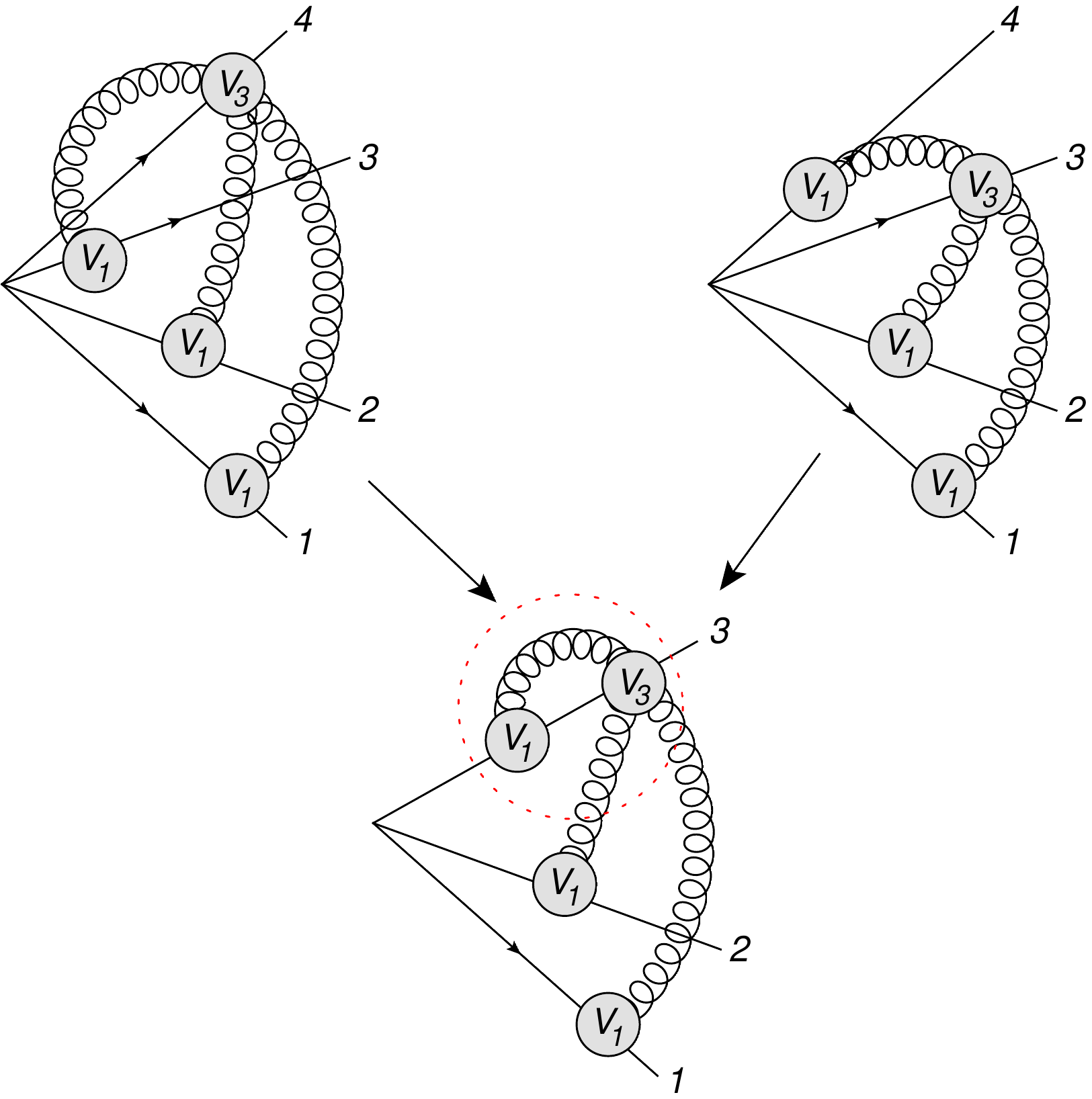}}
  \caption{Two different (1,1,1,3)-type webs
    $w_{1113}(\alpha_{14},\alpha_{24},\alpha_{34})$ (top left) and
    $w_{1113}(\alpha_{13},\alpha_{23},\alpha_{34})$ (top right), give a
    contribution to the (1,1,4) web in the limit that lines 3 and 4 are
    collinear. The red dashed circle indicates the symmetric contribution will
    be taken.}
  \label{fig:web1113to114}
\end{figure}

\begin{figure}
  \centering
  \scalebox{0.6}{\includegraphics{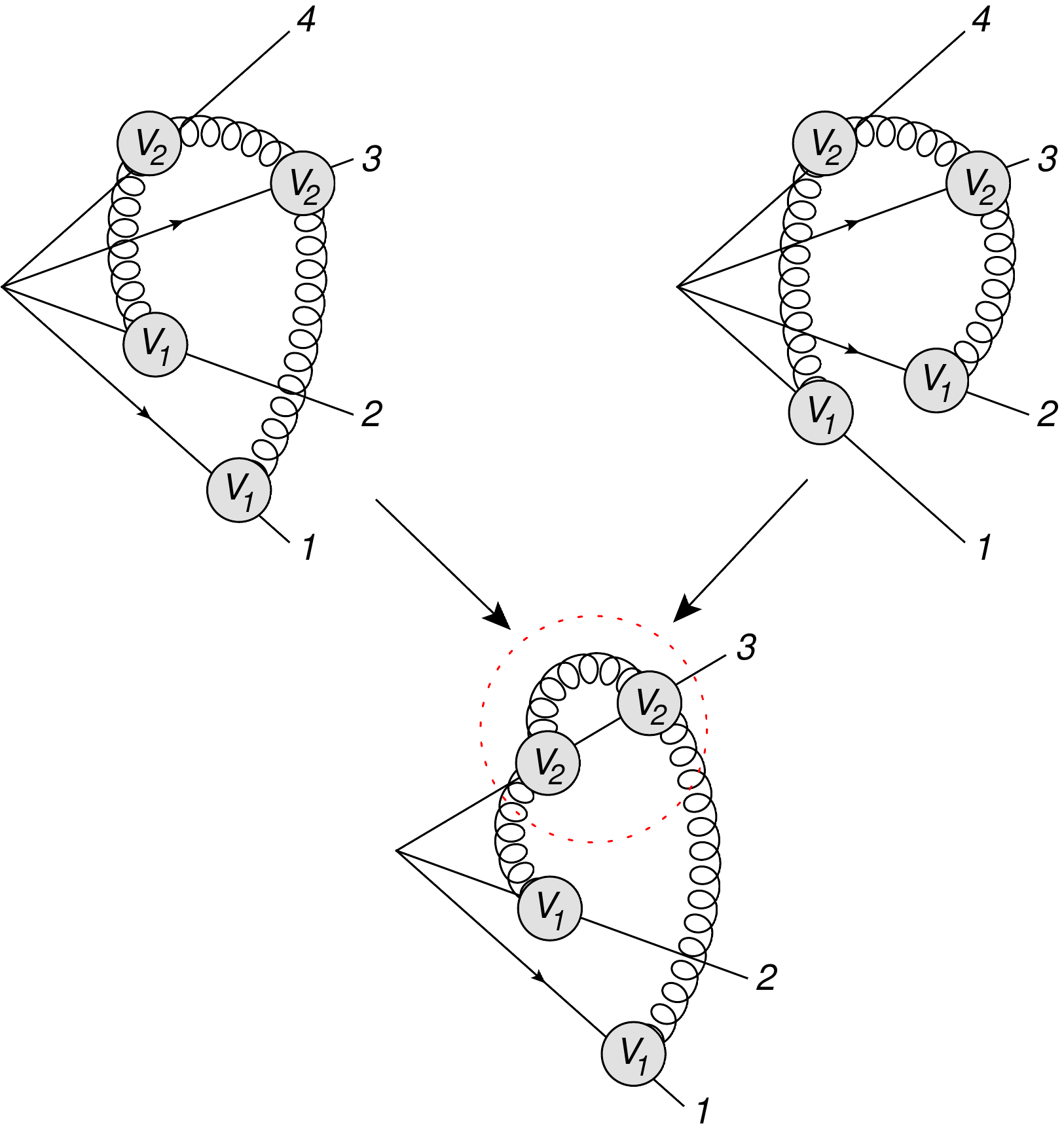}}
  \caption{Two different (1,1,2,2) webs
    $w_{1122}(\alpha_{24},\alpha_{34}, \alpha_{13})$ (top left) and
    $w_{1122}(\alpha_{14},\alpha_{34}, \alpha_{23})$ (top right), give a
    contribution to the (1,1,4) web in the limit that lines 3 and 4 are
    collinear. The red dashed circle indicates the symmetric contribution regarding the relative position of the two $V_2$  vertices along line 3 is taken.}
  \label{fig:web1122to114}
\end{figure}

To determine which four-line webs contribute to the~$c^{[3,3]}_3$ component of the (1,1,4) web upon applying collinear reduction, consider placing one of the two colour matrices appearing in the anticommutators in eq.~(\ref{two_ways_of_writing_c3}) on a fourth Wilson line instead. It is straightforward to see that the corresponding four-line web would be of the (1,1,1,3) type upon using eq.~(\ref{two_ways_of_writing_c3_1113}) and of the (1,1,2,2) type upon using eq.~(\ref{two_ways_of_writing_c3_1122}).   Furthermore, in each of these cases there are two distinct choices for the colour operator to be placed on line $4$. These considerations imply that in order to obtain the complete $c^{[3,3]}_3$ component of the (1,1,4) web upon taking 
the collinear limit $3|| 4$, one needs to sum up the four four-line webs shown in figures~\ref{fig:web1113to114} and~\ref{fig:web1122to114}:
the former presents the contributions of 
$w_{1113}(\alpha_{14},\alpha_{24},\alpha_{34})$ and
$w_{1131}(\alpha_{13},\alpha_{23},\alpha_{34})$, while the latter
depicts the two instances of the (1,1,2,2) web, namely $w_{1122}(\alpha_{24},\alpha_{34}, \alpha_{13})$ and $w_{1122}(\alpha_{14},\alpha_{34}, \alpha_{23})$, all of which contribute to $c^{[3,3]}_3$ when the lines $3$ and $4$ are identified.

We will now examine taking the limit in more detail. We first note that the collinear limit can only be consistently taken after summing up the four contributions above. Indeed, as we shall see, each of the four separate webs features singularities such as those in eq.~(\ref{F11reduce1}), owing to the fact that $r(\alpha_{ij})$ of eq.~(\ref{r_def}) diverges and the basis functions $M_{k,l,n}(\alpha_{ij})$ become complex in the limit where the would-be-boomerang gluon $\alpha_{ij}\to -1$. We will therefore postpone taking the limit $\alpha_{ij}\to-1$ until we have added all four contributions. Nonetheless, for clarity we consider the four webs in turn.

\begin{figure}[htb]
\begin{center}
\vspace*{20pt}
\scalebox{0.7}{\includegraphics{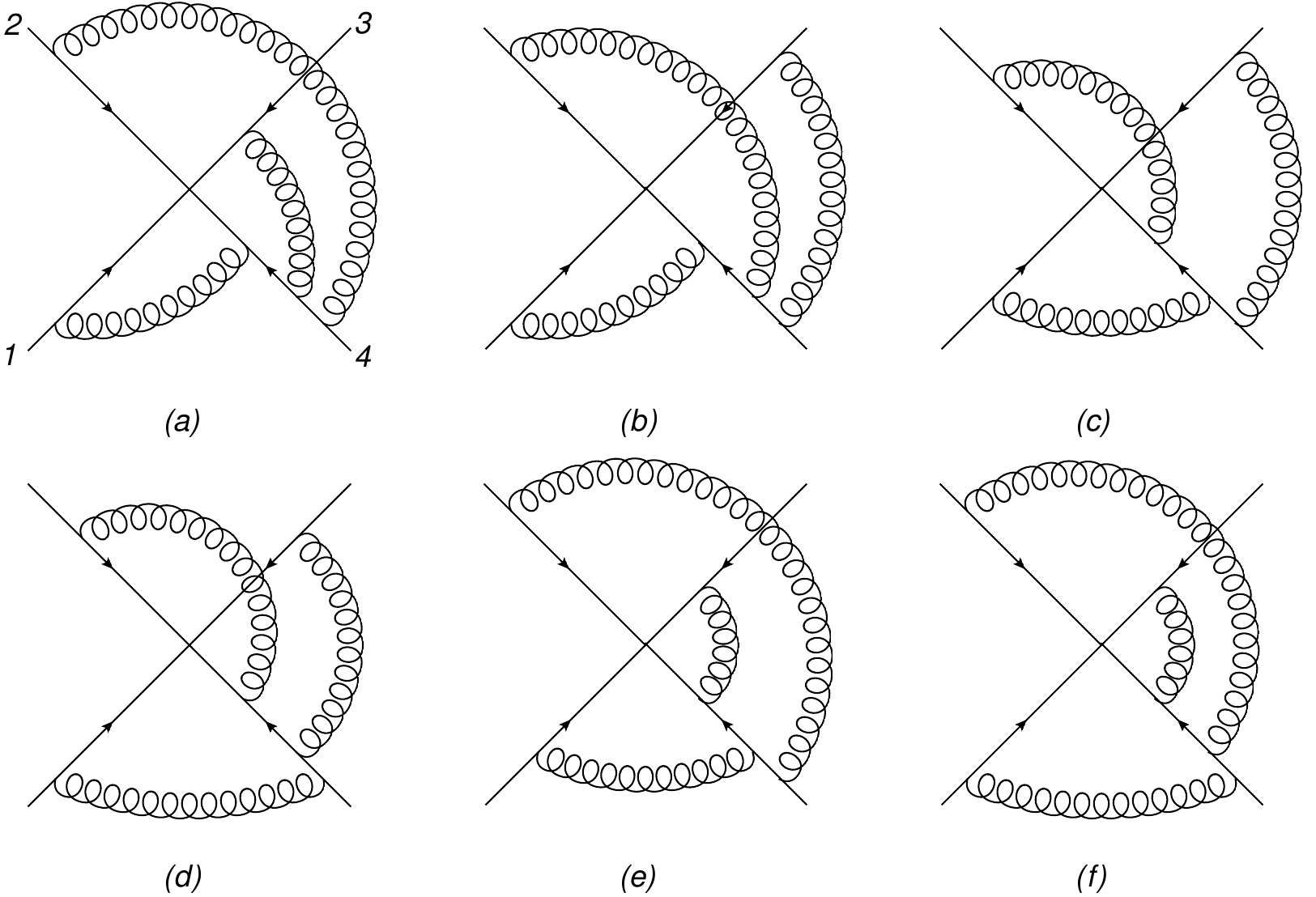}}
\caption{The (1,1,1,3) web.}
\label{fig:1113}
\end{center}
\end{figure}

Let us begin by considering the (1,1,1,3) of figure \ref{fig:1113}, which is shown in terms of effective vertices in figure~\ref{fig:web1113to114} (top left).

According to refs.~\cite{Gardi:2013saa,Falcioni:2014pka}, the integrated subtracted (1,1,1,3) web can be written as
%\begin{equation}
%\begin{split}
%\overline{w}_{(1,1,1,3)}^{(3)}(\alpha_{14}, \alpha_{24}, \alpha_{34}) =& -\frac{1}{6}T_1^a T_2^b T_3^c T_4^d \left( %\frac{1}{4\pi} \right) ^3 r(\alpha_{14}) r(\alpha_{24}) r(\alpha_{34}) \\
%& \times \left \{ f^{ade} f^{bce} G_{(1,1,1,3)}^{(3)}(\alpha_{14}, \alpha_{24}, \alpha_{34}) + f^{abe} f^{cde}  %G_{(1,1,1,3)}^{(3)}(\alpha_{24}, \alpha_{14}, \alpha_{34}) \right \},
%\end{split}
%\end{equation}
%where
%\beqa
%  && \hspace{-1cm} G_{(1,1,1,3)} \left( \alpha_{14}, \alpha_{24}, \alpha_{34} \right) \, = \,
%  \frac12 \, M_{2,0,0} (\alpha_{14}) M_{0,0,0} (\alpha_{24}) M_{0,0,0} (\alpha_{34})
%  \label{1113} \\ &&
%  + \, \frac12 \, M_{2,0,0} (\alpha_{34}) M_{0,0,0} (\alpha_{14}) M_{0,0,0} (\alpha_{24})
%  -  \, M_{2,0,0} (\alpha_{24}) M_{0,0,0} (\alpha_{14}) M_{0,0,0} (\alpha_{34})
%  \nonumber \\ &&
%  + \, M_{0,0,0} (\alpha_{14}) M_{1,0,0} (\alpha_{24}) M_{1,0,0} (\alpha_{34})
%  + \, M_{0,0,0} (\alpha_{34}) M_{1,0,0} (\alpha_{14}) M_{1,0,0} (\alpha_{24})
%  \nonumber \\ &&
%  - \, 2 \, M_{0,0,0} (\alpha_{24}) M_{1,0,0} (\alpha_{14}) M_{1,0,0} (\alpha_{34}) \, .
%  \nonumber
%\eeqa
\beqa
\label{1113gen}
  \overline{w}^{(3,-1)}_{(1,1,1,3)} \left( \alpha_{14}, \alpha_{24}, \alpha_{34} \right)
  & = & - \frac16 \, T_1^a T_2^b T_3^c T_4^d  \, \left( \frac{1}{4 \pi} \right)^3
  r(\alpha_{14}) \, r(\alpha_{24}) \, r(\alpha_{34}) \nonumber \\
  && \hspace{-4cm} \times \, \Big[ f^{ade} f^{ebc} \, G_{(1,1,1,3)} 
  \left( \alpha_{14}, \alpha_{24}, \alpha_{34} \right) + f^{ace} f^{ebd} \,
  G_{(1,1,1,3)} \left( \alpha_{24}, \alpha_{14}, \alpha_{34} \right) \Big] \, ,
\eeqa
where
\begin{align}
 \label{1113}
 \begin{split}
  & \hspace{-1cm} G_{(1,1,1,3)} \left( a_1, a_2, a_3 \right) \, = \, 
  \frac12 \, M_{2,0,0} (a_1) M_{0,0,0} (a_2) M_{0,0,0} (a_3)
  \\ &
  + \, \frac12 \, M_{2,0,0} (a_3) M_{0,0,0} (a_1) M_{0,0,0} (a_2)
  -  \, M_{2,0,0} (a_2) M_{0,0,0} (a_1) M_{0,0,0} (a_3)
   \\ &
  + \, M_{0,0,0} (a_1) M_{1,0,0} (a_2) M_{1,0,0} (a_3)
  + \, M_{0,0,0} (a_3) M_{1,0,0} (a_1) M_{1,0,0} (a_2)
  \\ &
  - \, 2 \, M_{0,0,0} (a_2) M_{1,0,0} (a_1) M_{1,0,0} (a_3) \, .
  \end{split}
\end{align}
We will now take the collinear limit, beginning with the colour factors.  We identify line 4 with line 3 and symmetrise:
\begin{equation}
    \begin{split}
    T_1^a T_2^b T_3^c T_4^d f^{ade} f^{ebc} \underset{3 || 4}{\longrightarrow} T_1^a T_2^b \frac{1}{2} \{T_3^c , T_3^d\}f^{ade}f^{ebc} = \frac{1}{2} [T_1^a,T_1^b] [T_2^b,T_2^c]\{T_3^a , T_3^c\} = \frac{1}{2} c_3^{[3,3]}, \\
    T_1^a T_2^b T_3^c T_4^d f^{ace} f^{ebd} \underset{3 || 4}{\longrightarrow} T_1^a T_2^b \frac{1}{2} \{T_3^c , T_3^d\}f^{ace}f^{ebd} = \frac{1}{2} [T_1^a,T_1^b] [T_2^b,T_2^c]\{T_3^a , T_3^c\} = \frac{1}{2} c_3^{[3,3]}.
    \end{split}
    \label{eq:col1113}
\end{equation}
Identifying lines 3 and 4 in the $\alpha$-variables and defining $\alpha=\alpha_{33}$, we find 
\begin{equation}
\begin{split}
\overline{w}_{(1,1,1,3)}^{(3,-1)}(\alpha_{14}, \alpha_{24}, \alpha_{34}) \underset{3 || 4}{\longrightarrow} &   - \frac{1}{12} c_3^{[3,3]}\left( \frac{1}{4\pi} \right)^3 r(\alpha_{13}) r(\alpha_{23}) r(\alpha) \\
& \times \Big[ G_{(1,1,1,3)} 
  \left( \alpha_{13}, \alpha_{23}, \alpha \right) +  G_{(1,1,1,3)} \left( \alpha_{23}, \alpha_{13}, \alpha \right) \Big]  \,,
\label{w1113reduce}
\end{split}
\end{equation}
where it is understood that the limit $\alpha\to -1$ will be taken once all the contributions are collected.

As shown in figure~\ref{fig:web1113to114} (top right), we must also consider the (1,1,3,1)
web, where three gluons are emitted form line 3. We may readily write the result as $\overline{w}_{(1,1,1,3)}(\alpha_{13},\alpha_{23},\alpha_{34})$ as it corresponds to the (1,1,1,3) web considered above with lines 3 and 4 swapped. Repeating the steps above yields
\begin{equation}
\begin{split}
\overline{w}_{(1,1,3,1)}^{(3,-1)}(\alpha_{13}, \alpha_{23}, \alpha_{34}) \underset{3 || 4}{\longrightarrow} &   - \frac{1}{12} c_3^{[3,3]}\left( \frac{1}{4\pi} \right)^3 r(\alpha_{13}) r(\alpha_{23}) r(\alpha) \\
& \times \Big[ G_{(1,1,1,3)} 
  \left( \alpha_{13}, \alpha_{23}, \alpha \right) +  G_{(1,1,1,3)} \left( \alpha_{23}, \alpha_{13}, \alpha \right) \Big],  
\label{w1131reduce}
\end{split}
\end{equation}
which is the same contribution as eq.~(\ref{w1113reduce}).

\begin{figure}
\begin{center}
\scalebox{0.9}{\includegraphics{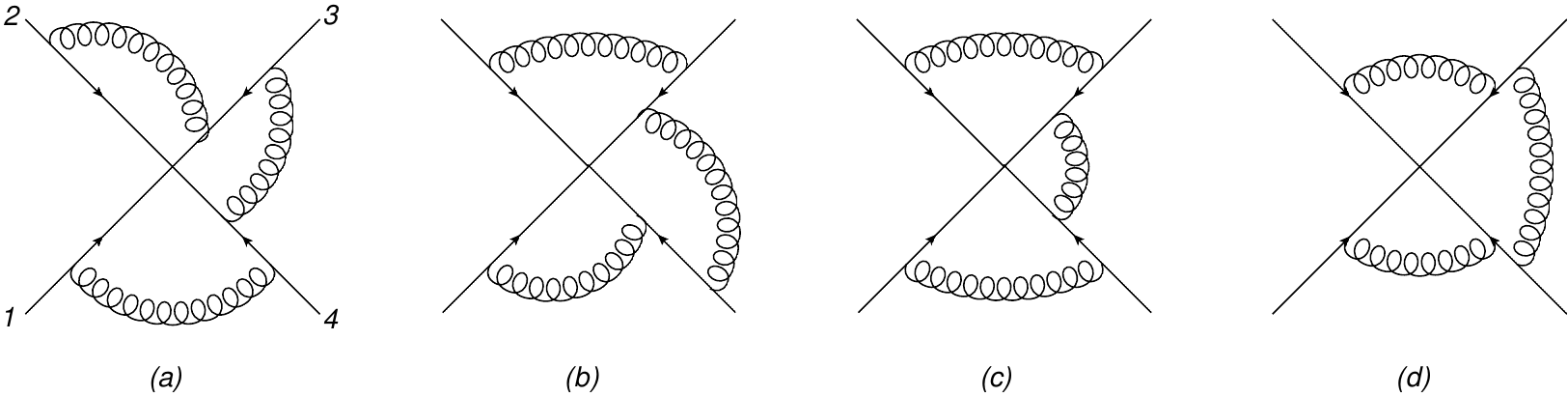}}
\caption{The (1,1,2,2) web.}
\label{fig:1122}
\end{center}
\end{figure}
Next, we must consider the (1,1,2,2) web of figure~\ref{fig:1122}, shown with
effective vertices in figure~\ref{fig:web1122to114}, top-left. The
subtracted web is given by a permutation of the (1,2,2,1) web~\cite{Gardi:2013saa,Falcioni:2014pka}:
\beqa
\label{1221gen}
  \overline{w}^{(3,-1)}_{(1,1,2,2)} \left( \alpha_{14}, \alpha_{34}, \alpha_{23} \right) & = &
   \frac16  \, f^{ade} f^{ebc} T_1^a T_2^b T_3^c T_4^d \,\,
  \left( \frac{1}{4 \pi} \right)^3 r(\alpha_{12}) \, r(\alpha_{23}) \, r(\alpha_{34})
  \nonumber \\
  && \hspace{2cm} \times \,
  G_{(1,1,2,2)} \left( \alpha_{14}, \alpha_{34}, \alpha_{23} 
  \right) \, ,
\eeqa
where
\begin{align}
\label{1221genG}
\begin{split}
  & \hspace{-1cm} G_{(1,1,2,2)} \left( a_1, a_2, a_3 \right) \, = \,  
  - \, \frac12 \, M_{2,0,0} (a_1)  M_{0,0,0} (a_2)  M_{0,0,0} (a_3) 
   \\  & 
  - \, \frac12 \, M_{2,0,0} (a_3) M_{0,0,0} (a_1) M_{0,0,0} (a_2)
  + M_{2,0,0} (a_2) M_{0,0,0}(a_1) M_{0,0,0}(a_3) 
   \\ &
  - \, M_{0,0,0} (a_1) M_{1,0,0} (a_2) M_{1,0,0} (a_3)
  - M_{0,0,0} (a_3) M_{1,0,0} (a_1) M_{1,0,0} (a_2) 
   \\ &
  + \, 2 \, M_{0,0,0} (a_2) M_{1,0,0} (a_1) M_{1,0,0} (a_3)
  - 4 \, M_{0,2,0} (a_2) M_{0,0,0} (a_1) M_{0,0,0} (a_3) \, . 
\end{split}
\end{align}
Note the second argument in this function, $a_2$, corresponds to the $\alpha$ between the two lines with two gluons attached; this is the gluon which is due to become a boomerang.  The result is symmetric between $a_1$ and $a_3$.
Using eq.~(\ref{eq:col1113}), we find
\begin{equation}
\begin{split}
\overline{w}_{(1,1,2,2)}^{(3,-1)}(\alpha_{14}, \alpha_{34}, \alpha_{23})
 \underset{3 || 4}{\longrightarrow}  &\,\, \frac{1}{12} c_3^{[3,3]} \left( \frac{1}{4\pi} \right)^3 r(\alpha_{13}) r(\alpha_{23}) r(\alpha) G_{(1,1,2,2)}\left( \alpha_{13}, \alpha, \alpha_{23} 
  \right)   .
\end{split}
\label{eq:w1122reduce}
\end{equation}
The second (1,1,2,2) web is a different permutation of the (1,2,2,1) web in refs.~\cite{Gardi:2013saa,Falcioni:2014pka} which corresponds to
$\overline{w}_{1122}(\alpha_{13},\alpha_{34},\alpha_{24})$ (figure~\ref{fig:web1122to114}). Repeating the steps above gives
\begin{equation}
\begin{split}
\overline{w}_{(1,1,2,2)}^{(3,-1)}(\alpha_{13}, \alpha_{34}, \alpha_{24})
 \underset{3 || 4}{\longrightarrow}  & \,\,\frac{1}{12} c_3^{[3,3]} \left( \frac{1}{4\pi} \right)^3 r(\alpha_{13}) r(\alpha_{23}) r(\alpha) G_{(1,1,2,2)}\left( \alpha_{13}, \alpha, \alpha_{23} 
  \right).
\end{split}
\label{eq:w1122sreduce}
\end{equation}

Finally, the (1,1,4) web should be obtained by 1/2 the sum of the four contributions in eqs.~(\ref{w1113reduce}), (\ref{w1131reduce}), (\ref{eq:w1122reduce}) and (\ref{eq:w1122sreduce}). The factor of 1/2 is due following step 4 in the rules at the start of this section, reflecting the fact that the (1,1,4) web has higher symmetry than the original webs, since the ends of the boomerang gluon now attach to the same Wilson line. The collinear reduction procedure thus gives
\begin{equation}
\label{11131122_114_limit}
    \begin{split}
 \overline{w}^{(3,-1)}_{(1,1,4);3}=\, & \frac12\, \lim_{3||4} \Bigg\{\overline{w}^{(3,-1)}_{(1,1,1,3)} \left( \alpha_{14}, \alpha_{24}, \alpha_{34} \right) 
 +\overline{w}_{(1,1,3,1)}^{(3,-1)}(\alpha_{13}, \alpha_{23}, \alpha_{34}) 
\\&\hspace*{50pt}+\overline{w}_{(1,1,2,2)}^{(3,-1)}(\alpha_{14}, \alpha_{34}, \alpha_{23})
+\overline{w}_{(1,1,2,2)}^{(3,-1)}(\alpha_{13}, \alpha_{34}, \alpha_{24})\Bigg\}
\\ =\, &  \frac12\,\lim_{\alpha\to -1}\Bigg\{
         -\frac{1}{6} c_3^{[3,3]}  \left( \frac{1}{4\pi} \right)^3 r(\alpha_{13}) r(\alpha_{23}) r(\alpha) \\
        & \qquad\qquad \times \Big[ G_{(1,1,1,3)}\left(\alpha_{13},\alpha_{23},\alpha\right) + G_{(1,1,1,3)}\left(\alpha_{23},\alpha_{13},\alpha\right) - G_{(1,1,2,2)}\left(\alpha_{13},\alpha,\alpha_{23}\right) \Big]\Bigg\} \\
        =\,&  \frac12\, \lim_{\alpha\to -1}\left\{-\frac{2}{3} c_3^{[3,3]}  \left( \frac{1}{4\pi} \right)^3 r(\alpha_{13}) r(\alpha_{23}) r(\alpha) M_{0,0,0}(\alpha_{13}) M_{000}(\alpha_{23}) M_{0,2,0}(\alpha)\right\}
        \\
        =\,& -c_3^{[3,3]}  \left( \frac{1}{4\pi} \right)^3 \frac{4\pi^2}{9} r(\alpha_{13}) r(\alpha_{23}) M_{0,0,0}(\alpha_{13}) M_{0,0,0}(\alpha_{23})\,,
    \end{split}
\end{equation}
where in the second step we inserted the expressions for the kinematic functions $G_{(1,1,1,3)}$ and $G_{(1,1,2,2)}$ from eqs.~(\ref{1113}) and (\ref{1221genG}) in terms of basis functions, observing a remarkable cancellation of all terms except for the one containing $M_{0,2,0}(\alpha)$, and in the last step we have taken the limit $\alpha\to -1$, using
\begin{equation}
   \lim_{\alpha\to -1} r(\alpha) \, M_{0,2,0}(\alpha) 
   = \lim_{\alpha\to -1} \frac{1+\alpha^2}{1-\alpha^2}\left(
   \frac23 \log^3(\alpha) + 4 \zeta_2\ln(\alpha)\right) = \frac43\pi^2\,.
\end{equation}
Note that the two terms in the brackets conspire to cancel the singularity, yielding together a finite, real limit. 

In summary we observe here multiple cancellations of potential singularities emanating from the pole of $r(\alpha)$, which is realised in the sum of webs in eq.~(\ref{11131122_114_limit}) both through the cancellation of a host of terms containing the basis functions $M_{0,0,0}(\alpha)$, $M_{1,0,0}(\alpha)$ and $M_{2,0,0}(\alpha)$ and through the finite limit of the remaining $r(\alpha)  M_{0,2,0}(\alpha)$ term. The final result in eq.~(\ref{11131122_114_limit}) exactly matches the $c_3^{[3,3]}$
contribution to the (1,1,4) web of eq.~(\ref{F114res}) obtained by a direct computation (see appendix~\ref{app:114calc}, with the result given in eq.~(\ref{F1143calc2}) there). 

As already discussed above, it is not possible to reproduce the fully
antisymmetric colour factor $c_4^{[3,3]}$ by collinear reduction. Thus,
this contribution to the (1,1,4) web cannot be checked by this
procedure. Furthermore, the (1,2,3) boomerang web -- which only has a
$c_4^{[3,3]}$ term -- is also unobtainable.

Next, let us consider the two-line webs of section~\ref{sec:2lines}. Of the three boomerang-web examples considered there, the (3,3), (5,1) and (2,4), only the last has a component that is accessible through collinear reduction, namely its contribution involving the $c_2^{[3,2]}$ colour factor of eq.~(\ref{colbasis2}), which involves anticommutators. All other contributions of these two-line webs are proportional to $c_1^{[3,2]}$  and are thus unobtainable through collinear reduction. 

The $c_2^{[3,2]}$ contribution to the (2,4) web was computed directly in eq.~(\ref{F242}). Using the notation of eq.~(\ref{wbar_calF_relation}) and the basis functions in appendix~\ref{app:functions}, the result can be expressed as:
\begin{equation}
\label{24_direct}
 \overline{w}^{(3,-1)}_{(2,4);2}(\alpha_{13})=\,
 \left(\frac{1}{4\pi}\right)^3\, c_2^{[3,2]} F^{(3)}_{(2,4);2}(\alpha_{13}) \,
= -\left(\frac{1}{4\pi}\right)^3\, c_2^{[3,2]} \,\frac{4}{9}\pi^2 
\Big(r(\alpha_{13}) M_{0,0,0}(\alpha_{13})\Big)^2\,,
\end{equation}
where the colour factor is defined in eq.~(\ref{colbasis2}) and for later convenience we renamed the Wilson lines as 1 and 3.

Let us now compare this to the collinear reduction of the (1,1,4) web, i.e. starting with the final result in eq.~(\ref{11131122_114_limit}) and considering the limit $1 || 2$. 
Let us first examine the colour factor $c_3^{[3,3]}$ defined in eq.~(\ref{colbasis3}).
We find that upon merging the lines 1 and 2, it becomes 
\begin{equation}
\label{114Colour_to_24}
   c_3^{[3,3]}  = - f^{cbe} f^{ade} T_1^a T_2^b    T_3^{\{c,d\}}\,\to\,  - f^{cbe} f^{ade} \frac12 T_1^{\{a,b\}}   
   T_3^{\{c,d\}} =  2 c_2^{[3,2]} \,,
\end{equation}
where in the final step we used the definition in eq.~(\ref{colbasis2}).
Upon applying collinear reduction to the (1,1,4) web of eq.~(\ref{11131122_114_limit}) we must  account for the extra symmetry of the (2,4) web due to the fact that the two non-boomerang gluons connect the same two Wilson lines (but distinct lines in the original (1,1,4) web) by including an extra factor $1/2$. We thus obtain, 
\begin{align}
\label{24_limit}
\begin{split}
  \overline{w}^{(3,-1)}_{(2,4);2}(\alpha_{13})
  \,&=\, 
  \frac12 \lim_{1||2}
  \overline{w}^{(3,-1)}_{(1,1,4);3}(\alpha_{13},\alpha_{23})\\
        \,&=\, \frac12 \lim_{1||2}\Bigg\{
        -c_3^{[3,3]}  \left( \frac{1}{4\pi} \right)^3 \frac{4\pi^2}{9} r(\alpha_{13}) r(\alpha_{23}) M_{0,0,0}(\alpha_{13}) M_{0,0,0}(\alpha_{23})\Bigg\}\,
        \\
        \,&=\, 
        -c_2^{[3,2]}  \left( \frac{1}{4\pi} \right)^3 \frac{4\pi^2}{9} \Big( r(\alpha_{13})  M_{0,0,0}(\alpha_{13}) \Big)^2\,,
        \end{split}
\end{align}
where in the final step we just used the reduced colour structure from eq.~(\ref{114Colour_to_24}) and identified $\alpha_{23}=\alpha_{13}$. In this case, no $\alpha$ variable tends to $-1$, as no new boomerang gluon is generated, so the limit is straightforward to take.  Evidently eq.~(\ref{24_limit}) agrees with the direct computation in eq.~(\ref{24_direct}), thus providing an additional check of the computations. 
\begin{figure}[h]
\begin{center}
\scalebox{0.8}{\includegraphics{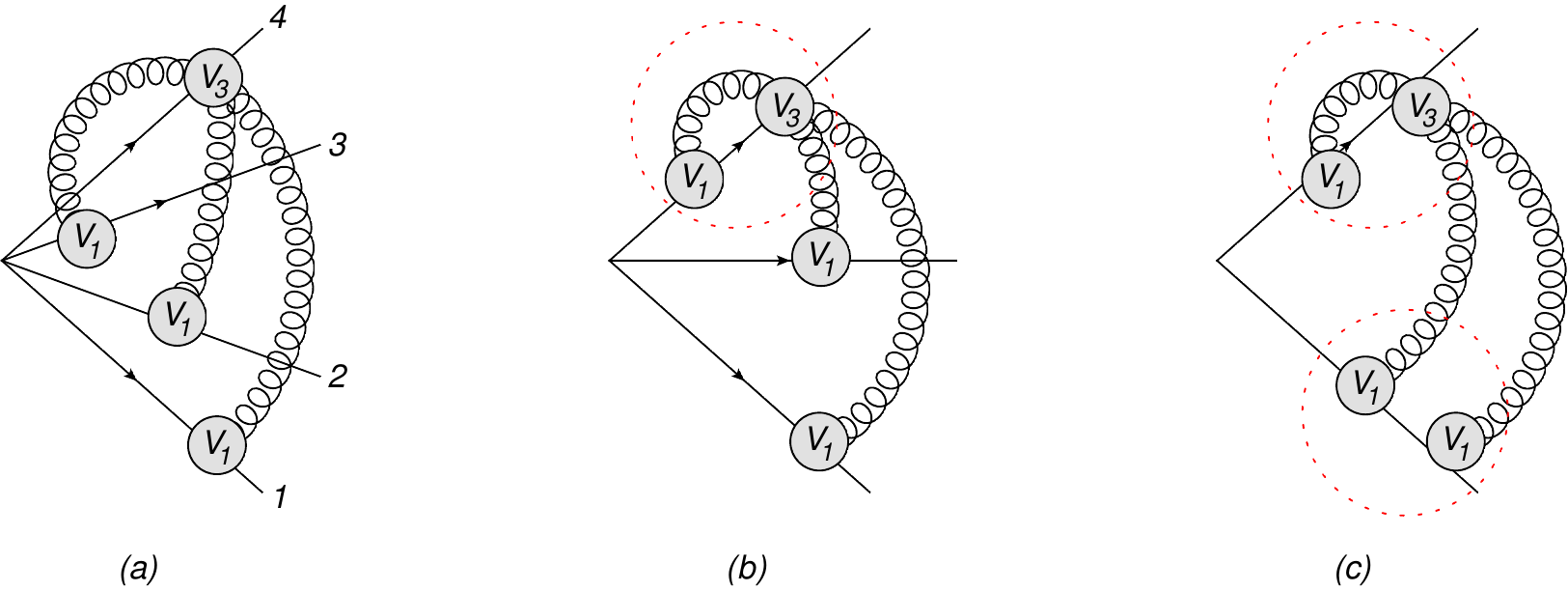}}\\
\scalebox{0.8}{\includegraphics{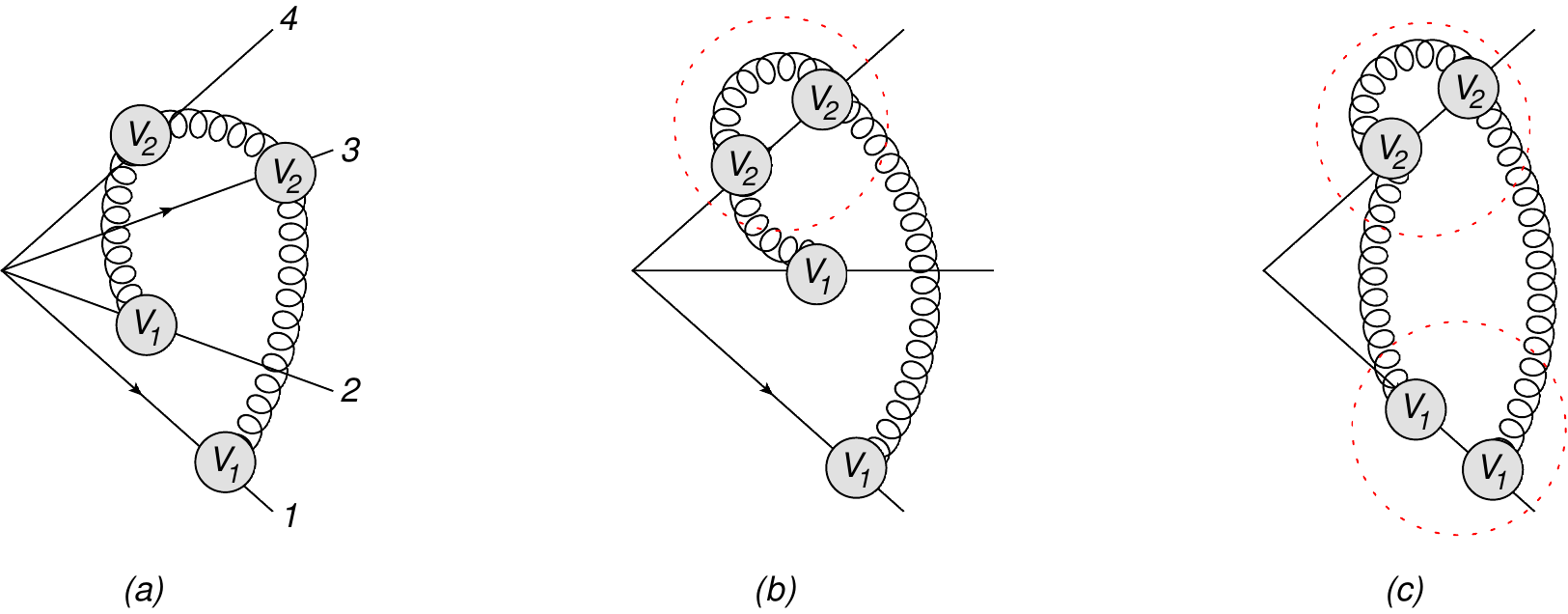}}
\caption{Upper diagrams:
(a) Effective vertex diagram for the (1,1,1,3) web; (b)
  Contribution to the (1,1,4) web obtained from collinear reduction of
  (a); (c) Contribution to the (2,4) web obtained from collinear
  reduction of (b). The red circle denotes symmetrisation of
  vertices.
  Lower diagrams: The same starting with the 1122 web.}
\label{fig:vertex1}
\end{center}
\end{figure}

To conclude this section it is useful to take another look at the collinear reduction process we have just completed, in which two consecutive collinear limits have been applied, first getting the $c_3^{[3,3]}$ component of the (1,1,4) web by taking $3||4$, and then getting the $c_2^{[3,2]}$ component of the (2,4) web by further taking $1||2$.  This process is summarised diagrammatically in figure~\ref{fig:vertex1} using the effective vertex formalism. The basic feature of this example of collinear reduction, is what is described by eq.~(\ref{two_ways_of_writing_c3}), namely that the very same colour structure of three-line webs emerges upon merging different configurations of effective vertices. Of course  the same follows for the two-line colour structure $c^{[3,2]}_2$, which can be expressed in two different ways as
\begin{subequations}
\label{two_ways_of_writing_c2}
\begin{align}
c^{[3,2]}_2 &= -\frac14 \left\{T_1^a, T_1^b\right\} \left\{T_3^c,\, C_{3,2}^{ ac,b}(3)\right\}\,,
\label{two_ways_of_writing_c2_1113}\\
c^{[3,2]}_2 &= \frac14 \left\{T_1^a, T_1^b\right\} \left\{C_{2,1}^{be}(3) , \,C_{2,1}^{ ea}(3)\right\}\,.
\label{two_ways_of_writing_c2_1122}
\end{align}
\end{subequations}
This interesting example therefore illustrates both the over-completeness of the colour basis of eq.~(\ref{colbasis}) when used to express boomerang webs and the non-trivial structure of the relations between webs spanning a different number of Wilson lines.

\section{Discussion}
\label{sec:discuss}

In this paper, we have have taken another step towards the calculation of the
multiparton soft anomalous dimension for massive Wilson lines at three-loop
order, continuing the programme of work developed in
refs.~\cite{Gardi:2010rn,Gardi:2011yz,Gardi:2013ita,Gardi:2013saa,Falcioni:2014pka}.
In our approach, the
logarithm of the soft function is calculated directly in terms of
Feynman diagrams known as {\it (multiparton) webs}. In this work we
have focused specifically on what we call {\it boomerang webs},
containing multiple gluon exchanges where at least one gluon has both its endpoints on the
same Wilson line. As in previous work on Multiple Gluon Expchange Webs (MGEWs), we set up the calculation in configuration space, and introduced a suitable exponential infrared regulator in order to isolate the ultraviolet divergences associated with the vertex where the Wilson lines meet. The latter are evaluated using dimensional regularization.
We have classified all boomerang webs through three-loop order, addressed new aspects of regularization and renormalization that arise in this class of webs and provided explicit results for their contributions to the soft anomalous dimension.

Beyond the significance of the results as components of the three-loop massive soft anomalous dimension, our study highlights several interesting aspects of webs and multi-loop computations.
First, we find that self-energy diagrams, including arbitrary clusters of such, completely decouple from other boomerang webs. 
Boomerang webs which span two or more Wilson lines, include a priori both self-energy subdiagrams and diagrams where the boomerang gluons straddle one or more emission vertices connecting to other Wilson lines. However, we have proven in full generality (section~\ref{sec:selfenergy}) that only the latter type contribute. That is, all web diagrams containing self-energy subdiagrams have a \emph{vanishing} exponentiated colour factor. 
This clearly simplifies the computation of such webs, since only a subset of the diagrams need to be computed. Furthermore, the boomerang gluons in these diagrams do not require any infrared regulator: their regularization is guaranteed by that associated with the non-boomerang gluons which they straddle. This is shown in section~\ref{sec:kinboom} and illustrated in a variety of examples in section~\ref{sec:calculate}.
 
Our proof for the complete decoupling of self-energy diagrams from multi-line boomerang webs utilises the \emph{replica trick} and relies on
the combinatorial properties of the web mixing matrix, as well as the colour algebra (namely that self-energy-type subdiagrams are diagonal in colour space). 
These general observations suggest that there may be further interesting insights
to be gained about the structure of web mixing matrices, and also
about what happens when one combines their combinatorial properties
with the known colour algebra of a non-Abelian gauge theory.
Recently, the calculation of these matrices was systematically extended to the four-loop order~\cite{Agarwal:2020nyc,Agarwal:2021him}, so there is clear scope for further progress in this area.

A significant part of our study here has been dedicated to the properties of the kinematic functions arising in boomerang webs. Our findings are consistent with the 
conjecture~\cite{Falcioni:2014pka,Gardi:2013saa} that MGEWs spanning several Wilson lines are expressible as sums of products of harmonic polylogarithms of individual cusp angles $\alpha_{ij}$, multiplied by a unique rational function $r(\alpha_{ij})=\frac{1+\alpha_{ij}^2}{1-\alpha_{ij}^2}$ for every gluon exchange between the lines $i$ and $j$. As reviewed in section~\ref{sec:MGEWkin}, the appearance of these functions can be most easily understood in configuration space, where a suitable choice of variables leads to factorization of the gluon propagator, and the integral can be recast in a $d\log$ form. This was first noted in the context of the angle-dependent cusp anomalous dimension~\cite{Henn:2013wfa}, and then conjectured to apply to any MGEW~\cite{Falcioni:2014pka,Gardi:2013saa}. 
The fact that the dependence on several kinematic variables does not lead to new types of singularities is a highly non-trivial feature, which is special to MGEWs.
Moreover, as reviewed here in section~\ref{sec:basisfunctions}, MGEWs were conjectured to be expressible in terms of a restricted class of harmonic polylogarithms, defined (see eq.~(\ref{eq:Mbasis})) through a single integral over a product of three types of logarithms~\cite{Falcioni:2014pka}. 
In this paper we extended the class of webs for which this conjecture was tested. 
We found that while the main properties still hold, namely boomerang webs are still expressible\footnote{We emphasise that the applicability of the basis of functions of~\cite{Falcioni:2014pka} to boomerang webs is not a priori obvious. Indeed, there are instances, such as the (2,4) web, where the form of the integral is rather different to eq.~(\ref{eq:Mbasis}), and yet the final result can be recast in terms of such functions.} as sums of products of the same basis of functions, one salient feature is lost, namely boomerang webs are no more functions of uniform, maximal transcendental weight. Instead, they always display a weight-drop of at least one unit for every boomerang gluon (with no rational factor) and furthermore, often feature mixed weight,  as summarised in Table~\ref{tab:weights_3loop}. 

The mechanism leading to the weight-drop and the mixed weight is analysed in some detail in section~\ref{sec:kinboom}. This may be of broader interest, well beyond the context of webs, because a general understanding of transcendental weight in perturbative computations is lacking. In particular, while it has been observed that certain quantities in ${\cal N}=4$ Super-Yang-Mills (SYM) feature uniform, maximal weight, and moreover are equal to the maximal weight terms on the corresponding quantities in QCD, these properties are not general, and the underlying mathematical reasons for these relations and for the complex mixed-weight structure in QCD, remain elusive. 
Nevertheless, from the perspective of comparing results in different gauge theories it is not surprising to see that non-boomerang MGEWs have uniform maximal weight, while boomerang webs feature a weight-drop and mixed weight. Indeed, only the latter contribute to the renormalization of the coupling, thus affecting the QCD result, while not the ${\cal N}=4$ SYM one.  

Our finding that, despite the mixed weight, boomerang webs can still be expressed in terms of the same basis of functions provides further evidence that the conjecture above
holds for all MGEWs. This is interesting because it is already known that going beyond MGEWs, the basis of transcendental functions must be extended (and likewise the set of rational functions they accompany) as seen explicitly in the calculation of the angle-dependent cusp anomalous dimension in QCD at three loops~\cite{Grozin:2014hna,Grozin:2015kna}, and in QED~\cite{Bruser:2020bsh} at four loops.

In section \ref{sec:collinear}, we have generalised and applied the collinear reduction
procedure developed in ref.~\cite{Falcioni:2014pka}, motivated by the
effective vertex formalism of ref.~\cite{Gardi:2013ita}. 
First we have seen that collinear limits generating boomerang gluons involve taking the limit $\alpha_{ij}\to -1$, which may be rather subtle in dimensional regularization.
For the self-energy web, the direct computation requires $\epsilon>\frac12$, and it therefore cannot be recovered from the expanded (1,1) web. We have then seen that in boomerang webs, in which boomerang gluons straddle other emissions along the Wilson line, order-by-order treatment in $\epsilon$ is in fact possible, as all singularities are regularised by small positive values of $\epsilon$. This guarantees the validity in principle of the collinear reduction process.

Next we examined the non-trivial example of the (1,1,4) web, demonstrating that it can be recovered from  results for non-boomerang four-line webs. Specifically, we have shown that upon considering the \emph{complete set} of four-line webs (specifically, the (1,1,1,3), (1,1,3,1) webs and two instances of the (1,1,2,2) web) whose collinear limit contribute to the (1,1,4) web, the limit exists, and through a rather intricate set of cancellations, reproduces the result of the direct calculation.  This provides a strong check of our results. 

The fact that different four-line webs, with different compositions of effective colour vertices contribute together is an interesting feature, which is a reflection of the fact that the basis in eq.~(\ref{colbasis}) becomes over-complete when contracting pairs of adjoint indices of different effective colour vertices on the same line, as needed when forming boomerang webs.
While we showed that collinear reduction may be used to compute (or check) certain components of boomerang webs, we stress that this procedure does not constrain those contributions to lower-line webs involving a fully antisymmetric colour factor, such as e.g. $c_4^{[3,3]}$ in eq.~(\ref{colbasis3}) or $c_1^{[3,2]}$ in eq.~(\ref{colbasis2}), and thus cannot be generally used as a replacement of direct computations.

Work towards calculating the remaining contributions to the soft
anomalous dimension is ongoing, using a range of different techniques.  In this paper, we have completed another subset of the remaining calculations while also proving a general result on the decoupling of self-energy diagrams and deepening our understanding of the analytic structure of webs and their collinear limits.

%%%%%%%%%%%%%%%%%%%%%%%%%

\section*{Acknowledgments}

We thank Claude Duhr and Calum Milloy for useful discussions at early stages of this project. EG and CDW are
supported by the UK Science and Technology Facilities Council (STFC). JMS is
supported by a Royal Society University Research Fellowship and the ERC Starting
Grant 715049 ``QCDforfuture''.  CDW is very grateful to the Higgs Centre for
Theoretical Physics, for generous hospitality on multiple occasions.

\appendix
\addtocontents{toc}{\protect\setcounter{tocdepth}{1}}

\section{Results for lower-order webs}
\label{app:lowerorder}

%\Einan{I now added colour factors and checked sign and normalization for the three webs we quote here. I unified the notation and qrote it always in terms of $\kappa$ and a Gamma function, since this is the way it's used when combined with 3-loop webs. Note that for the $(1,1,2)$ I consider leg 3 as the one connected twice (also earlier this was so but no colour factor was quoted). This determines the sign.}

Here, we collect some useful results for one- and two-loop webs,
obtained in refs.~\cite{Gardi:2013saa,Falcioni:2014pka} using the approach described
in section~\ref{sec:review}. First, there is the single gluon exchange
web connecting lines $i$ and $j$ at one-loop order. This web has a colour factor $c^{[1,2]}=T_1 \cdot T_2$ and a kinematic factor
\begin{align}
\label{F1res}
\begin{split}
\mathcal{F}^{(1)} (\alpha_{12} , \mu^2/m^2 , \epsilon)
\,&=\,  \kappa \Gamma(2\epsilon) \int_{0}^{1} dx p_\epsilon(x, \alpha_{12})\\
\,&=\,  \kappa \Gamma(2\epsilon) 
\, \frac{r(\alpha_{12})}{\epsilon}
%\\&\qquad\,\times\,
\left(
 {}_2F_1\left(1,  2\epsilon, 1 + \epsilon, \frac{\alpha_{12}}{1 + \alpha_{12}}\right)
 -
{}_2F_1\left(1,  2\epsilon, 1 + \epsilon, \frac{1}{1 + \alpha_{12}}\right)
\right)
\\
\,&=\,  \kappa \Gamma(2\epsilon)  \,2 r(\alpha_{12}) \ln( \alpha_{12})+\,{\cal O}(\epsilon^0)\,
\\
\,&=\,
-\frac{g_s^2}{16\pi^2}\,\frac{2}{\epsilon}\,r(\alpha_{12})\, \ln(\alpha_{12}) \,+\,{\cal O}(\epsilon^0)\,,
\end{split}
\end{align}
where in the first line we used $p_\epsilon(x,\alpha)$ of eq.~(\ref{propafu2}), in the second we performed the integral keeping the exact $\epsilon$ dependence in terms of  Gauss hypergeometric functions, and expressed the rational function using $r(\alpha)$ of eq.~(\ref{r_def})
and in the third and fourth lines we expanded in~$\epsilon$ keeping only the singular term.
Higher-order terms, ${\cal O}(\epsilon^k)$ for $k\geq 0$, will be needed for the renormalization of higher-order webs. These  follow simply from
expansion of the propagator function $p_\epsilon$ under the integral in the first line of (\ref{F1res}) according to eq.~(\ref{expP}), or alternatively from the $\epsilon$ expansion of the hypergeometric functions in the second line of eq.~(\ref{F1res}).

\begin{figure}[htb]
\begin{center}
\scalebox{0.6}{\includegraphics{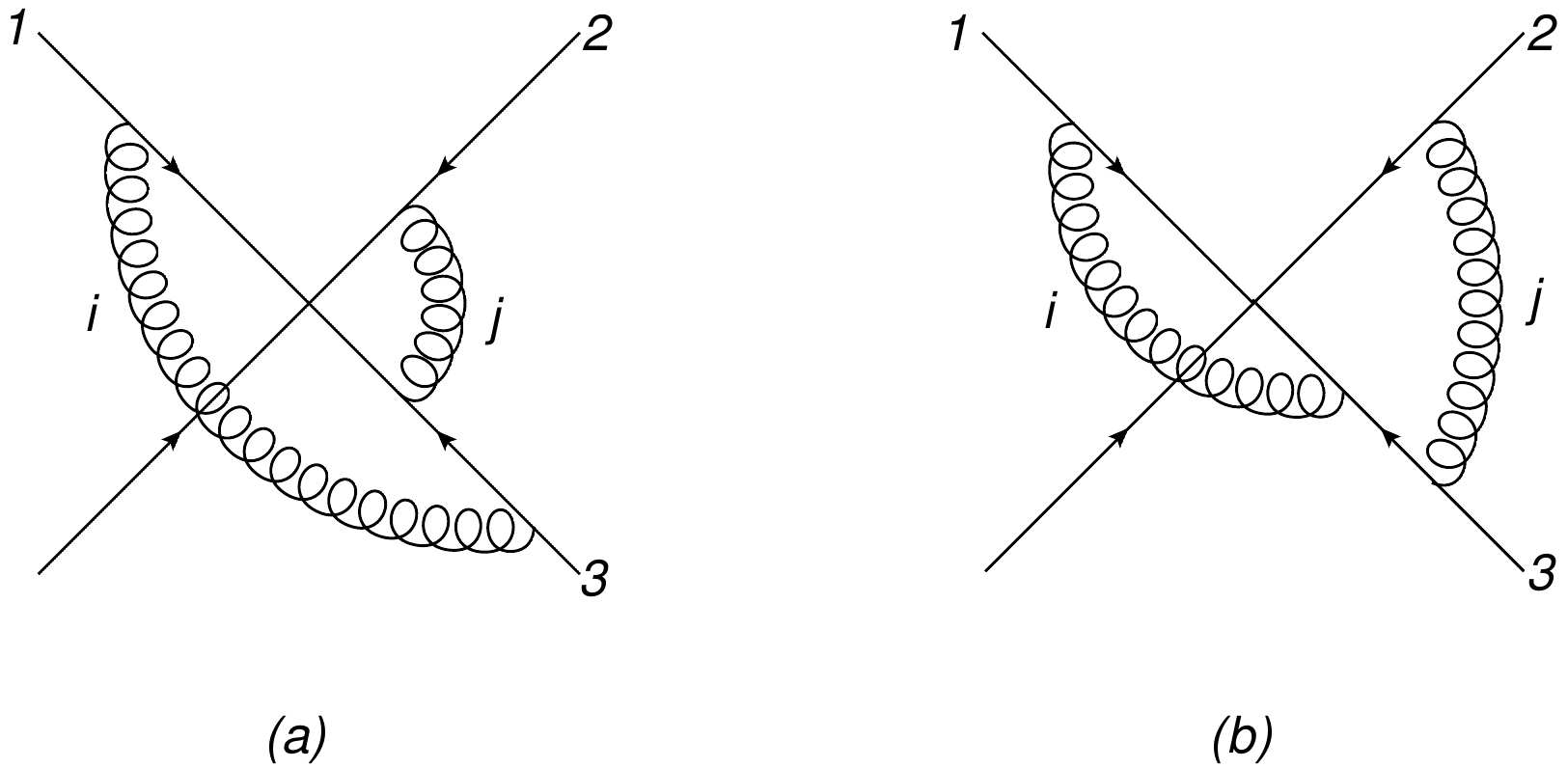}}
\caption{The (1,1,2) web.}
\label{fig:121}
\end{center}
\end{figure}

Next, consider the two-loop three-line (1,1,2) web of figure~\ref{fig:121}.
This web contributes through a single connected colour factor proportional to the structure constant $f^{abc}$. Considering a (1,1,2) web with two attachments on line $3$ and single attachments on each of the lines $1$ and $2$, the colour factor is $c^{[2,3]}=\frac12 if^{abc}T_1^aT_2^bT_3^c$ and the corresponding kinematic factor (for the non-subtracted web)  is (see ref.~\cite{Gardi:2013saa})
\begin{align}
\label{F121form}
\begin{split}
\mathcal{F}_{(1,1,2)}^{(2)}(\alpha_{13},\alpha_{23})
&= \kappa^2 \Gamma(4\epsilon) \frac{1}{2\epsilon} \int_0^1 dydz\,
  p_\epsilon(y,\alpha_{23}) p_\epsilon(z,\alpha_{13}) 
\\
& \qquad \left\{ \left( \frac{z}{y} \right)^{2\epsilon} {_2}F_1\left(4\epsilon,
  2\epsilon;1+2\epsilon; - \frac{z}{y}\right)
-\left(\frac{y}{z} \right)^{2\epsilon}
  {_2}F_1\left(4\epsilon,2\epsilon;1+2\epsilon;-\frac{y}{z}\right) \right\}
   \\
&=
2 \kappa^2 \Gamma(4\epsilon)\int_{0}^{1} dy
dz\, p_\epsilon(z, \alpha_{13}) p_\epsilon(y, \alpha_{23})\Big [ \ln
  \left( \frac{z}{y}\right) + \,\epsilon\,\Big \{ 4\text{Li}_2 \left( -\frac{z}{y}
  \right) \\
&\quad+ \ln ^2 \left( \frac{z}{y} \right) + \frac{\pi^2}{3} \Big
  \} + \mathcal{O}(\epsilon^2) \Big ].
\end{split}
\end{align}

\begin{figure}[htb]
\begin{center}
\scalebox{0.6}{\includegraphics{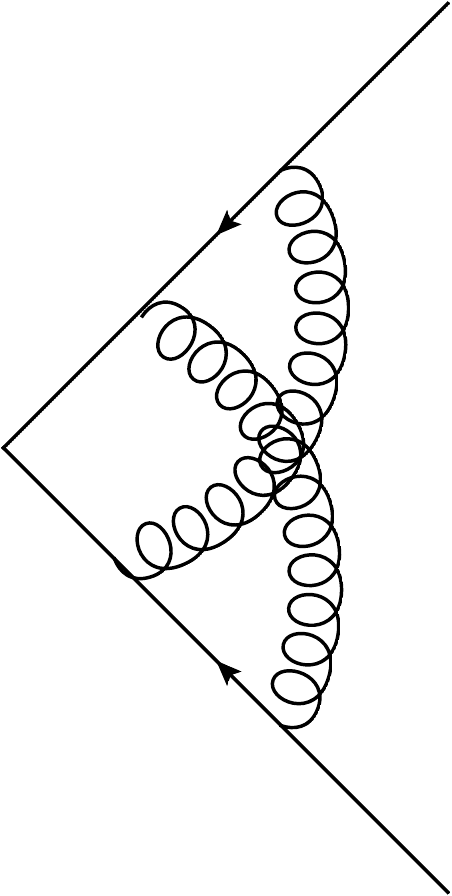}}
\caption{The (2,2) web: the crossed-gluon diagram denoted by $X$ in the text is the only diagram in the (2,2) web with a non-zero exponentiated colour factor.}
\label{fig:cross}
\end{center}
\end{figure}

Finally, let us consider the two-loop two-line web consisting of two gluons exchanged between lines $1$ and $2$. This web has two diagrams, but only one of them, with the gluons crossed (see figure~\ref{fig:cross}), has a non-vanishing exponentiated colour~\hbox{factor~$c^{[2,2]}=\frac{N_c}{2}\,T_1 \cdot T_2$.}
The corresponding kinematic factor was computed long ago~\cite{Korchemsky:1987wg}. In our  notations it is reported in eq.~(4.7) in~\cite{Falcioni:2014pka} and it reads:
\begin{equation}
\mathcal{F}^{(2,-1)}_X = \kappa^2\Gamma(4\epsilon)\int_{0}^{1}dy\,dz \, p_0(y,\alpha_{12})p_0(z,\alpha_{12}) \text{ln}\Big(\frac{1-z}{1-y}\frac{y}{z}\Big)\theta(y>z),
\label{F2-1}
\end{equation}
at ${\cal  O}(\epsilon^{-1})$  and
\begin{equation}
\begin{split}
\mathcal{F}^{(2,0)}_X &= \kappa^2\Gamma(4\epsilon)\int_{0}^{1}dy dz\,  p_0(y,\alpha_{12})p_0(z,\alpha_{12}) \Big[\text{ln}^2\Big(\frac{1-z}{1-y}\Big) - \text{ln}^2\Big(\frac{z}{y}\Big) -4\,\text{Li}_2\Big(-\frac{z}{y}\Big) \\
&+ 4\text{Li}_2\Big(-\frac{1-z}{1-y}\Big) + \text{ln}\Big(\frac{1-z}{1-y}\frac{y}{z}\Big) \text{ln}\big(q(y,\alpha_{12})q(z,\alpha_{12})\big)\Big]\theta(y>z).
\label{F20}
\end{split}
\end{equation}
and ${\cal O}(\epsilon^0)$.

\section{Basis functions and their symbols}
\label{app:functions}

In this appendix, we present explicit forms for the basis functions of
eq.~(\ref{eq:Mbasis}), together with their symbols, as taken from
ref.~\cite{Falcioni:2014pka}. We only include the functions up to weight 4, as higher-weights do not appear in the webs computed here at ${\cal O}(1/\epsilon)$. 
We refer the reader to ref.~\cite{Falcioni:2014pka} for a more complete list.

For the functions themselves, we have
\begin{itemize}

\item Weight one.
    \begin{align}
          \begin{split}
          M_{0,0,0}(\alpha) \, = \, 2 \log(\alpha) \, .
          \end{split}
    \end{align}

\item Weight two.
    \begin{align}
          \begin{split}
          M_{1,0,0}(\alpha) \, = \, 2 \, \text{Li}_2 (\alpha^2) + 4 \log(\alpha)
          \log \left(1 - \alpha^2 \right) - 2 \log^2 (\alpha) - 2 \, \zeta_2 .
          \end{split}
    \end{align}

\item Weight three.
    \begin{align}
          \begin{split}
          M_{0,0,2}(\alpha) \, = \, \frac{8}{3} \log^3(\alpha) \, ,
          \end{split}
    \end{align}
    \begin{align}
          \begin{split}
          M_{0,1,1}(\alpha) \, = \, 2 \, \text{Li}_3 (\alpha^2) - 2 \log(\alpha)
          \bigg[ \text{Li}_2 (\alpha^2) + \frac{\log^2 (\alpha)}{3}  + \zeta_2 \bigg]
          - 2 \, \zeta_3 \, ,
          \end{split}
    \end{align}
    \begin{align}
          \begin{split}
          M_{0,2,0}(\alpha) \, = \, \frac{2}{3} \log^3 (\alpha) + 4 \, \zeta_2 \,
          \log(\alpha) \, ,
          \end{split}
    \end{align}
    \begin{align}
          \begin{split}
          M_{2,0,0}(\alpha) & = - \, 4 \, \bigg[ \text{Li}_3 (\alpha^2) + 2 \text{Li}_3
          \left(1 - \alpha^2 \right) \bigg] - 8 \log \left(1 - \alpha^2 \right) \log^2(\alpha) \\
          & + \, \frac{8}{3} \, \log^3(\alpha) + 8 \, \zeta_2 \, \log(\alpha)
          + 4 \, \zeta_3 \, .
          \end{split}
    \end{align}

\item Weight four.
    \begin{align}
	  \begin{split}
	  M_{3,0,0} (\alpha) & = \, 12 \, \bigg[ \text{Li}_4 (\alpha^2) - 4 \, \text{Li}_4
	  \left(1 - \alpha^2 \right) \bigg] - 24 \, S_{2,2} (\alpha^2) \\ &
	  - \, 24 \, \log \left(1 - \alpha^2 \right) \, \text{Li}_3 (\alpha^2)
	  - 24 \, \log^2 \left(1 - \alpha^2 \right) \, \log^2 (\alpha) \\ &
	  + \, 16 \, \log \left(1 - \alpha^2 \right) \, \log^3(\alpha) - \, 4 \, \log^4 (\alpha) \\ &
	  - \, 24 \, \zeta_2 \, \log(\alpha) \,
	  \log \left[ \frac{\alpha}{\left(1 - \alpha^2 \right)^2} \right] \\ &
	  + \, 24 \, \zeta_3 \, \log \left[ \alpha \left(1 - \alpha^2 \right) \right] -
	  \, 6 \, \zeta_4 \, ,
	  \end{split}
    \end{align}
    \begin{align}
	  \begin{split} \hspace{-3mm}
	  M_{1,2,0} (\alpha) & = \, 4 \, \text{Li}_4 (\alpha^2) - 4 \, \log (\alpha) \,
	  \text{Li}_3 (\alpha^2) + 2 \, \log^2(\alpha) \, \text{Li}_2 (\alpha^2) \\ &
	  + \, \frac{4}{3} \, \log^3 (\alpha) \, \log \left(1 - \alpha^2 \right)
	  - \, \frac{2}{3} \, \log^4(\alpha) \\ &
	  + \, \zeta_2 \, \Big[ 8 \, \log(\alpha) \log \left(1 - \alpha^2 \right) +
	  4 \, \text{Li}_2 (\alpha^2) - 6 \, \log^2 (\alpha) \Big] \\ &
	  + \, 4 \, \zeta_3 \, \log(\alpha) - 14 \, \zeta_4 \, ,
	  \end{split}
    \end{align}
    \begin{align}
	  \begin{split}
	  M_{1,0,2} (\alpha) & = \, 4 \, \text{Li}_4 (\alpha^2) - 8 \, \log(\alpha) \,
	  \text{Li}_3 (\alpha^2) + 8 \, \log^2 (\alpha) \, \text{Li}_2 (\alpha^2) \\ &
	  + \frac{16}{3} \, \log^3(\alpha) \, \log \left(1 - \alpha^2 \right)
	  - \, \frac{4}{3} \, \log^4 (\alpha) - 4 \, \zeta_4 \, ,
	  \end{split}
    \end{align}
    \begin{align}
          \begin{split} \hspace{-3mm}
          M_{1,1,1} (\alpha) & = - \, 4 \, \text{Li}_4 (\alpha^2) + 4 \, S_{2,2} (\alpha^2) +
          2 \, \log \left[ \alpha \left(1 - \alpha^2 \right)^2 \right] \, \text{Li}_3 (\alpha^2) \\ &
          + \,4 \, \log (\alpha) \, \text{Li}_3 \left(1 - \alpha^2 \right) \\ &
          - \, \frac{4}{3} \, \log^2 (\alpha) \, \log \left(1 - \alpha^2 \right) \bigg[
          \log (\alpha) - 3 \, \log \left(1 - \alpha^2 \right) \bigg] \\ &
          - \, 8 \, \zeta_2 \, \log(\alpha)\log(1-\alpha^2)
          - 2 \, \zeta_3 \, \log \left[ \alpha \left(1 - \alpha^2 \right)^2 \right] + 3 \, \zeta_4 \, ,
          \end{split}
    \end{align}
\begin{align}
          M_{0,2,1} (\alpha) & = \frac23\pi^2\log^2(\alpha)+\frac23 \log^4(\alpha)\,,
 \end{align}
 \begin{align}
 \label{M003}
          M_{0,0,3} (\alpha) & = 4 \log^4(\alpha)\,.
 \end{align}
\end{itemize}
%Here we make use of harmonic polylogarithm functions, as defined in ref.~\cite{Remiddi:1999ew}. 
One can easily check that 
\begin{equation}
\label{M021Relation}
    M_{0,2,1} (\alpha) = \frac12 M_{0,0,0}(\alpha)M_{0,2,0}(\alpha)
\end{equation}
which is one example of the relations in~eq.~(\ref{con2}). The symbol of each function may be found
in table~\ref{tab:basis}.
\begin{table}[h]
\begin{center}
\begin{tabular}{|c|c|c|}
  \hline
  \multicolumn{3}{ | c | }{$M_{k, l,n} (\alpha)$} \\
  \hline
  w & Name & symbol \\
  \hline \hline
  1 & $M_{0,0,0}$ & $ 2 \, (\otimes \alpha )$ \\
  \hline \hline
  2 & $M_{1,0,0}$ & $ - 4 \,\alpha \otimes \eta$  \\
  \hline \hline
  \multirow{4}{*}{3}
  & $M_{0,0,2}$ & $ 16 \, \alpha \otimes \alpha \otimes \alpha $ \\
  \cline{2-3}
  & $M_{0,1,1}$ & $ - 4 \, \alpha \otimes \eta \otimes \alpha$  \\
  \cline{2-3}
  & $M_{0,2,0}$ & $ 4 \, \alpha \otimes \alpha \otimes \alpha$  \\
  \cline{2-3}
  & $M_{2,0,0}$ & $ 16 \, \alpha \otimes \eta \otimes \eta$  \\
  \hline \hline
  \multirow{6}{*}{4}
  & $M_{1,0,2}$ & $ - 32 \, \alpha \otimes \alpha \otimes \alpha \otimes \eta$  \\
  \cline{2-3}
  & $M_{1,1,1}$ & $ - 16 \, \alpha \otimes \alpha \otimes \alpha \otimes \alpha
                                 + 8 \, \alpha \otimes \eta \otimes \alpha \otimes \eta
                                 + 8 \, \alpha \otimes \eta \otimes \eta \otimes \alpha$  \\
  \cline{2-3}
  & $M_{1,2,0}$ & $ - 8 \, \alpha \otimes \alpha \otimes \alpha \otimes \eta
                                - 8 \, \alpha \otimes \eta \otimes \alpha \otimes \alpha$  \\
  \cline{2-3}
  & $M_{3,0,0}$ & $ - 96 \, \alpha \otimes \eta \otimes \eta \otimes \eta$ 
  \\
   \cline{2-3}
  & $M_{0,2,1}$ & $ 16 \, \alpha \otimes \alpha \otimes \alpha \otimes \alpha$ 
   \\
   \cline{2-3}
  & $M_{0,0,3}$ & $ 96 \, \alpha \otimes \alpha \otimes \alpha \otimes \alpha$ \\
  \hline \hline
\end{tabular}
\caption{Symbols of the all linearly independent functions of the MGEW basis
of \eqn{eq:Mbasis} up to weight $4$.}
\label{tab:basis}
\end{center}
\end{table}

\section{Gluon emission vertex counterterm}
\label{app:counterterm}

\begin{figure}[hb]
\begin{center}
\scalebox{0.6}{\includegraphics{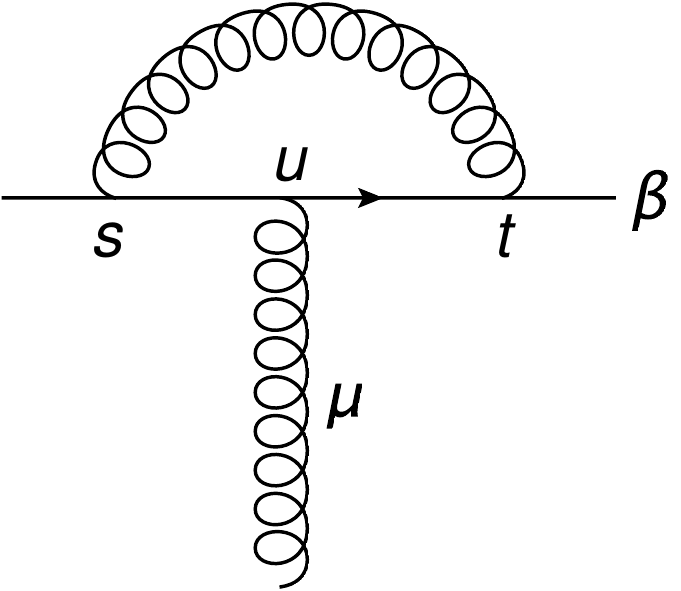}}
\caption{Diagram used for the calculation of the vertex counterterm,
  where $\beta$ is the 4-velocity of the Wilson line, and $\mu$ the
  Lorentz index of the emitted gluon.}
\label{fig:vertexloop}
\end{center}
\end{figure}

In this appendix, we calculate the counterterm for the vertex coupling
a gluon to a Wilson line. Consider the diagram of
figure~\ref{fig:vertexloop},
which shows a gluon being emitted from a
Wilson line at distance parameter $u$, dressed by a boomerang gluon
whose endpoints have distances $s$ and $t$. To identify the singularity associated with the boomerang gluon we must integrate over $s$ and $t$,
keeping~$u$ fixed.
Applying the Feynman rules of eqs.~(\ref{propdef}, \ref{FRemit}), but
removing the exponential regulator for the boomerang gluon as
discussed in section~\ref{sec:13}, we obtain
%\begin{equation}
%  C(a) = T_1^b T_1^a T_1^b T_2^a = \left(C_{R_1}-\frac{1}{2}N_c\right)T_1\cdot T_2,
%\label{Cmushroom}
\begin{equation}
ig_s\,\mu^{\epsilon}\,T_1^b T_1^a T_1^b \beta^\mu \int_0^\infty  du
e^{-imu\sqrt{\beta^2-i\varepsilon}} I(u),
\label{CTcalc1}
\end{equation}
where
\begin{align}
I(u)&={\cal N} g_s^2\mu^{2\epsilon}\,\beta^2\int_0^u ds\int_u^\infty dt
\left[-(t\beta-s\beta)^2\right]^{\epsilon-1}\notag\\
&=-{\cal N}g_s^2\mu^{2\epsilon}\, (-\beta^2)^\epsilon
\int_0^u ds\int_u^\infty dt\,(t-s)^{2\epsilon-2} \nonumber
\\
&=-{\cal N}g_s^2\mu^{2\epsilon}\, (-\beta^2)^\epsilon u^{2\epsilon}
\frac{1}{2\epsilon\,(1-2\epsilon)}\,,
\label{Iudef}
\end{align}
where in the last line we performed the integrals over $s$ and $t$ assuming $0<\epsilon<\frac12$.
%The integral can be straightforwardly carried out to give
%\begin{equation}
%\int_0^u ds\int_u^\infty dt
%\left[-(s-t)^2+i\varepsilon\right]^{\epsilon-1}=
%\frac{u^{2\epsilon}}{2\epsilon(1-2\epsilon)}=\frac{1}{2\epsilon}
%+{\cal O}(\epsilon^0),
%\label{Icalc}
%\end{equation}
We conclude that upon neglecting terms of ${\cal O}(\epsilon^0)$ and above, eq.~(\ref{CTcalc1}) assumes the form of a usual emission vertex from a Wilson line, multiplied by the factor
\begin{equation}
\label{Mushroom_sing_result}
-\frac{g_s^2}{8\pi^2\epsilon}\,  \left(C_{R_1}-\frac{1}{2}N_c\right)+{\cal O}(\epsilon^0)\,,
\end{equation}
where we manipulated the colour factor in eq.~(\ref{CTcalc1}) as in eq.~(\ref{Cmushroom}).
Equation~(\ref{Mushroom_sing_result}) represents the singularity associated with shrinking the boomerang gluon surrounding the emission vertex to a point. To remove this local
singularity, we must introduce a pure counterterm that is the negative
of this result, i.e.
\begin{equation}
Z_{v}=1+Z_{v}^{(1)} \left(C_{R_1}-\frac{1}{2}N_c\right) +\ldots\,,\qquad  \quad
Z_{v}^{(1)}  =+\frac{g_s^2}{8\pi^2\epsilon}\,\, \,,
\label{Zgs1}
\end{equation}
%\begin{equation}
%Z_{v}^{(1)}=+\frac{g_s^2}{8\pi^2\epsilon}\,\,  \left(C_{R_1}-\frac{1}{2}N_c\right)\,,
%\label{Zgs1}
%\end{equation}
where the subscript $v$ stands for the gluon-emission \emph{vertex}.

\section{Calculation of web mixing matrices}
\label{app:Rcalc}

In this appendix we review the replica-trick based algorithm developed in
ref.~\cite{Gardi:2010rn} for calculating the web mixing matrix of
eq.~(\ref{Wnidef}) for a given web. This algorithm is also heavily
used in section~\ref{sec:selfenergy} to prove that self-energy graphs
do not contribute in the overall expression for a boomerang web.

Given a web $W$, we may separate its soft gluon part (i.e. the part of the diagrams 
remaining after the Wilson lines are removed) into a set of $n_c$
connected pieces. We now consider a theory in which there are $N$
non-interacting copies of the gluon fields\footnote{One must also
  replicate any additional matter that can couple to the gluons off
  the Wilson lines, although this is irrelevant for this paper.},
which may connect with the same Wilson lines. Then one may associate a
{\it replica index} $i\in[1,N]$ with each connected subdiagram, such
that these are completely independent. The exponentiated colour factor
of eq.~(\ref{ECFs}) for a given diagram $D$ is then obtained as
follows:
\begin{enumerate}
\item One considers a particular hierarchy $h$ of the $n_c$ replica
  number assignments for all connected pieces of $D$.
\item For each $h$, one reorders the gluon attachments on each Wilson
  line, so that replica indices are increasing (they may be equal, but not ever decrease) along the direction of
  the appropriate 4-velocity $\beta_i$. This reordering leads to a new diagram,
  whose colour factor is labelled by ${\cal R}[D|h]$ in
  ref.~\cite{Gardi:2010rn}.
\item The contribution to the colour factor in the replicated theory
  from each ordering is defined to be
\begin{displaymath}
M_N(h){\cal R}[D|h],
\end{displaymath}
where $M_N(h)$ is the multiplicity of the hierarchy $h$ (we will see
an example in what follows).
\item Finally, one must sum over all possible hierarchies, and take
  the ${\cal O}(N)$ part of the total colour factor thus obtained:
\begin{equation}
\tilde{C}(D)=\left[\sum_h M_N(h){\cal R}[D|h]\right]_{{\cal O}(N)}.
\label{Ctildedef}
\end{equation}
\end{enumerate}
As an illustration of this procedure, let us consider the (1,1,2) web
of figure~\ref{fig:121}. This has two connected pieces (each a single
gluon exchange), so that $n_c=2$. Assigning replica indices $i$ and
$j$ to them, there are three possible hierarchies $h$, which are
listed in table~\ref{tab:121} along with their multiplicities
$M_N(h)$.
\begin{table}[h]
\begin{center}
\begin{tabular}{c|c|c|c|c}
$h$ & ${\cal R}[a|h]$ & ${\cal R}[b|h]$ & $M_N(h)$ & ${\cal
  O}(N)$ part of $M_N(h)$ \\
\hline
$i=j$ & $C(a)$ & $C(b)$ & $N$ & 1\\
$i<j$ & $C(a)$ & $C(a)$ & $\frac12 N(N-1)$ & $-\frac12$ \\
$i>j$ & $C(b)$ & $C(b)$ & $\frac12 N(N-1)$ & $-\frac12$
\end{tabular}
\caption{Replica analysis of the (1,1,2) web of figure~\ref{fig:121}.}
\label{tab:121}
\end{center}
\end{table}
Now consider diagram $(a)$. If the replica indices are equal, then
reordering of the gluons according to their replica indices has no
effect, so that the same diagram is obtained. For the hierarchy $i<j$,
the gluons are already correctly ordered according to replica index,
so that again the same diagram is obtained. Finally, for the hierarchy
$i>j$, the gluons get reordered, producing diagram $(b)$. Adding the
colour factors for each hierarchy weighted according to multiplicity,
one obtains
\begin{equation}
\tilde{C}(a)=C(a)-\frac12\left(C(a)+C(b)\right)=
\frac12 \left(C(a)-C(b)\right).
\label{Ca121}
\end{equation}
Repeating this analysis for diagram $(b)$, one obtains the web mixing
matrix (from eq.~(\ref{ECFs}))
\begin{equation}
R_{(1,1,2)}=\frac12\left(\begin{array}{rr} 1 & -1 \\ -1 & 1
\end{array}\right).
\label{R121}
\end{equation}
Further examples of this technique can be found throughout
section~\ref{sec:selfenergy}.

\section{Calculation of the (1,1,4) web}
\label{app:114calc}

\begin{figure}[htb]
\begin{center}
\scalebox{0.6}{\includegraphics{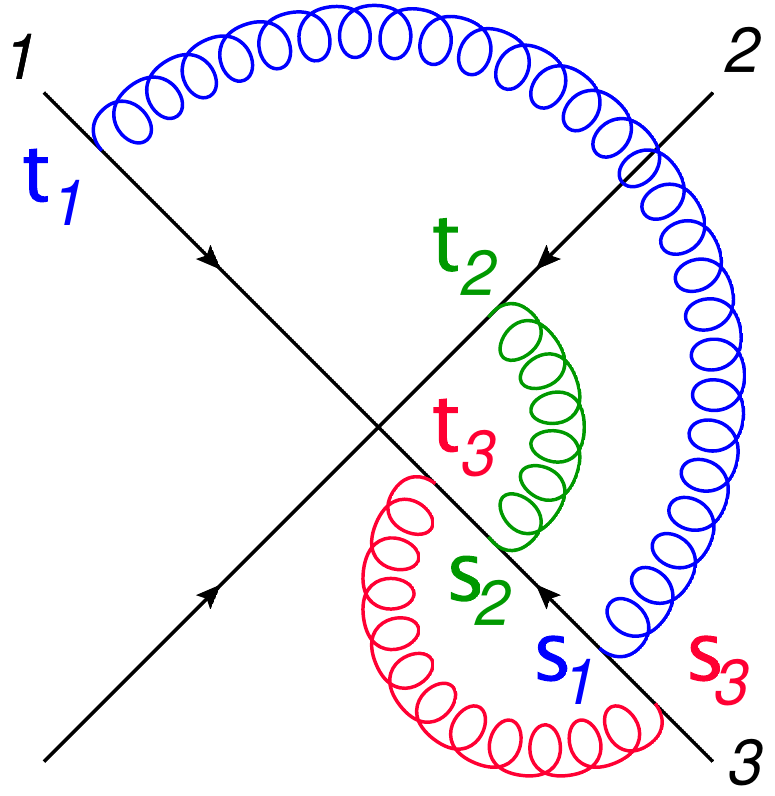}}
\caption{Diagram $(a)$ of the (1,1,4) web of fig.~\ref{fig:114}. The parameters indicating the position of each gluon emission and absorption along the lines are indicated by $s_i$ and $t_i$ for $i=1$ to $3$.}
\label{fig:114a}
\end{center}
\end{figure}

Here we present the calculation of the (1,1,4) web integrals in
section~\ref{sec:114}. Considering the 12 diagrams in fig.~\ref{fig:114}, only
the first six, denoted $(a)$ through $(f)$, enter eq.~(\ref{W114}), while the
remaining six involve self-energy subdiagrams. Our goal here is to compute the kinematic functions ${\cal F}_D$ for diagrams 
$D=a$ through $f$.
Specifically, we wish to express these functions as in eq.~(\ref{kerneldef}) so our first task will be to determine the integration kernels  $\phi_D$  for these diagrams.
We will illustrate the method using diagram $(a)$ shown in figure~\ref{fig:114a} and then present the final results for the others.

As explained in the main text, the exponential regulator of eq.~(\ref{FRemit}) is used on all but the boomerang gluon attachments, leading to the following integral for diagram $(a)$:
\begin{equation}
\begin{split}
\mathcal{F}_a (\alpha_{13} ,\alpha_{23} )
=\, & g_s^6 \bar\mu^{6\epsilon} \mathcal{N}^3(\beta_1 \cdot \beta_3)(\beta_2 \cdot \beta_3)\beta_3^2 \int_{0}^{\infty} ds_1 ds_2 ds_3 dt_1 dt_2 dt_3 \\
&(-(s_1 \beta_3 -t_1 \beta_1)^2 + i\varepsilon)^{\epsilon -1}
(-(s_2 \beta_3 -t_2 \beta_2)^2 + i\varepsilon)^{\epsilon -1} (-(s_3 \beta_3 -t_3 \beta_3)^2 + i\varepsilon)^{\epsilon -1} \\ &e^{-im(s_1+s_2)\sqrt{\beta_3^2-i\varepsilon}-imt_1\sqrt{\beta_1^2-i\varepsilon}-imt_2\sqrt{\beta_2^2-i\varepsilon}} \theta(s_3>s_1)\theta(s_1>s_2)\theta(s_2>t_3)
\end{split}
\end{equation}
where $\mathcal{N}$ is defined in eq.~(\ref{Ndef}) and for  brevity we do not indicate the dependence of $\mathcal{F}_a$ on $\bar{\mu}^2/m^2$ and $\epsilon$ as arguments.
Let us begin by noting (see figure \ref{fig:114a}) that all $s_i$ variables run along line~$3$ while the $t_i$ variables each runs along the respective line $i$, and thus we rescale the variables according to
$s_i\sqrt{\beta_3^2-i\varepsilon} =\sigma_i$ and
$t_i\sqrt{\beta_i^2-i\varepsilon} =\tau_i$
to get:
\begin{equation}
\begin{split}
\mathcal{F}_a(\alpha_{13} ,\alpha_{23})
= \,&g_s^6 \bar\mu^{6\epsilon} \mathcal{N}^3\frac{\gamma_{13}\gamma_{23}}{4} \int_{0}^{\infty} d\sigma_1 d\sigma_2 d\sigma_3 d\tau_1 d\tau_2 d\tau_3\,\, \theta(\sigma_3>\sigma_1)\theta(\sigma_1>\sigma_2)\theta(\sigma_2>\tau_3)\\
&(-\sigma_1^2 -\tau_1^2 +\gamma_{13}\sigma_1\tau_1 + i\varepsilon)^{\epsilon -1}
(-\sigma_2^2 -\tau_2^2 +\gamma_{23}\sigma_2\tau_2 + i\varepsilon)^{\epsilon -1}\\
&(-\sigma_3^2 -\tau_3^2 +2\sigma_3\tau_3 + i\varepsilon)^{\epsilon -1}
e^{-im(\sigma_1 + \sigma_2 + \tau_1 + \tau_2)} \,.
\end{split}
\end{equation}
Now we perform a change of variables,
$\lambda_i = \sigma_i + \tau_i$, for $i=1$ through $3$ along with
\[
z = \frac{\sigma_1}{\sigma_1 + \tau_1},\qquad
y = \frac{\sigma_2}{\sigma_2 + \tau_2},\qquad
x = \frac{\sigma_3}{\sigma_3 + \tau_3}\,.
\]
Next we perform the integral over the boomerang parameter $\lambda_3$ to obtain
\begin{equation}
\begin{split}
\mathcal{F}_a(\alpha_{13} ,\alpha_{23})
=\, &-g_s^6 \bar\mu^{6\epsilon} \mathcal{N}^3\frac{\gamma_{13}\gamma_{23}}{4}
\frac{e^{3i\pi\epsilon}}{2\epsilon}\int_{0}^{\infty} d\lambda_1 d\lambda_2
\int_{0}^{1} dx dy dz \ \lambda_1^{2\epsilon-1}\lambda_2^{2\epsilon-1} \\
&\quad ((2x-1)^2 )^{\epsilon -1} P_{\epsilon}(y, \gamma_{23})P_{\epsilon}(z,
\gamma_{13})\,\,e^{-im(\lambda_1 +\lambda_2)} \\
& \quad\left[ \left( \frac{y\lambda_2}{1-x} \right)^{2\eps} - \left(
    \frac{z\lambda_1}{x}\right)^{2\epsilon} \right] \theta(\lambda_1z>\lambda_2y)\theta(\lambda_2xy>\lambda_1z (1-x)) .
\end{split}
\end{equation}
We can now combine the remaining length parameters of the exponentially regulated gluons,
\begin{align*}
&\lambda = \lambda_1 + \lambda_2
&w = \frac{\lambda_1}{\lambda_1 + \lambda_2}
\end{align*}
and perform the integral over $\lambda$ to obtain
\begin{equation}
\begin{split}
\mathcal{F}_a(\alpha_{13} ,\alpha_{23})
= &-g_s^6 \left(\frac{\bar\mu^2}{m^2}\right)^{3\epsilon} \mathcal{N}^3 \frac{\Gamma(6\epsilon)}{8\epsilon} \int_{0}^{1} dw dy dz \int_{\frac12}^{1} dx  \\
&\quad \left [ \left (\frac{(1-w)y}{1-x} \right )^{2\epsilon} -\left
    (\frac{wz}{x} \right ) ^{2\epsilon} \right ]
w^{2\epsilon-1}(1-w)^{2\epsilon-1}((2x-1)^2 )^{\epsilon -1} \\
&\quad p_{\epsilon}(y, \gamma_{23})p_{\epsilon}(z, \gamma_{13}) \theta
\left(\frac{w}{1-w} > \frac{y}{z}\right) \theta \left(\frac{y}{z}\frac{x}{1-x} > \frac{w}{1-w} \right)\\
= &\ \kappa^3 \Gamma(6\epsilon)  \int_{0}^{1} dy dz\int_{\frac{1}{2}}^{1}dx \, ((2x-1)^2 )^{\epsilon -1} p_\epsilon(y, \gamma_{23})p_\epsilon(z, \gamma_{13}) \phi_a(x,y,z;\epsilon)\,.
\end{split}
\end{equation}
Here we have defined the kernel $\phi_a$ as:
\begin{equation}
\begin{split}
\phi_a (x,y,z;\epsilon) =\,-\frac{1}{\epsilon}\,& \int_{0}^{1} dw \left [ \left (\frac{x}{wz} \right ) ^{-2\epsilon}- \left (\frac{1-x}{(1-w)y} \right )^{-2\epsilon} \right ] w^{2\epsilon-1}(1-w)^{2\epsilon-1} \\
&\quad  \theta
\left(\frac{w}{1-w} > \frac{y}{z}\right) \theta \left(\frac{y}{z}\frac{x}{1-x} > \frac{w}{1-w} \right)\\
=\,-\frac{1}{\epsilon}\,&\int_{0}^{1} dw \left [ \left (\frac{x}{z} \right ) ^{-2\epsilon}w^{4\epsilon-1}(1-w)^{2\epsilon-1}- \left (\frac{1-x}{y} \right ) ^{-2\epsilon} w^{2\epsilon-1}(1-w)^{4\epsilon-1}\right ] \\
&\quad  \theta
\left(\frac{w}{1-w} > \frac{y}{z}\right) \theta \left(\frac{y}{z}\frac{x}{1-x} > \frac{w}{1-w} \right).
\end{split}
\end{equation}
Upon changing the integration variable to $u = \frac{w}{1-w}$,
the integral becomes:
\begin{equation}
\begin{split}
\phi_a (x,y,z;\epsilon)=&  \,-\frac{1}{\epsilon}\,\left (\frac{x}{z} \right ) ^{-2\epsilon}\!
\int_{\frac{y}{z}}^{\frac{y}{z}\frac{x}{1-x}} du
{u}^{4\epsilon-1}(1+{u})^{-6\epsilon}   - \frac{1}{\epsilon}\left
  (\frac{1-x}{y} \right )^{-2\epsilon}\!
\int_{\frac{y}{z}}^{\frac{y}{z}\frac{x}{1-x}} du {u}^{2\epsilon-1}(1+{u})^{-6\epsilon} \\
=& \,-\frac{1}{\epsilon}\,\left (\frac{x}{z} \right ) ^{-2\epsilon} \Bigg \{
\int_{0}^{\frac{y}{z}\frac{x}{1-x}} du\, {u}^{4\epsilon-1}(1+{u})^{-6\epsilon}
-\int_{0}^{\frac{y}{z}} du\, {u}^{4\epsilon-1}(1+{u})^{-6\epsilon} \Bigg \} \\
&\,+\frac{1}{\epsilon}\,\left (\frac{1-x}{y} \right )^{-2\epsilon}\Bigg \{
\int_{0}^{\frac{y}{z}\frac{x}{1-x}} du\, {u}^{2\epsilon-1}(1+{u})^{-6\epsilon}
-\int_{0}^{\frac{y}{z}} du\, {u}^{2\epsilon-1}(1+{u})^{-6\epsilon} \Bigg \} .
\end{split}
\end{equation}
We can then perform this integral over ${u}$ and expand in $\epsilon$ to find:
\begin{equation}\label{eq:phia114}
\begin{split}
&\phi_a (x,y,z;\epsilon)= \\&=-\frac{1}{4\epsilon^2}\left( \frac{y^2}{xz} \right)^{2\epsilon} \left[
  \left( \frac{x}{1-x} \right)^{4\epsilon}\!
  {_2}F_1\left(6\epsilon,4\epsilon;1+4\epsilon;-\frac{y}{z}\frac{x}{1-x}\right) -
  {_2}F_1\left(6\epsilon,4\epsilon;1+4\epsilon;-\frac{y}{z}\right) \right] \\
& +\frac{1}{2\epsilon^2}\left( \frac{y^2}{(1-x)z} \right)^{2\epsilon} \left[
  \left( \frac{x}{1-x} \right)^{2\epsilon}\!
  {_2}F_1\left(6\epsilon,2\epsilon;1+2\epsilon;-\frac{y}{z}\frac{x}{1-x}\right) -
  {_2}F_1\left(6\epsilon,2\epsilon;1+2\epsilon;-\frac{y}{z}\right) \right]  \\
&= \ln^2 \Big (\frac{x}{1-x} \Big ) -\,\epsilon\, \Bigg \{12
\text{Li}_3\left(-\frac{y}{z}\right) -12 \text{Li}_3\left(-\frac{x y}{(1-x)
    z}\right) \\
& \qquad \qquad+12 \ln \left(\frac{x}{1-x}\right) \text{Li}_2\left(-\frac{y}{z}\right)+ 2 \ln^2 \Big (\frac{x}{1-x} \Big ) \ln \left( \frac{z(1-x)}{y^2} \right) \Bigg \}  + {\cal O}(\epsilon^2).
\end{split}
\end{equation}
We now follow the same method to obtain results for the remaining diagrams $(b)$ to $(f)$ of the $(1, 1, 4)$ web. The respective kernels read:
%%%%%%%%%%%%
\begin{align}
\label{kernels114b}
\begin{split}
&\phi_b (x,y,z;\epsilon)= \\&=-\frac{1}{2\epsilon^2}\left( \frac{y^2}{xz} \right)^{2\epsilon} \left[
  {_2}F_1\left(6\epsilon,2\epsilon;1+2\epsilon;-\frac{y}{z}\right) - \left( \frac{1-x}{x} \right)^{2\epsilon}
  {_2}F_1\left(6\epsilon,2\epsilon;1+2\epsilon;-\frac{y}{z}\frac{1-x}{x}\right) \right]\\
& +\frac{1}{4\epsilon^2}\left( \frac{y^2}{(1-x)z} \right)^{2\epsilon} \left[
  {_2}F_1\left(6\epsilon,4\epsilon;1+4\epsilon;-\frac{y}{z}\right) -\! \left(\frac{x}{1-x} \right)^{4\epsilon}\!
  {_2}F_1\left(6\epsilon,4\epsilon;1+4\epsilon;-\frac{y}{z}\frac{1-x}{x}\right) \right]  \\
&= \ln^2 \Big (\frac{x}{1-x} \Big )  -\, \epsilon\, \Bigg \{12
\text{Li}_3\left(-\frac{z}{y}\right)-12 \text{Li}_3\left(-\frac{xz}{(1-x)y}\right) \\
& \qquad\qquad +12 \ln \left(\frac{x}{1-x}\right)
\text{Li}_2\left(-\frac{z}{y}\right)  +2 \ln^2 \Big (\frac{x}{1-x} \Big ) \ln
\left(\frac{y(1-x)}{z^2}\right) \Bigg \}+ {\cal O}(\epsilon^2) ;
\end{split}
\end{align}
\begin{align}
\label{kernels114c}
\begin{split}
&\phi_c (x,y,z;\epsilon)=\\&= \frac{1}{4\epsilon^2} \left( \frac{y^2}{x z} \right)^{2\epsilon}
          \left[2\, {_2}F_1\left(6\epsilon,2\epsilon;1+2\epsilon;-\frac{y}{z}\right)- 2
          \left(\frac{1-x}{x}\right)^{2\epsilon}
          {_2}F_1\left(6\epsilon,2\epsilon;1+2\epsilon;-\frac{y}{z}\frac{1-x}{x}\right)
          \right. \\
  & \qquad \qquad \qquad\left. + \left(\frac{1-x}{x}\right)^{2\epsilon}
    {_2}F_1\left(6\epsilon,4\epsilon;1+4\epsilon;-\frac{y}{z}\frac{1-x}{x}\right) -
    {_2}F_1\left(6\epsilon,4\epsilon;1+4\epsilon;-\frac{y}{z}\right) \right]\\
&=\frac{1}{2\epsilon}\ln \Big (\frac{x}{1-x} \Big )
+\frac{1}{2} \ln \Big (\frac{x}{1-x} \Big ) \ln \Big (\frac{(1-x) y^4}{x^3 z^2} \Big )  \\
&\quad-\,\epsilon\, \Bigg \{ 12 \text{Li}_3\left(-\frac{(1-x) y}{x z}\right)-12
\text{Li}_3\left(-\frac{y}{z}\right)-\frac13 \ln^3\left(\frac{x}{1-x}\right) \\
& \qquad \qquad \qquad
-\ln\left(\frac{x}{1-x}\right) \ln\left( \frac{xz}{y^2}\right)
\ln\left(\frac{x^2 z}{y^2(1-x)}\right)\Bigg \}+ {\cal O}(\epsilon^2);
\end{split}
\end{align}
\begin{align}
\label{kernels114d}
\begin{split}
&\phi_d (x,y,z;\epsilon)= \\&= \frac{1}{4\epsilon^2} \left( \frac{y^2}{z(1-x)} \right)^{2\epsilon}
          \left[2\, {_2}F_1\left(6\epsilon,2\epsilon;1+2\epsilon;-\frac{y}{z}\right)- 2
          \left(\frac{1-x}{x}\right)^{4\epsilon}
          {_2}F_1\left(6\epsilon,2\epsilon;1+2\epsilon;-\frac{y}{z}\frac{1-x}{x}\right)
          \right. \\
  & \qquad \qquad \qquad\left. + \left(\frac{1-x}{x}\right)^{4\epsilon}
    {_2}F_1\left(6\epsilon,4\epsilon;1+4\epsilon;-\frac{y}{z}\frac{1-x}{x}\right) -
    {_2}F_1\left(6\epsilon,4\epsilon;1+4\epsilon;-\frac{y}{z}\right) \right]
\\
&=   \frac{1}{\epsilon}\ln \Big (\frac{x}{1-x} \Big ) + 2 \ln \Big (\frac{x}{1-x} \Big ) \ln \Big (\frac{y^2}{xz} \Big )
-\,\epsilon\,\Bigg \{ 12 \text{Li}_3\left(-\frac{(1-x) y}{x z}\right)\\
& \quad-12
\text{Li}_3\left(-\frac{y}{z}\right) -\frac23 \ln^3\left(\frac{x}{1-x}\right)
%\\& \qquad \qquad \qquad
-2\ln\left(\frac{x}{1-x}\right)\ln^2\left( \frac{xz}{y^2} \right)
\Bigg \}+ {\cal O}(\epsilon^2);
\end{split}
\end{align}
\begin{align}
\label{kernels114e}
\begin{split}
&\phi_e (x,y,z;\epsilon)=\\&=  \frac{1}{4\epsilon^2} \left( \frac{z^2}{y(1-x)} \right)^{2\epsilon}
          \left[2\, {_2}F_1\left(6\epsilon,2\epsilon;1+2\epsilon;-\frac{z}{y}\right)- 2
          \left(\frac{1-x}{x}\right)^{4\epsilon}
          {_2}F_1\left(6\epsilon,2\epsilon;1+2\epsilon;-\frac{z}{y}\frac{1-x}{x}\right)
          \right. \\
  & \qquad \qquad \qquad\left. + \left(\frac{1-x}{x}\right)^{4\epsilon}
    {_2}F_1\left(6\epsilon,4\epsilon;1+4\epsilon;-\frac{z}{y}\frac{1-x}{x}\right) -
    {_2}F_1\left(6\epsilon,4\epsilon;1+4\epsilon;-\frac{z}{y}\right) \right]
\\
&=   \frac{1}{\epsilon}\ln \Big (\frac{x}{1-x} \Big ) + 2 \ln \Big (\frac{x}{1-x} \Big ) \ln \Big (\frac{z^2}{xy} \Big )   -\,\epsilon\,\Bigg \{ 12 \text{Li}_3\left(-\frac{(1-x) z}{x y}\right)-12
\text{Li}_3\left(-\frac{z}{y}\right)
\\& \qquad \qquad \qquad
-\frac23 \ln^3\left(\frac{x}{1-x}\right)
-2\ln\left(\frac{x}{1-x}\right)\ln^2\left( \frac{xy}{z^2} \right)
\Bigg \}+ {\cal O}(\epsilon^2);
\end{split}
\end{align}
\begin{align}
\label{kernels114f}
\begin{split}
&\phi_f (x,y,z;\epsilon)=\\
&=\frac{1}{4\epsilon^2} \left( \frac{z^2}{x y} \right)^{2\epsilon}
          \left[2\, {_2}F_1\left(6\epsilon,2\epsilon;1+2\epsilon;-\frac{z}{y}\right)- 2
          \left(\frac{1-x}{x}\right)^{2\epsilon}
          {_2}F_1\left(6\epsilon,2\epsilon;1+2\epsilon;-\frac{z}{y}\frac{1-x}{x}\right)
          \right. \\
  & \qquad \qquad \qquad\left. + \left(\frac{1-x}{x}\right)^{2\epsilon}
    {_2}F_1\left(6\epsilon,4\epsilon;1+4\epsilon;-\frac{z}{y}\frac{1-x}{x}\right) -
    {_2}F_1\left(6\epsilon,4\epsilon;1+4\epsilon;-\frac{z}{y}\right) \right]\\
&=\frac{1}{2\epsilon}\ln \Big (\frac{x}{1-x} \Big )
+\frac{1}{2} \ln \Big (\frac{x}{1-x} \Big ) \ln \Big (\frac{(1-x) z^4}{x^3 y^2} \Big ) -\,\epsilon\, \Bigg \{ 12 \text{Li}_3\left(-\frac{(1-x) z}{x y}\right)-12
\text{Li}_3\left(-\frac{z}{y}\right)
\\
& \qquad \qquad \qquad
-\frac13 \ln^3\left(\frac{x}{1-x}\right)
-\ln\left(\frac{x}{1-x}\right) \ln\left( \frac{xy}{z^2}\right)
\ln\left(\frac{x^2 y}{z^2(1-x)}\right)\Bigg \}+ {\cal O}(\epsilon^2)   .
\end{split}
\end{align}
It is important to note that the overall degree of divergence in $\epsilon$ of $\phi_a$ and $\phi_b$ is lower than that of the remaining diagrams: this is related to the fact that the latter diagrams are composed of two  subdiagrams that may be shrunk to the origin separately, while the former can only be shrunk upon taking all gluons to the origin simultaneously.  One can also see that each of the $\phi_D$ functions vanish at $x=\frac12$.  This is an example of the general behaviour of the $x_l$ integral for a boomergang gluon which straddle one or more other gluon emissions, as discussed around eq.~(\ref{x_l_integral}).

According to eq.~(\ref{W114}), the contribution to the colour factor
$c_3^{[3,3]}$ requires the combination
\begin{displaymath}
\phi_3=\frac12\left[\phi_a+\phi_b\right],
\end{displaymath}
such that expanding the results of eqs.~(\ref{eq:phia114}) and (\ref{kernels114b}) gives the coefficient
of $\epsilon^{-1}$ to be
\begin{align}
\mathcal{F}_{(1,1,4);3}^{(3,-1)}(\alpha_{13},\alpha_{23}) & = -\frac{1}{6}\left( \frac{g_s^2}{8\pi^2}\right)^3\int_{\frac{1}{2}}^{1} dx \ln ^2 \left( \frac{x}{1-x}\right) [2x-1]^{-2} \int_{0}^{1} dy dz \,p_0(y, \alpha_{23}) p_0(z, \alpha_{13}).
\label{F1143calc}
\end{align}
Using the result
\begin{equation}
\int_{\frac{1}{2}}^{1} dx \ln ^2 \left( \frac{x}{1-x}\right)
    [2x-1]^{-2} = \frac{\pi^2}{3},
\label{xint1}
\end{equation}
one obtains
\begin{align}
\mathcal{F}_{(1,1,4);3}^{(3,-1)}(\alpha_{13},\alpha_{23}) & = -\frac{\pi^2}{18}\left( \frac{g_s^2}{8\pi^2}\right)^3\int_{0}^{1} dy dz \,p_0(y, \alpha_{23}) p_0(z, \alpha_{13}).
\label{F1143calc2}
\end{align}

Next, we consider the kinematic
contribution to the colour factor $c_4^{[3,3]}$, for which
eq.~(\ref{eq:Ctilde114}) dictates we need the combination
\begin{align}
\phi_4(x,y,z;\epsilon)&=\frac12\left[- \phi_a + \phi_b + \phi_c + \phi_d - \phi_e -\phi_f
\right]\notag\\
&=-\frac92\ln\left(\frac{x}{1-x}\right)\ln\left(\frac{z}{y}\right)
+ \frac12 B(x,y,z)\epsilon+{\cal O}(\epsilon^2),
\label{phi4def}
\end{align}
where we have expanded in $\epsilon$ in the second line and defined
\begin{align}
B(x,y,z)
% &= 12 \text{Li}_3\left(\frac{(x-1) z}{x y}\right)-12 \text{Li}_3\left(\frac{(x-1) y}{x z}\right)+24 \ln \left(\frac{x}{1-x}\right) \text{Li}_2\left(-\frac{y}{z}\right) \notag\\
% &+3 \ln ^2(1-x) \ln \left(\frac{y}{z}\right)-15 \ln ^2(x) \ln \left(\frac{y}{z}\right)+12 \ln (1-x) \ln (y) \ln (x z) \notag\\
% &-12 \ln (x) \ln (z) \ln (y(1-x))+15 \ln \left(\frac{x}{1-x}\right) \ln ^2(y)-2 \pi ^2 \ln (y(1-x))\notag\\
% &-3 \ln \left(\frac{x}{1-x}\right) \ln ^2(z) +2 \pi ^2 \ln (x z)+6 \ln ^2(y) \ln (z)-6 \ln (y) \ln ^2(z)\notag\\
% &-2 \ln ^3(y)+2 \ln ^3(z) \notag\\
&= \ln\left( \frac{y}{z} \right) \left(  3\ln^2\left(\frac{x}{1-x}\right) -9
  \ln\left(\frac{x}{1-x}\right)\ln\left(\frac{x^2}{yz}\right) -2
  \ln^2\left(\frac{y}{z} \right) -2\pi^2 \right) \notag \\
& \qquad +12\ln\left(\frac{x}{1-x}\right) \left( \text{Li}_2\left( -\frac{y}{z}\right)
  - \text{Li}_2\left( -\frac{z}{y}\right) \right)   \\
  & \qquad -12 \left( \text{Li}_3\left(-\frac{(1-x)y}{xz}\right) -
  \text{Li}_3\left(-\frac{(1-x)z}{xy}\right) \right). \notag
% \\
% &= 12 \text{Li}_3\left(-\frac{(1-x) z}{x y}\right)-12 \text{Li}_3\left(-\frac{(1-x) y}{x z}\right)+24 \ln \left(\frac{x}{1-x}\right) \text{Li}_2\left(-\frac{y}{z}\right) \notag\\
% & \qquad + 3\ln^2\left( \frac{x}{1-x} \right)\ln\left( \frac{y}{z}\right) -2\ln^3\left(
%   \frac{y}{z} \right) + 2\pi^2 \ln\left( \frac{xz}{(1-x)y} \right) \\
% & \qquad+
%   3\ln\left(\frac{x}{1-x}\right) \ln\left( \frac{y}{z}\right) \ln\left(
%   \frac{y^5 z}{x^6} \right).\notag
\label{Bdef}
\end{align}
This can be simplified further using relations between dilogarithms, but we
leave it in this form to keep the antisymmetry between $y$ and $z$ manifest.  $\phi_4$ enters the unrenormalized kinematic factor accompanying
$c_4^{[3,3]}$ through the integral
\begin{equation}
\mathcal{F}^{(3)}_{(1,1,4);4}(\alpha_{13},\alpha_{23}) =\kappa^3\Gamma(6\epsilon) \int_{\frac{1}{2}}^{1} dx \int_{0}^{1} dy dz [(2x-1)^2]^{\epsilon-1}p_\epsilon(y, \alpha_{23})p_\epsilon(z, \alpha_{13})\phi_4(x,y,z;\epsilon).
\label{F4calc2}
\end{equation}
One can check that $\lim_{x\to\frac12}B(x,y,z)=0$ and hence the singularity at $x=\frac12$ is
integrable for $\epsilon>0$ and we find
\begin{align}
  \label{eq:postBintegration}
    \mathcal{F}^{(3)}_{(1,1,4);4}&(\alpha_{13},\alpha_{23}) \notag \\
=& \,\kappa^3 \Gamma(6\epsilon) \int_{0}^{1} dy dz
    p_\epsilon(y, \alpha_{23})p_\epsilon(z, \alpha_{13}) \notag \\
    & \quad  \times \left[ -\frac{9}{4\epsilon}
      \ln\left( \frac{z}{y} \right) - 6\text{Li}_2\left(
          -\frac{z}{y} \right) +6 \text{Li}_2\left( -\frac{y}{z} \right) -
        \frac94 \ln\left( \frac{z}{y} \right) (2+\ln(yz)) + {\cal
          O}(\epsilon)\right]  \notag \\
       =\, & \left( \frac{g_s^2}{8\pi^2} \right)^3 \int_{0}^{1} dy dz\,  p_0(y,
    \alpha_{23}) p_0(z, \alpha_{13}) \\
    & \quad \times \left[ \frac{3}{8\epsilon^2}\ln\left( \frac{z}{y} \right) +
      \frac{1}{8\epsilon} \left( 8\text{Li}_2\left(
           -\frac{z}{y} \right) -8 \text{Li}_2\left( -\frac{y}{z} \right)
          \right. \right. \notag  \\
  & \qquad  \left. \left. +3\ln\left( \frac{z}{y} \right) \left(2+\ln(yz) + \ln
    \left( q(y,\alpha_{23}) q(z,\alpha_{13}) \right) +3\ln\left( \frac{\mu^2}{m^2}\right) \right) \right) \right] + {\cal
          O}(\epsilon^0) . \notag
\end{align}
We note that the antisymmetry between $y$ and $z$ remains manifest in this expression.

% Putting
% things together, we find that the subtraction terms assume the form
% \begin{align}
% \Delta W^{(3)}_{(1,1,4)}&=\left(\frac{\alpha_s}{4\pi}\right)^3
% \frac12 \int_{0}^{1} dy dz p_0(y,
% \alpha_{23})p_0(z, \alpha_{13}) \Big \{  2 \ln q(y, \alpha_{23})
% \ln y+2 \ln q(y, \alpha_{23}) \ln z\notag\\
% &-\ln ^2q(y, \alpha_{23})+4 \ln q(y, \alpha_{23})-2
% \ln q(z, \alpha_{13}) \ln y-2 \ln q(z, \alpha_{13}) \ln z\notag\\
% &+\ln ^2q(z, \alpha_{13})-4 \ln q(z, \alpha_{13})+2 \ln ^2(y)+4 \ln y-2
% \ln ^2z-4 \ln z  \Big \}.
% \label{DelW114}
% \end{align}
% Combining this with the web itself and relabelling the integration
% variables, one finds the subtracted web kernel of eq.~(\ref{G114res}).

\section{Steps in the calculation of the (2,4) web}
\setcounter{tocdepth}{0}
\label{app:24calc}

The calculation of the (2,4) web is presented in section~\ref{sec:24}. Here we collect some intermediate results, first considering the expansion of the Appell $F_1$ function appearing in
eq.~(\ref{phia24c}) (appendix~\ref{app:Appell}) and then evaluating the polylogarithmic integrals in eq.~(\ref{FtoG24}) (appendix~\ref{app:Polylog24}).

\subsection{Expansion of the Appell $F_1$ function entering the (2,4) web}
\label{app:Appell}

Here, we explain how to expand the Appell $F_1$ function appearing in
eq.~(\ref{phia24c}), as a series in the dimensional regularisation
parameter $\epsilon$. Starting with the well-known one-dimensional integral representation 
%(see e.g.~\cite{Kniehl:2011ym})
\begin{equation}
F_1(a,b,b',c;x,y) = \frac{\Gamma(c)}{\Gamma(a)\Gamma(c-a)} \int_{0}^{1}du \frac{u^{a-1}(1-u)^{c-a-1}}{(1-ux)^{b}(1-uy)^{b'}},
\label{F1int}
\end{equation}
one has
\begin{equation}
F_1(2\epsilon,6\epsilon,-2\epsilon,1+2\epsilon; x, y) = \frac{\Gamma(1+2\epsilon)}{\Gamma(2\epsilon)\Gamma(1)} \int_{0}^{1}du \frac{u^{2\epsilon-1}}{(1-ux)^{6\epsilon}(1-uy)^{-2\epsilon}}.
\label{1}
\end{equation}
Application of the Feynman parameter trick
\begin{equation}
\frac{1}{A^{\alpha}B^{\beta}} = \frac{\Gamma(\alpha+\beta)}{\Gamma(\alpha)\Gamma(\beta)} \int_{0}^{1} dt \frac{t^{\alpha-1}(1-t)^{\beta-1}}{(At+B(1-t))^{\alpha+\beta}}
\label{Feynman}
\end{equation}
gives %\Chris{(CDW: I interchanged $x\leftrightarrow y$ in the equation below, which can be seen to be correct by comparing with eq.~(\ref{Feynman}).)}
\begin{equation}
\begin{split}
F_1(2\epsilon,6\epsilon,-2\epsilon,1+2\epsilon; x, y)= \frac{\Gamma(1+2\epsilon)\Gamma(4\epsilon)}{\Gamma(2\epsilon)\Gamma(-2\epsilon)\Gamma(6\epsilon)}&\int_{0}^{1}dt \hspace{0.1cm} t^{-2\epsilon-1}(1-t)^{6\epsilon-1} \int_{0}^{1}du \\
&u^{2\epsilon-1} [1-u(ty+(1-t)x)]^{-4\epsilon}.
\end{split}
\end{equation}
The $u$ integral produces a hypergeometric function, which can be
expanded in $\epsilon$ using \texttt{HypExp}, such that one obtains
\begin{equation}
\begin{split}
F_1(2\epsilon,6\epsilon,-2\epsilon,1+2\epsilon; x, y) &= \frac{1}{2\epsilon}\frac{\Gamma(1+2\epsilon)\Gamma(4\epsilon)}{\Gamma(2\epsilon)\Gamma(-2\epsilon)\Gamma(6\epsilon)}\int_{0}^{1}dt \hspace{0.1cm} t^{-2\epsilon-1}(1-t)^{6\epsilon-1}\\
&+ 4\epsilon \frac{\Gamma(1+2\epsilon)\Gamma(4\epsilon)}{\Gamma(2\epsilon)\Gamma(-2\epsilon)\Gamma(6\epsilon)}\int_{0}^{1}dt \hspace{0.1cm} t^{-2\epsilon-1}(1-t)^{6\epsilon-1} \text{Li}_2(Q) +\cdots
\end{split}
\label{F1calc}
\end{equation}
where %\Chris{(CDW: The original result for $Q$ below was incorrect, but has now been fixed.)}
\begin{equation}
Q=ty+(1-t)x.
\label{Qdef}
\end{equation}
The integral in the first term in eq.~(\ref{F1calc}) yields a complete beta
function. For the second term, the $t$ integral produces a pole in
$\epsilon$ for $t\rightarrow 0,1$, in which cases the argument of the
dilogarithm reduces to $y$ and $x$ respectively. One then obtains
\begin{equation}
F_1(2\epsilon,6\epsilon,-2\epsilon,1+2\epsilon, x, y) =
1 - \epsilon^2 \Big(4 \text{Li}_2(y) - 12 \text{Li}_2(x)\Big)+{\cal O}(\epsilon^3).
\label{F1exp}
\end{equation}

\subsection{Evaluation of the (2,4) web polylogarithmic integrals}
\label{app:Polylog24}

Our task in this appendix is to explicitly evaluate the $y$ and $z$ integrals in eq.~(\ref{FtoG24}) with the kernel of  ${\cal G}^{(3)}_{(2,4);1}(y,z,\alpha_{12})$ given in eq.~(\ref{G24}). As these integrals do not directly lend themselves to the form of eq.~(\ref{eq:Mbasis}), we will first evaluate them in terms of Goncharov polylogarithms, and then show that the result can be expressed in terms of the basis functions.  

As a first step, one may straightforwardly integrate those terms in
the kernel ${\cal G}^{(3)}_{(2,4);1}$ that depend only upon $y$ or $z$
individually (where the transformation $z\rightarrow 1-z$,
$y\rightarrow 1-y$ is useful in the latter case). One then obtains
\begin{align}
F^{(3)}_{(2,4);1}(\alpha_{12})&=\frac43\bigg\{r^2(\alpha_{12})
 \Big[8M_{0,1,1}(\alpha_{12})
- 2M_{1,1,1}(\alpha_{12}) - 2M_{0,2,1}(\alpha_{12})\Big]+I(\alpha_{12})\bigg\},
\label{F324calc1}
\end{align}
where we have denoted the remaining integrals $I(\alpha)=I_1(\alpha)+I_2(\alpha)+I_3(\alpha)$ with
\begin{align}
I_1(\alpha)&=2\int_{0}^{1}dy\,dz\,p_0(y,\alpha)p_0(z,\alpha)
\text{Li}_2\Big(\frac{z}{y}\frac{1-y}{1-z}\Big)
\theta(y>z)\notag\\
I_2(\alpha)&=\int_{0}^{1}dy\,dz\,p_0(y,\alpha)p_0(z,\alpha)
\text{ln}\Big(\frac{z}{1-z}\Big)
\text{ln}\Big(\frac{q(y,\alpha)}{y^2}\Big)\theta(y>z)\notag\\
I_3(\alpha)&=-\int_{0}^{1}dy\,dz\,p_0(y,\alpha)p_0(z,\alpha)
\text{ln}\Big(\frac{y}{1-y}\Big) \text{ln}\Big(
\frac{q(z,\alpha)}{z^2}\Big)\theta(y>z).
\label{I123}
\end{align}
To calculate $I_1$, we may first rewrite the dilogarithm function as a
Goncharov polylogarithm~\cite{Goncharov:1998kja}:
\begin{equation}
\text{Li}_2\Big(\frac{z}{y}\frac{1-y}{1-z}\Big) =
-\text{G}_{0,1}\Big(\frac{z}{y}\frac{1-y}{1-z}\Big).
\label{di}
\end{equation}
The right-hand side has a non-trivial function of $y$ and $z$ in the
final argument, and one may replace this with a sum of simpler
Goncharov polylogarithms involving only $y$ or $z$~\cite{Duhr:2019tlz}:
\begin{equation}
\begin{split}
\text{Li}_2\Big(\frac{z}{y}\frac{1-y}{1-z}\Big) =& -\text{G}_1(y)\big(\text{G}_0(z)-\text{G}_1(z)\big) + \text{G}_0(y)\big(\text{G}_0(z)-\text{G}_1(z)\big) -\text{G}_{0,0}(y)\\
&-\text{G}_{0,0}(z)+ \text{G}_{0,1}(z)+\text{G}_{0,z}(y)+\text{G}_{1,0}(y)+\text{G}_{1,0}(z)\\
&-\text{G}_{1,1}(z)-\text{G}_{1,z}(y) + 2\,\zeta_2 - i\pi \big(\text{G}_1(y)+\text{G}_0(z)-\text{G}_0(y)-\text{G}_1(z)\big).
\label{fin}
\end{split}
\end{equation}
Having rewritten the dilogarithm according to eq.~(\ref{fin}),
we may use partial fractioning to rewrite the propagator functions in
eq.~(\ref{I123}) to be linear in the integration variable, as in
eq.~(\ref{p0_part_fracs}).  Each term in the integral can now be
carried out using standard methods, and one finds
\begin{equation}
\begin{split}
I_1(\alpha)
=\,& \,2r^2(\alpha) \bigg[- \frac{\pi^4}{36} -2\text{G}_0(\alpha)\zeta_3- \pi^2 \Big(\frac{2}{3}\text{G}_{0,-1}(\alpha) + \frac{2}{3}\text{G}_{0,1}(\alpha)\Big) + 8\text{G}_{0,-1,-1,0}(\alpha)\\
&- 8\text{G}_{0,-1,0,0}(\alpha) + 8\text{G}_{0,-1,1,0}(\alpha) - 8\text{G}_{0,0,-1,0}(\alpha)- 8\text{G}_{0,0,1,0}(\alpha) \\
&+ 8\text{G}_{0,1,-1,0}(\alpha) - 8\text{G}_{0,1,0,0}(\alpha)+ 8\text{G}_{0,1,1,0}(\alpha)  \bigg]\, .
\end{split}
\label{I1res}
\end{equation}
For the other integrals in eq.~(\ref{I123}), one may apply similar
techniques to obtain
\begin{align}
\text{ln}\Big(\frac{z}{1-z}\Big) \text{ln}\Big(\frac{q(y,\alpha)}{y^2}\Big) =\,& \big(\text{G}_0(z)-\text{G}_1(z)\big)\bigg(\text{G}_{{1}/({1-\alpha})}(y) + \text{G}_{{\alpha}/({\alpha-1})}(y)- 2\,\text{G}_0(y)\bigg),
\label{yf}
\end{align}
and similarly for $y\leftrightarrow z$. We then find
\begin{equation}
\begin{split}
I_2(\alpha)+I_3(\alpha) =\,& r^2(\alpha) \bigg[-\zeta_4+8\,\zeta_3\Big(\text{G}_{-1}(\alpha)+\text{G}_1(\alpha)\Big)
\\&
+8\zeta_2\Big(\text{G}_{-1,0}(\alpha)+2\text{G}_{0,-1}(\alpha)-\text{G}_{0,0}(\alpha)+2\text{G}_{0,1}(\alpha)+\text{G}_{1,0}(\alpha)\Big)\\
& -16\text{G}_{-1,0,-1,0}(\alpha)+16\text{G}_{-1,0,0,0}(\alpha)-16\text{G}_{-1,0,1,0}(\alpha)-32\text{G}_{0,-1,-1,0}(\alpha)\\
&+32\text{G}_{0,-1,0,0}(\alpha)-32\text{G}_{0,-1,1,0}(\alpha)+48\text{G}_{0,0,-1,0}(\alpha)-32\text{G}_{0,0,0,0}(\alpha)\\
&+48\text{G}_{0,0,1,0}(\alpha)-32\text{G}_{0,1,-1,0}(\alpha)+32\text{G}_{0,1,0,0}(\alpha)-32\text{G}_{0,1,1,0}(\alpha)\\
&-16\text{G}_{1,0,-1,0}(\alpha)+16\text{G}_{1,0,0,0}(\alpha)-16\text{G}_{1,0,1,0}(\alpha)\bigg]\, ,
\end{split}
\end{equation}
so that the total contribution from all integrals in eq.~(\ref{I123})
is
\begin{equation}
\begin{split}
I(\alpha) =\,& r^2(\alpha)
\Big[-6\,\zeta_4+4\,\zeta_3\Big(2\text{G}_{-1}(\alpha)+2\text{G}_{1}(\alpha)-\text{G}_0(\alpha)\Big)\\
&+8\zeta_2\Big(\text{G}_{-1,0}(\alpha)+\text{G}_{0,-1}(\alpha)-\text{G}_{0,0}(\alpha)+\text{G}_{0,1}(\alpha)+\text{G}_{1,0}(\alpha)\Big) \\
&-16\text{G}_{-1,0,-1,0}(\alpha)+16\text{G}_{-1,0,0,0}(\alpha)-16\text{G}_{-1,0,1,0}(\alpha)-16\text{G}_{0,-1,-1,0}(\alpha)\\
&+16\text{G}_{0,-1,0,0}(\alpha)-16\text{G}_{0,-1,1,0}(\alpha)+32\text{G}_{0,0,-1,0}(\alpha)-32\text{G}_{0,0,0,0}(\alpha)\\
&+32\text{G}_{0,0,1,0}(\alpha)-16\text{G}_{0,1,-1,0}(\alpha)+16\text{G}_{0,1,0,0}(\alpha)-16\text{G}_{0,1,1,0}(\alpha)\\
&-16\text{G}_{1,0,-1,0}(\alpha)+16\text{G}_{1,0,0,0}(\alpha)-16\text{G}_{1,0,1,0}(\alpha)\Big]\,.
\label{Itot}
\end{split}
\end{equation}
To express this result in terms of basis functions, we may take its
symbol, obtaining
\begin{equation}
\mathcal{S}\Big(\frac{I(\alpha)}{r^2(\alpha)}\Big)  = - 16\, \alpha \otimes \eta \otimes \alpha \otimes \eta - 16\, \alpha \otimes \eta \otimes \eta \otimes \alpha,
\label{itot}
\end{equation}
where $\eta$ has been defined in eq.~(\ref{alphabet}). Comparing this
with the symbols of the basis functions in table~\ref{tab:basis}, we construct the ansatz
\begin{equation}
\begin{split}
I(\alpha) =\,& r^2(\alpha) \Big[ A\,M_{0,0,0}^4(\alpha) + B\,M_{0,0,0}(\alpha)M_{0,0,2}(\alpha)
%\\&
+C\,M_{0,0,0}(\alpha)M_{0,2,0}(\alpha)
- 2\,M_{1,1,1}(\alpha)
\Big] \, .
\label{Iansatz}
\end{split}
\end{equation}
Fitting the coefficients using the expressions for the basis functions given
in appendix~\ref{app:functions} we find
\begin{equation}
A=-\frac{1}{24},\quad B=0,\quad C=-\frac12\,.
\label{ABC}
\end{equation}
The final result in eq.~(\ref{F241_main_text}) follows upon substituting the coefficients eq.~(\ref{ABC}) into eq.~(\ref{Iansatz}) and using the latter in eq.~(\ref{F324calc1}) along with the relations in eqs.~(\ref{M021Relation}) and~(\ref{M003}).
%\begin{align}
%F^{(3)}_{(2,4);1}(\alpha_{12})&=\frac43r^2(\alpha_{12})\Big[8M_{0,1,1}(\alpha_{12})
%- 4M_{1,1,1}(\alpha_{12}) - 2M_{0,2,1}(\alpha_{12}) - \frac{1}{24}
%\,M_{0,0,0}^4(\alpha_{12}) \notag\\
%&\qquad \qquad \qquad - \frac{1}{2}\,M_{0,0,0}(\alpha_{12})M_{0,2,0}(\alpha_{12})\Big].
%\label{F241}
%\end{align}

\bibliography{refs.bib}

\end{document}